%% file: main.tex
\documentclass[aps,prd,twocolumn,floatfix,10pt,amssymb,amsfont,amsmath,superscriptaddress,float,nofootinbib,longbibliography]{revtex4-2}

\usepackage{preamb}

\input{tikz}

\begin{document}

\title{Dualities in one-dimensional quantum lattice models:\\ symmetric Hamiltonians and matrix product operator intertwiners}

\author{Laurens Lootens}
\email{laurens.lootens@ugent.be}
\affiliation{\it Department of Physics and Astronomy, Ghent University, Krijgslaan 281, 9000 Gent, Belgium}
\author{Clement Delcamp}
\affiliation{\it Max Planck Institute for the Physics of Complex Systems, N\"othnitzer Stra{\ss}e 38, 01187 Dresden, Germany}
\author{Gerardo Ortiz}
\affiliation{\it Department of Physics, Indiana University, Bloomington, IN 47405, United States of America}
\author{Frank Verstraete}
\affiliation{\it Department of Physics and Astronomy, Ghent University, Krijgslaan 281, 9000 Gent, Belgium}

\begin{abstract}
\noindent
We present a systematic recipe for generating and classifying duality transformations in one-dimensional quantum lattice systems. Our construction emphasizes the role of global symmetries, including those described by (non)-abelian groups but also more general categorical symmetries. These symmetries can be realized as matrix product operators which allow the extraction of a fusion category that characterizes the algebra of all symmetric operators commuting with the symmetry. Known as the bond algebra, its explicit realizations are classified by module categories over the fusion category. A duality is then defined by a pair of distinct module categories giving rise to dual realizations of the bond algebra, as well as dual Hamiltonians. Symmetries of dual models are in general distinct but satisfy a categorical Morita equivalence. A key novelty of our categorical approach is the explicit construction of matrix product operators that intertwine dual bond algebra realizations at the level of the Hilbert space, and in general map local order operators to non-local string-order operators. We illustrate this approach for known dualities such as Kramers-Wannier, Jordan-Wigner, Kennedy-Tasaki and the IRF-vertex correspondence, a new duality of the $t$-$J_z$ chain model, and dualities in models with the exotic Haagerup symmetry. Finally, we comment on generalizations to higher dimensions.

\end{abstract}

\maketitle

\input{_Introduction.tex}

\input{_Hamiltonian.tex}

\input{_Examples.tex}

\input{_Discussion.tex}

\bigskip\bigskip\noindent
{\bf Acknowledgements:} We are grateful to Paul Fendley, Romain Couvreur, Rui-Zhen Huang, Alex Turzillo, Victor Albert, J\"urgen Fuchs and Christoph Schweigert for interesting discussions and useful comments, and to Jacob Bridgeman for providing the categorical data of the Haagerup module categories. This work is supported by the Research Foundation Flanders (FWO) via grant nr. G087918N and G0E1820N. LL is supported by a PhD grant from the FWO. G.O acknowledges support from the US Department of Energy grant DE-SC0020343.

\bigskip \bigskip
\input{main.bbl}

\end{document}

%% file: tikz.tex
\pgfdeclarelayer{backA}
\pgfdeclarelayer{middleA}
\pgfdeclarelayer{frontA}
\pgfdeclarelayer{backB}
\pgfdeclarelayer{middleB}
\pgfdeclarelayer{frontB}
\pgfdeclarelayer{backC}
\pgfdeclarelayer{middleC}
\pgfdeclarelayer{frontC}
\pgfsetlayers{backC,middleC,frontC,main,backB,middleB,frontB,backA,middleA,frontA}

\tikzset{->1-/.style={decoration={
			markings,
			mark=at position #1 with {\arrow{Triangle[fill=black, width=3.5pt, length=3.5pt]}}},postaction={decorate}}}
			
\tikzset{->2-/.style={decoration={
			markings,
			mark=at position #1 with {\arrow{Triangle[fill=BrickRed, width=3.5pt, length=3.5pt]}}},postaction={decorate}}}
\tikzset{->3-/.style={decoration={
			markings,
			mark=at position #1 with {\arrow{Triangle[fill=BlueViolet!80, width=3.5pt, length=3.5pt]}}},postaction={decorate}}}
			
\tikzset{
	every picture/.append style={
		line join=round,
		line cap=round,
		line width = 0.7pt
	}
}

\tikzset{mask point/.style={ transform shape, sloped, 
minimum width=20pt, minimum height=4pt, inner sep=0pt, ultra thin, font=\tiny}}

\tikzset{
    clip even odd rule/.code={\pgfseteorule},
    invclip/.style={clip,insert path=[clip even odd rule]{
    [reset cm](-\maxdimen,-\maxdimen)rectangle(\maxdimen,\maxdimen)
    }}} 
    
\newcommand\clipIntersection[1]{
    (#1.north west) -- (#1.north east) -- (#1.south east) -- (#1.south west) -- (#1.north west)
}

\tikzstyle{dot}=[dash pattern= on 0.6pt off 2.1pt]
\tikzstyle{obj}=[line width = 0.7pt, line join=round, line cap=round, shorten >= 1pt, shorten <= 1pt]
\tikzstyle{speObj}=[line width = 0.7pt, line join=round, line cap=join, shorten >= 1pt, shorten <= 1pt, color=BrickRed]
\tikzstyle{mod}=[color=BlueViolet!80,line width = 0.7pt, line join=round, line cap=round]
\tikzstyle{mor}=[draw=none, fill=gray!22]

\newcommand\clipAroundNode[2]{
    \begin{pgfscope}
        \begin{scope}[overlay]
            \path[invclip] \clipIntersection{#1};#2
        \end{scope}
    \end{pgfscope}
}

\newcommand\clipAroundThreeNodes[4]{
    \begin{pgfscope}
        \begin{scope}[overlay]
            \path[invclip] \clipIntersection{#1} \clipIntersection{#2} \clipIntersection{#3};#4
        \end{scope}
    \end{pgfscope}
}

\newcommand{\middleCoo}[6]{
    \path (#1) --coordinate[pos=0.5](#3) (#2);
    \path (#1) --coordinate[pos={0.5-\sp}](#4) (#2);
    \path (#1) --coordinate[pos={0.5+\sp}](#5) (#2);
}

\newcommand{\barycentre}[8]{
    \middleCoo{#1}{#2}{A}{}{}{}{#8};
    \middleCoo{#2}{#3}{B}{}{}{}{#8};
    \middleCoo{#1}{#3}{C}{}{}{}{#8};
    \path[name path=PA#3] (A) -- (#3);
    \path[name path=PB#1] (B) -- (#1);
    \path[name path=PC#2] (C) -- (#2);
    \path [name intersections={of=PA#3 and PB#1,by={#4}}];
    \path (#1) --coordinate[pos=0.84](#5) (#4);
    \path (#2) --coordinate[pos=0.84](#6) (#4);
    \path (#3) --coordinate[pos=0.84](#7) (#4);
}

\newcommand{\barycentreSq}[9]{
    \middleCoo{#1}{#2}{A1}{A2}{A3}{};
    \middleCoo{#3}{#4}{B1}{B2}{B3}{};
    \middleCoo{#1}{#3}{C1}{C2}{C3}{};
    \middleCoo{#2}{#4}{D1}{D2}{D3}{};
    \middleCoo{A1}{B1}{#5}{}{};
    \path[name path=PA2B2] (A2) -- (B2);
    \path[name path=PA3B3] (A3) -- (B3);
    \path[name path=PC2D2] (C2) -- (D2);
    \path[name path=PC3D3] (C3) -- (D3);       
    \path [name intersections={of=PA2B2 and PC2D2,by={#6}}];
    \path [name intersections={of=PA2B2 and PC3D3,by={#8}}];
    \path [name intersections={of=PA3B3 and PC2D2,by={#7}}];
    \path [name intersections={of=PA3B3 and PC3D3,by={#9}}];
}

\newcommand{\centerarc}[5]{
    \draw[#1] ($(#2)+({#5*cos(#3)},{#5*sin(#3)})$) arc (#3:#4:#5);
}

\newcommand{\flowChart}[0]{
	\begin{tikzpicture}[scale=0.6,baseline={([yshift=-1ex]current bounding box.center)}]
    	\node (a) at (0,0) {$\mathbb H_0$};
    	\node (b1) at (3,3) {$\mc C_0,\mc M_0$};
    	\node (b2) at (3,0) {$\mc C_0',\mc M_0'$};
    	\node (b3) at (3,-3) {$\phantom{\mc C_0,\mc M_0}$}; 
        \path (b2) -- (b3) node[font=\Large, midway, sloped] {$\dots$};
    	
    	\draw[->,>=Triangle] (a) -- (b1.west);
    	\draw[->,>=Triangle] (a) -- (b2.west);
    	\draw[->,>=Triangle,dotted] (a) -- (b3.west);
    	
    	\node (c1) at (6,3) {$\mc D$};
    	\node (c2) at (6,0) {$\mc D'$};
    	\draw[->,>=Triangle] (b1) -- (c1);
    	\draw[->,>=Triangle] (b2) -- (c2);
    	
    	\node (d11) at (9,4) {$\mc C_1,\mc M_1$};
    	\node (d12) at (9,3) {$\mc C_2,\mc M_2$};
    	\node (d13) at (9,2) {$\phantom{\mc C_2,\mc M_2}$};
    	\path (d12) -- (d13) node[font=\footnotesize, pos=.5, sloped] {$\dots \;\;$};
    	\draw[->,>=Triangle] (c1) -- (d11.west);
    	\draw[->,>=Triangle] (c1) -- (d12.west);
    	\draw[->,>=Triangle,dotted] (c1) -- (d13.west);
    	
    	\node (d21) at (9,1) {$\mc C'_1,\mc M'_1$};
    	\node (d22) at (9,0) {$\mc C'_2,\mc M'_2$};
    	\node (d23) at (9,-1) {$\phantom{\mc C'_2,\mc M'_2}$};
    	\path (d22) -- (d23) node[font=\footnotesize, pos=.5, sloped] {$\dots \;\;$};
    	\draw[->,>=Triangle] (c2) -- (d21.west);
    	\draw[->,>=Triangle] (c2) -- (d22.west);
    	\draw[->,>=Triangle,dotted] (c2) -- (d23.west);
    	
    	\node (e11) at (12,4) {$\mathbb H_1$};
    	\node (e12) at (12,3) {$\mathbb H_2$};
    	\draw[->,>=Triangle] (d11) -- (e11);
    	\draw[->,>=Triangle] (d12) -- (e12);
    	
    	\node (e21) at (12,1) {$\mathbb H_1'$};
    	\node (e22) at (12,0) {$\mathbb H_2'$};
    	\draw[->,>=Triangle] (d21) -- (e21);
    	\draw[->,>=Triangle] (d22) -- (e22);

	\end{tikzpicture}
}

\newcommand{\heuristics}[3]{
	\begin{tikzpicture}[scale=0.6,baseline={([yshift=-1ex]current bounding box.center)}]
	    \def\l{1.2};
        \ifthenelse{#1 = 1}{
            \coordinate (a) at (0,0);
            \draw[] (0,0) -- (0,-\l);
            \draw[] (0,0) -- ({cos(30)*\l},{+sin(30)*\l});
            \draw[] (0,0) -- (-{cos(30)*\l},{+sin(30)*\l});
            \node[draw=black, fill=white, circle,inner sep=3.6pt] at (0,0) {};
            \node[] at (0,0) {${\sss #2}$};
            \centerarc{#3}{a}{-10}{-170}{0.6*\l};
            \ifthenelse{\equal{#3}{speObj}}{
                \node[regular polygon,regular polygon sides=4,draw,fill=BrickRed!30,inner sep=1.6pt] at (0,-0.6*\l) {};
            }{}
            \ifthenelse{\equal{#3}{mod}}{
                \node[regular polygon,regular polygon sides=4,draw,fill=BlueViolet!30,inner sep=1.6pt] at (0,-0.6*\l) {};
            }{}

        }{}
        \ifthenelse{#1 = 2}{
        	\coordinate (a) at (0,0);
            \draw[] (0,0) -- (0,-\l);
            \draw[] (0,0) -- ({cos(30)*\l},{sin(30)*\l});
            \draw[] (0,0) -- (-{cos(30)*\l},{sin(30)*\l});
            \node[draw=black, fill=white, circle,inner sep=3.6pt] at (0,0) {};
            \node[] at (0,0) {${\sss #2}$};
            \centerarc{#3}{a}{190}{-10}{0.6*\l};
            \ifthenelse{\equal{#3}{speObj}}{
                \node[regular polygon,regular polygon sides=4,draw,fill=BrickRed!30,inner sep=1.6pt,rotate=30] at ({0.6*cos(30)*\l},{0.6*sin(30)*\l}) {};
                \node[regular polygon,regular polygon sides=4,draw,fill=BrickRed!30,inner sep=1.6pt,rotate=60] at ({-0.6*cos(30)*\l},{0.6*sin(30)*\l}) {};
            }{}
            \ifthenelse{\equal{#3}{mod}}{
                \node[regular polygon,regular polygon sides=4,draw,fill=BlueViolet!30,inner sep=1.6pt,rotate=30] at ({0.6*cos(30)*\l},{0.6*sin(30)*\l}) {};
                \node[regular polygon,regular polygon sides=4,draw,fill=BlueViolet!30,inner sep=1.6pt,rotate=60] at ({-0.6*cos(30)*\l},{0.6*sin(30)*\l}) {};
            }{}
        }{}
        \ifthenelse{#1 = 0}{
            \coordinate (a) at (0,0);
            \draw[] (0,0) -- (0,1.3*\l);
            \draw[] (0,0) -- ({cos(30)*\l},{-sin(30)*\l});
            \draw[] (0,0) -- (-{cos(30)*\l},{-sin(30)*\l});
            \draw[] (0,1.3*\l) -- ($({cos(30)*\l},{+sin(30)*\l})+(0,1.3*\l)$);
            \draw[] (0,1.3*\l) -- ($(-{cos(30)*\l},{+sin(30)*\l})+(0,1.3*\l)$);
            \node[draw=black, fill=white, circle,inner sep=3.6pt] at (0,0) {};
            \node[] at (0,0) {${\sss \tilde #2}$};
            \node[draw=black, fill=white, circle,inner sep=3.6pt] at (0,1.3*\l) {};
            \node[] at (0,1.3*\l) {${\sss #2}$};
        }{}
	\end{tikzpicture}
}

\newcommand{\heuristicsIsing}[4]{
	\begin{tikzpicture}[scale=0.6,baseline={([yshift=-1ex]current bounding box.center)}]
	    \def\l{1.2};
	    \ifthenelse{\equal{#4}{XX}}{
	        \def\inep{1pt};
	    }{
	        \def\inep{1pt};
	    }
        \ifthenelse{#1 = 1}{
            \coordinate (a) at (0,0);
            \draw[] (0,0) -- (0,-\l);
            \draw[] (0,0) -- ({cos(30)*\l},{+sin(30)*\l});
            \draw[] (0,0) -- (-{cos(30)*\l},{+sin(30)*\l});
            \node[draw=black, fill=white, circle,inner sep=3.6pt] at (0,0) {};
            \node[] at (0,0) {${\sss #2}$};
            \ifthenelse{\equal{#3}{speObj}}{
                \node[rectangle,draw,fill=BrickRed!30,inner sep=\inep] at (0,-0.6*\l) {${\sss #4}$};
            }{}
            \ifthenelse{\equal{#3}{mod}}{
                \node[regular polygon,regular polygon sides=4,draw,fill=BlueViolet!30,inner sep=\inep] at (0,-0.6*\l) {${\sss #4}$};
            }{}

        }{}
        \ifthenelse{#1 = 2}{
            \draw[] (0,0) -- (0,-\l);
            \draw[] (0,0) -- ({cos(30)*\l},{sin(30)*\l});
            \draw[] (0,0) -- (-{cos(30)*\l},{sin(30)*\l});
            \node[draw=black, fill=white, circle,inner sep=3.6pt] at (0,0) {};
            \node[] at (0,0) {${\sss #2}$};
            \ifthenelse{\equal{#3}{speObj}}{
                \node[rectangle,draw,fill=BrickRed!30,inner sep=\inep,rotate=-60] at ({0.6*cos(30)*\l},{0.6*sin(30)*\l}) {${\sss #4}$};
                \node[rectangle,draw,fill=BrickRed!30,inner sep=\inep,rotate=60] at ({-0.6*cos(30)*\l},{0.6*sin(30)*\l}) {${\sss #4}$};
            }{}
            \ifthenelse{\equal{#3}{mod}}{
                \node[regular polygon,regular polygon sides=4,draw,fill=BlueViolet!30,inner sep=\inep,rotate=30] at ({0.6*cos(30)*\l},{0.6*sin(30)*\l}) {${\sss #4}$};
                \node[regular polygon,regular polygon sides=4,draw,fill=BlueViolet!30,inner sep=\inep,rotate=60] at ({-0.6*cos(30)*\l},{0.6*sin(30)*\l}) {${\sss #4}$};
            }{}
        }{}
	\end{tikzpicture}
}

\newcommand{\heuristicsProjector}[0]{
	\begin{tikzpicture}[scale=1,baseline={([yshift=-1ex]current bounding box.center)}]
    	\def\r{8pt};
    	\def\l{3.5};
    	\def\sp{0.16};
    	\coordinate (a1) at (0,0);
        \coordinate (a2) at (\l/2,0);
    	\coordinate (a3) at (0,\l/2);
    	\coordinate (a4) at (\l/2,\l/2);

        \clip (-0.55,\l/4-0.55) rectangle (3*\l/2+0.55,\l/2+0.3);
        
    	\middleCoo{a1}{a2}{m12}{m12-1}{m12-2}{\sp};
    	\middleCoo{a1}{a3}{m13}{m13-1}{m13-3}{\sp};
    	\middleCoo{a2}{a4}{m24}{m24-2}{m24-4}{\sp};
    	\middleCoo{a3}{a4}{m34}{m34-3}{m34-4}{\sp};
    	\barycentreSq{a1}{a2}{a3}{a4}{b1234}{b1234-1}{b1234-2}{b1234-3}{b1234-4};
    	
    	\foreach \x in {0,\l/2,\l}{
        	\draw ($(b1234)+(\x,-0.2)$) -- ($(m34)+(\x,0.25)$);
            \draw[rounded corners=\r] 
        	    ($(m13-3)+(\x,0)$) -- ($(b1234-3)+(\x,0)$) -- ($(m34-3)+(\x,0.25)$)
        	    ($(m24-4)+(\x,0)$) -- ($(b1234-4)+(\x,0)$) -- ($(m34-4)+(\x,0.25)$);
        }
        \draw[dash pattern= on 0.2pt off 2.3pt] 
            (m13-3) -- ($(m13-3)+(-0.3,0)$)
            ($(m24-4)+(\l,0)$) -- ($(m24-4)+(\l+0.3,0)$);
        \node[yshift=-4pt] at (\l/4,\l/4-0.2) {${\sss \msf i-\frac{3}{2}}$};
        \node[yshift=-4pt] at (\l/4+\l/2,\l/4-0.2) {${\sss \msf i-\frac{1}{2}}$};
        \node[yshift=-4pt] at (\l/4+\l,\l/4-0.2) {${\sss \msf i+\frac{1}{2}}$};

        \node[above] at (m13-3) {${\sss \msf i-2}$};
        \node[above] at ($(m13-3)+(\l/2,0)$) {${\sss \msf i-1}$};
        \node[above] at ($(m13-3)+(\l,0)$) {${\sss \msf i}$};
        \node[above] at ($(m13-3)+(3*\l/2,0)$) {${\sss \msf i+1}$};
        
        \node[yshift=-3.5pt, rectangle, fill=white, draw=black, inner sep=1.2pt] at ($(m34)+(0,0)$) {${\sss \;\, \mathbb P_\mc G \;\,}$};
        \node[yshift=-3.5pt, rectangle, fill=white, draw=black, inner sep=1.2pt] at ($(m34)+(\l/2,0)$) {${\sss \;\, \mathbb P_\mc G \;\,}$};
        \node[yshift=-3.5pt, rectangle, fill=white, draw=black, inner sep=1.2pt] at ($(m34)+(\l,0)$) {${\sss \;\, \mathbb P_\mc G \;\,}$};
	\end{tikzpicture}
}

\newcommand{\heuristicsKWMPO}[6]{
	\begin{tikzpicture}[scale=1,baseline={([yshift=-1ex]current bounding box.center)}]
        \def\x{0.5};
        \draw[draw,fill=BlueViolet!30] (0,0) rectangle (\x,\x);
        \draw (\x/2,0) -- ($(\x/2,0)+(0,-0.6)$);
        \draw (0,\x/2) -- ($(0,\x/2)+(-0.6,0)$);
        \draw (\x,\x/2) -- ($(\x,\x/2)+(0.6,0)$);
        \foreach \s in {0.25,0.5,0.75}{
            \draw (\s*\x,\x) -- ($(\s*\x,\x)+(0,0.6)$);
        }
        \node[left] at ($(0,\x/2)+(-0.6,0)$) {${\sss #1}$};
        \node[below] at ($(\x/2,0)+(0,-0.6)$) {${\sss #2}$};
        \node[right] at ($(\x,\x/2)+(0.6,0)$) {${\sss #3}$};
        \node[left] at ($(0.25*\x,\x)+(0,0.6)$) {${\sss #4}$};
        \node[above] at ($(0.5*\x,\x)+(0,0.6)$) {${\sss #5}$};  \node[right] at ($(0.75*\x,\x)+(0,0.6)$) {${\sss #6}$};
	\end{tikzpicture}
}

\newcommand{\heuristicsJWMPO}[4]{
	\begin{tikzpicture}[scale=1,baseline={([yshift=-1ex]current bounding box.center)}]
        \def\x{0.5};
        \draw[draw,fill=BlueViolet!30] (0,0) rectangle (\x,\x);
        \draw (\x/2,0) -- ($(\x/2,0)+(0,-0.6)$);
        \draw (0,\x/2) -- ($(0,\x/2)+(-0.6,0)$);
        \draw (\x,\x/2) -- ($(\x,\x/2)+(0.6,0)$);
        \draw (0.5*\x,\x) -- ($(0.5*\x,\x)+(0,0.6)$);
        \node[left] at ($(0,\x/2)+(-0.6,0)$) {${\sss #1}$};
        \node[below] at ($(\x/2,0)+(0,-0.6)$) {${\sss #2}$};
        \node[right] at ($(\x,\x/2)+(0.6,0)$) {${\sss #3}$};
        \node[above] at ($(0.5*\x,\x)+(0,0.6)$) {${\sss #4}$};
	\end{tikzpicture}
}

\newcommand{\monoidalAssociator}[1]{
	\begin{tikzpicture}[scale=0.6,baseline={([yshift=-1ex]current bounding box.center)}]
	    \def\l{1.5};
        \ifthenelse{#1 = 1}{
            \draw[obj] (0,0) -- (0,1);
            \draw[obj] (0,1) -- (-\l,3);
            \draw[obj] (0,1) -- (\l,3);
            \draw[obj] (-\l/2,2) -- (0,3);
            \node[mor, circle, inner sep=3.3pt] at (-\l/2,2) {};
            \node at (-\l/2,2) {${\sss i}$};
            \node[mor, circle, inner sep=3.3pt] at (0,1) {};
            \node at (0,1) {${\sss l}$};
            \node at (-0.6,1.3) {${\sss \mu}$};
        }{}
        \ifthenelse{#1 = 2}{
            \draw[obj] (0,0) -- (0,1);
            \draw[obj] (0,1) -- (-\l,3);
            \draw[obj] (0,1) -- (\l,3);
            \draw[obj] (\l/2,2) -- (0,3);
            \node[mor, circle, inner sep=3.3pt] at (0,1) {};
            \node at (0,1) {${\sss k}$};
            \node[mor, circle, inner sep=3.3pt] at (\l/2,2) {};
            \node at (\l/2,2) {${\sss j}$};
            \node at (0.6,1.3) {${\sss \nu}$};
        }{}
        \node[above] at (-\l,3) {${\sss \alpha}$};
        \node[above] at (0,3) {${\sss \beta}$};
        \node[above] at (\l,3) {${\sss \gamma}$};
        \node[below] at (0,0) {${\sss \delta}$};
	\end{tikzpicture}
}

\newcommand{\moduleAssociator}[1]{
	\begin{tikzpicture}[scale=0.6,baseline={([yshift=-1ex]current bounding box.center)}]
        \ifthenelse{#1 = 1}{
            \draw[mod] (0,0) -- (0,3);
            \draw[obj] (1,3) -- (0,2);
            \draw[obj] (2,3) -- (0,1);
            \node[mor, circle, inner sep=3.3pt] at (0,2) {};
            \node at (0,2) {${\sss i}$};
            \node[mor, circle, inner sep=3.3pt] at (0,1) {};
            \node at (0,1) {${\sss l}$};
            \node[left] at (0,1.5) {${\sss C}$};
        }{}
        \ifthenelse{#1 = 2}{
            \draw[mod] (0,0) -- (0,3);
            \draw[obj, name path=P1] (2,3) -- (0,1);
            \path[name path=P2] (1,3) -- (1,0);
            \path [name intersections={of=P1 and P2,by={sp}}];
            \draw[obj] (1,3) -- (sp);
            \node[mor, circle, inner sep=3.3pt] at (sp) {};
            \node at (sp) {${\sss j}$};
            \node[mor, circle, inner sep=3.3pt] at (0,1) {};
            \node at (0,1) {${\sss k}$};
            \node at (0.6,1.2) {${\sss \gamma}$};
        }{}
        \node[above] at (0,3) {${\sss A}$};
        \node[above] at (1,3) {${\sss \alpha}$};
        \node[above, yshift=-1pt] at (2,3) {${\sss \beta}$};
        \node[below] at (0,0) {${\sss B}$};
	\end{tikzpicture}
}

\newcommand{\gauging}[1]{
	\begin{tikzpicture}[scale=1,baseline={([yshift=-.5ex]current bounding box.center)}]                 
	    \def\l{3.5}   ;
        \ifthenelse{#1 = 1}{
            \draw[->1-=0.7] (0,0) -- (0,\l/4);
            \draw[->1-=0.5] (\l/4,-\l/4) -- (\l/4,0);
            \draw[->1-=0.5] (-\l/4,-\l/4) -- (-\l/4,0);
            \draw[] (\l/4+0.3,0) -- (\l/4,0);
            \draw[->1-=0.55] (\l/4,0) -- (0,0);
            \draw[->1-=0.66] (0,0) -- (-\l/4,0);
            \draw (-\l/4,0) -- (-\l/4-0.3,0);
            \node[mor, circle, inner sep=3pt] at (0,0) {};
            \node at (0,0) {${\sss 1}$};
            \node[draw=none, fill=black, circle, inner sep=1.5pt] at (\l/4,0) {};
            \node[draw=none, fill=black, circle, inner sep=1.5pt] at (-\l/4,0) {};
            \node[yshift=-5pt,xshift=11pt] at (0,\l/4) {${\sss g_1g_2}$};
            \node[xshift=10pt, yshift=5pt] at (-\l/4,-\l/4) {${\sss g_1^{-1}}$};
            \node[yshift=5pt,xshift=8pt] at (\l/4,-\l/4) {${\sss g_2}$};
        }{}
        \ifthenelse{#1 = 2}{
            \draw[->1-=0.7] (0,0) -- (0,\l/4);
            \draw[->1-=0.7] (\l/2,0) -- (\l/2,\l/4);
            \draw[->1-=0.5] (\l/4,-\l/4) -- (\l/4,0);
            \draw[->1-=0.5] (-\l/4,-\l/4) -- (-\l/4,0);
            \draw[] (\l/2+0.4,0) -- (\l/2,0);
            \draw[->1-=0.65] (\l/2,0) -- (\l/4,0);
            \draw[->1-=0.55] (\l/4,0) -- (0,0);
            \draw[->1-=0.65] (0,0) -- (-\l/4,0);
            \draw (-\l/4,0) -- (-\l/4-0.3,0);
            \node[mor, circle, inner sep=3pt] at (0,0) {};
            \node at (0,0) {${\sss 1}$};
            \node[mor, circle, inner sep=3pt] at (\l/2,0) {};
            \node at (\l/2,0) {${\sss 1}$};
            \node[draw=none, fill=black, circle, inner sep=1.5pt] at (\l/4,0) {};
            \node[draw=none, fill=black, circle, inner sep=1.5pt] at (-\l/4,0) {};
            \node[yshift=16pt, draw=black, fill=white, circle,inner sep=0pt] at (0,0) {${\sss R_h}$};
            \node[yshift=16pt, draw=black, fill=white, circle,inner sep=0pt] at (\l/2,0) {${\sss L_h}$};
        }{}
        \ifthenelse{#1 = 3}{
            \draw[->1-=0.7] (0,0) -- (0,\l/4);
            \draw[->1-=0.7] (\l/2,0) -- (\l/2,\l/4);
            \draw[->1-=0.5] (\l/4,-\l/4) -- (\l/4,0);
            \draw[->1-=0.5] (-\l/4,-\l/4) -- (-\l/4,0);
            \draw[] (\l/2+0.4,0) -- (\l/2,0);
            \draw[->1-=0.65] (\l/2,0) -- (\l/4,0);
            \draw[->1-=0.55] (\l/4,0) -- (0,0);
            \draw[->1-=0.65] (0,0) -- (-\l/4,0);
            \draw (-\l/4,0) -- (-\l/4-0.3,0);
            \node[mor, circle, inner sep=3pt] at (0,0) {};
            \node at (0,0) {${\sss 1}$};
            \node[mor, circle, inner sep=3pt] at (\l/2,0) {};
            \node at (\l/2,0) {${\sss 1}$};
            \node[draw=none, fill=black, circle, inner sep=1.5pt] at (\l/4,0) {};
            \node[draw=none, fill=black, circle, inner sep=1.5pt] at (-\l/4,0) {};
            \node[yshift=-14pt, draw=black, fill=white, circle,inner sep=0pt] at (\l/4,0) {${\sss R_h}$};
        }{}        
	\end{tikzpicture}
}

\newcommand{\gaugingHigherD}[1]{
	\begin{tikzpicture}[scale=1,baseline={([yshift=-.5ex]current bounding box.center)},z={(0.37,-0.45)}]                 
	    \def\l{3.5};
	    \begin{pgfonlayer}{frontA}
            \draw[->1-=0.6] (\l/4,0,\l/2) -- (\l/4,0,3*\l/4);
            \draw (\l/4,0,3*\l/4) -- (\l/4,0,3*\l/4+0.5);
            \draw[->1-=0.7] (\l/4,0,3*\l/4) -- (\l/4,\l/4,3*\l/4) node[mask point, pos=0.6](mk3){};
            \node[mor, circle, inner sep=3pt] at (\l/4,0,3*\l/4) {};
            \node at (\l/4,0,3*\l/4) {${\sss 1}$}; 
            \ifthenelse{#1 = 1}{
                \node[yshift=16pt, draw=black, fill=white, circle,inner sep=0pt] at (0,0,\l/2) {${\sss R_h}$};
                \node[yshift=16pt, draw=black, fill=white, circle,inner sep=0pt] at (\l/2,0,\l/2) {${\sss L_h}$};
                \node[yshift=16pt, draw=black, fill=white, circle,inner sep=0pt] at (\l/4,0,3*\l/4) {${\sss R_h}$};
                \node[yshift=16pt, draw=black, fill=white, circle,inner sep=0pt] at (\l/4,0,\l/4) {${\sss L_h}$};
            }{
            }
	    \end{pgfonlayer}
        \begin{pgfonlayer}{frontB}
            \clipAroundNode{mk3}{
                \draw[->1-=0.7] (0,0,\l/2) -- (0,\l/4,\l/2) node[mask point, pos=0.8](mk4){};
                \draw[->1-=0.7] (\l/2,0,\l/2) -- (\l/2,\l/4,\l/2);
                \draw[->1-=0.5] (\l/4,-\l/4,\l/2) -- (\l/4,0,\l/2);
                \draw[->1-=0.5] (-\l/4,-\l/4,\l/2) -- (-\l/4,0,\l/2);
                \draw[] (\l/2+0.4,0,\l/2) -- (\l/2,0,\l/2);
                \draw[->1-=0.45] (\l/2,0,\l/2) -- (\l/4,0,\l/2);
                \draw[->1-=0.55] (\l/4,0,\l/2) -- (0,0,\l/2) node[mask point, pos=0.8](mk5){};
                \draw[->1-=0.65] (0,0,\l/2) -- (-\l/4,0,\l/2);
                \draw (-\l/4,0,\l/2) -- (-\l/4-0.3,0,\l/2);
                \node[mor, circle, inner sep=3pt] at (0,0,\l/2) {};
                \node at (0,0,\l/2) {${\sss 1}$};
                \node[mor, circle, inner sep=3pt] at (\l/2,0,\l/2) {};
                \node at (\l/2,0,\l/2) {${\sss 1}$};
                \node[draw=none, fill=black, circle, inner sep=1.5pt] at (\l/4,0,\l/2) {};
                \node[draw=none, fill=black, circle, inner sep=1.5pt] at (-\l/4,0,\l/2) {};
            }
            \ifthenelse{#1 = 1}{
            }{
                \node[yshift=-14pt, draw=black, fill=white, circle,inner sep=0pt] at (\l/4,0,\l/2) {${\sss R_h}$};
            }
        \end{pgfonlayer}
        \begin{pgfonlayer}{middleB}
            \draw (-\l/4,0,-0.4) -- (-\l/4,0,0);
            \draw[->1-=0.6] (-\l/4,0,0) -- (-\l/4,0,\l/4);
            \draw[->1-=0.7] (-\l/4,0,\l/4) -- (-\l/4,0,\l/2);
            \draw (-\l/4,0,\l/2) -- (-\l/4,0,\l/2+0.4);
            \draw[->1-=0.7] (-\l/4,0,\l/4) -- (-\l/4,\l/4,\l/4) node[mask point, pos=0.6](mk1){};
            \node[mor, circle, inner sep=3pt] at (-\l/4,0,\l/4) {};
            \node at (-\l/4,0,\l/4) {${\sss 1}$};
            \draw (\l/4,0,-0.4) -- (\l/4,0,0);
            \draw[->1-=0.6] (\l/4,0,0) -- (\l/4,0,\l/4);
            \draw[->1-=0.7] (\l/4,0,\l/4) -- (\l/4,0,\l/2);
            \draw[->1-=0.7] (\l/4,0,\l/4) -- (\l/4,\l/4,\l/4) node[mask point, pos=0.6](mk2){};
            \node[mor, circle, inner sep=3pt] at (\l/4,0,\l/4) {};
            \node at (\l/4,0,\l/4) {${\sss 1}$};
        \end{pgfonlayer}
        \begin{pgfonlayer}{backB}
            \clipAroundThreeNodes{mk1}{mk4}{mk5}{
                \draw[->1-=0.7] (0,0) -- (0,\l/4);
                \draw[->1-=0.5] (\l/4,-\l/4) -- (\l/4,0);
                \draw[->1-=0.5] (-\l/4,-\l/4) -- (-\l/4,0);
                \draw[->1-=0.68] (\l/4,0) -- (0,0);
                \draw[->1-=0.45] (0,0) -- (-\l/4,0);
                \draw (-\l/4,0) -- (-\l/4-0.3,0);
                \draw (\l/4,0) -- (\l/4+0.24,0);
                \node[mor, circle, inner sep=3pt] at (0,0) {};
                \node at (0,0) {${\sss 1}$};
                \node[draw=none, fill=black, circle, inner sep=1.5pt] at (\l/4,0) {};
                \node[draw=none, fill=black, circle, inner sep=1.5pt] at (-\l/4,0) {};
            }
        \end{pgfonlayer}
	\end{tikzpicture}
}

\newcommand{\pent}[1]{
	\begin{tikzpicture}[scale=1,baseline={([yshift=-.5ex]current bounding box.center)}]
    	\def\r{8pt};
    	\def\l{3.5};
    	\def\sp{0.08};
    	
    	\ifthenelse{#1=1}{
            \def\shix{1.1};
            \def\shiy{-2};
        	\clip (0.65,-2.15) rectangle (4, 1.8);

            \def\x{0};
            \def\y{0};
            \coordinate (a1) at ($(0,0)+(\x,\y)$);
            \coordinate (a2) at ($(\l,0)+(\x,\y)$);
            \coordinate (a3) at ($(\l/2,{sqrt(3)*\l/2})+(\x,\y)$);   	            	
        	\middleCoo{a1}{a2}{m12}{m12-1}{m12-2}{\sp};
        	\middleCoo{a2}{a3}{m23}{m23-2}{m23-3}{\sp};
        	\middleCoo{a1}{a3}{m13}{m13-1}{m13-3}{\sp};
        	\barycentre{a1}{a2}{a3}{b123}{b123-1}{b123-2}{b123-3}{\sp};	
            \draw[obj,->1-=0.6] (m12) --node[pos=0.5,xshift=-5pt] {${\sss \mu}$} (b123);
            \draw[obj,->1-=0.6] (b123) --node[pos=0.35,yshift=4pt, sloped] {${\sss \alpha}$} (m13);
            \draw[obj,->1-=0.6] (b123) --node[pos=0.35,yshift=4pt, sloped] {${\sss \beta}$} (m23);
            \draw[mor] 
                (m12-1) to[out=90, in=90, looseness=1.5] (m12-2)
                (m13-1) to[out=-30, in=-30, looseness=1.5] (m13-3)
                (m23-2) to[out=-150, in=-150, looseness=1.5] (m23-3);
            \node[mor, circle, inner sep=4pt] at (b123) {};
            \draw[mod, rounded corners=\r] 
                (m12-1) -- (b123-1) -- (m13-1)
                (m12-2) -- (b123-2) -- (m23-2)
                (m23-3) -- (b123-3) -- (m13-3);
            \node[yshift=-5pt, xshift=-8pt] at (b123-1) {${\sss A}$};
            \node[yshift=-5pt, xshift=8pt] at (b123-2) {${\sss C}$};
            \node[yshift=9pt] at (b123-3) {${\sss B}$}; 
            \node[] at (b123) {${\sss i}$};
            \node[yshift=3.5pt] at (m12) {${\sss p}$};
            \node[yshift=-1.5pt, xshift=2.5pt] at (m13) {${\sss a}$};
            \node[yshift=-1.5pt, xshift=-2.5pt] at (m23) {${\sss b}$};
            
            \def\x{\shix};
            \def\y{\shiy};
            \coordinate (a1) at ($(0,0)+(\x,\y)$);
            \coordinate (a2) at ($(\l,0)+(\x,\y)$);
            \coordinate (a3) at ($(\l/2,{sqrt(3)*\l/2})+(\x,\y)$);   	            	
        	\middleCoo{a1}{a2}{m12}{m12-1}{m12-2}{\sp};
        	\middleCoo{a2}{a3}{m23}{m23-2}{m23-3}{\sp};
        	\middleCoo{a1}{a3}{m13}{m13-1}{m13-3}{\sp};
        	\barycentre{a1}{a2}{a3}{b123}{b123-1}{b123-2}{b123-3}{\sp};      	
            \draw[obj,->1-=0.6] (m12) --node[pos=0.5,xshift=-5pt] {${\sss \delta}$} (b123);
            \draw[obj,->1-=0.6] (b123) --node[pos=0.35,yshift=4pt, sloped] {${\sss \mu}$} (m13);
            \draw[obj,->1-=0.6] (b123) --node[pos=0.35,yshift=4pt, sloped] {${\sss \gamma}$} (m23);
            \draw[mor] 
                (m12-1) to[out=90, in=90, looseness=1.5] (m12-2)
                (m13-1) to[out=-30, in=-30, looseness=1.5] (m13-3)
                (m23-2) to[out=-150, in=-150, looseness=1.5] (m23-3);
            \node[mor, circle, inner sep=4pt] at (b123) {};
            \draw[mod, rounded corners=\r] 
                (m12-1) -- (b123-1) -- (m13-1)
                (m12-2) -- (b123-2) -- (m23-2)
                (m23-3) -- (b123-3) -- (m13-3);
            \node[yshift=-5pt, xshift=-8pt] at (b123-1) {${\sss A}$};
            \node[yshift=-5pt, xshift=8pt] at (b123-2) {${\sss D}$};
            \node[yshift=9pt] at (b123-3) {${\sss C}$}; 
            \node[] at (b123) {${\sss l}$};
            \node[yshift=3.5pt] at (m12) {${\sss d}$};
            \node[yshift=-1.5pt, xshift=2.5pt] at (m13) {${\sss p}$};
            \node[yshift=-1.5pt, xshift=-2.5pt] at (m23) {${\sss c}$};

            \draw[mor] (m13-3) -- (m13-1) to[out=150, in=-90, looseness=1] ($(m12-1)+(-\shix,-\shiy)$) -- ($(m12-2)+(-\shix,-\shiy)$) to[in=150, out=-90, looseness=1] (m13-3);
            \draw[mod] (m13-3) to[out=150, in=-90, looseness=1] ($(m12-2)+(-\shix,-\shiy)$);
            \draw[mod] (m13-1) to[out=150, in=-90, looseness=1] ($(m12-1)+(-\shix,-\shiy)$);
        }{}
        
    	\ifthenelse{#1=2}{
            \def\shix{-1.1};
            \def\shiy{-2};
         	\clip (-0.5,-2.15) rectangle (2.8, 1.8);
    	    
    	   \def\x{0};
    	   \def\y{0};
            \coordinate (a1) at ($(0,0)+(\x,\y)$);
            \coordinate (a2) at ($(\l,0)+(\x,\y)$);
            \coordinate (a3) at ($(\l/2,{sqrt(3)*\l/2})+(\x,\y)$);	    	            	
        	\middleCoo{a1}{a2}{m12}{m12-1}{m12-2}{\sp};
        	\middleCoo{a2}{a3}{m23}{m23-2}{m23-3}{\sp};
        	\middleCoo{a1}{a3}{m13}{m13-1}{m13-3}{\sp};
        	\barycentre{a1}{a2}{a3}{b123}{b123-1}{b123-2}{b123-3}{\sp};
            \draw[obj,->1-=0.6] (m12) --node[pos=0.5,xshift=-5pt] {${\sss \nu}$} (b123);
            \draw[obj,->1-=0.6] (b123) --node[pos=0.35,yshift=4pt, sloped] {${\sss \beta}$} (m13);
            \draw[obj,->1-=0.6] (b123) --node[pos=0.35,yshift=4pt, sloped] {${\sss \gamma}$} (m23);
            \draw[mor] 
                (m12-1) to[out=90, in=90, looseness=1.5] (m12-2)
                (m13-1) to[out=-30, in=-30, looseness=1.5] (m13-3)
                (m23-2) to[out=-150, in=-150, looseness=1.5] (m23-3);
            \node[mor, circle, inner sep=4pt] at (b123) {};
            \draw[mod, rounded corners=\r] 
                (m12-1) -- (b123-1) -- (m13-1)
                (m12-2) -- (b123-2) -- (m23-2)
                (m23-3) -- (b123-3) -- (m13-3);
            \node[yshift=-5pt, xshift=-8pt] at (b123-1) {${\sss B}$};
            \node[yshift=-5pt, xshift=8pt] at (b123-2) {${\sss D}$};
            \node[yshift=9pt] at (b123-3) {${\sss C}$}; 
            \node[] at (b123) {${\sss j}$};
            \node[yshift=3.5pt] at (m12) {${\sss q}$};
            \node[yshift=-1.5pt, xshift=2.5pt] at (m13) {${\sss b}$};
            \node[yshift=-1.5pt, xshift=-2.5pt] at (m23) {${\sss c}$};
            
            \def\x{\shix};
            \def\y{\shiy};
            \coordinate (a1) at ($(0,0)+(\x,\y)$);
            \coordinate (a2) at ($(\l,0)+(\x,\y)$);
            \coordinate (a3) at ($(\l/2,{sqrt(3)*\l/2})+(\x,\y)$);	    	            	
            \middleCoo{a1}{a2}{m12}{m12-1}{m12-2}{\sp};
            \middleCoo{a2}{a3}{m23}{m23-2}{m23-3}{\sp};
            \middleCoo{a1}{a3}{m13}{m13-1}{m13-3}{\sp};
            \barycentre{a1}{a2}{a3}{b123}{b123-1}{b123-2}{b123-3}{\sp};
            \draw[obj,->1-=0.6] (m12) --node[pos=0.5,xshift=-5pt] {${\sss \delta}$} (b123);
            \draw[obj,->1-=0.6] (b123) --node[pos=0.35,yshift=4pt, sloped] {${\sss \alpha}$} (m13);
            \draw[obj,->1-=0.6] (b123) --node[pos=0.35,yshift=4pt, sloped] {${\sss \nu}$} (m23);
            \draw[mor] 
            (m12-1) to[out=90, in=90, looseness=1.5] (m12-2)
            (m13-1) to[out=-30, in=-30, looseness=1.5] (m13-3)
            (m23-2) to[out=-150, in=-150, looseness=1.5] (m23-3);
            \node[mor, circle, inner sep=4pt] at (b123) {};
            \draw[mod, rounded corners=\r] 
            (m12-1) -- (b123-1) -- (m13-1)
            (m12-2) -- (b123-2) -- (m23-2)
            (m23-3) -- (b123-3) -- (m13-3);
            \node[yshift=-5pt, xshift=-8pt] at (b123-1) {${\sss A}$};
            \node[yshift=-5pt, xshift=8pt] at (b123-2) {${\sss D}$};
            \node[yshift=9pt] at (b123-3) {${\sss B}$}; 
            \node[] at (b123) {${\sss k}$};
            \node[yshift=3.5pt] at (m12) {${\sss d}$};
            \node[yshift=-1.5pt, xshift=2.5pt] at (m13) {${\sss a}$};
            \node[yshift=-1.5pt, xshift=-2.5pt] at (m23) {${\sss q}$};
            
            \draw[mor] (m23-3) -- (m23-2) to[out=30, in=-90, looseness=1] ($(m12-2)+(-\shix,-\shiy)$) -- ($(m12-1)+(-\shix,-\shiy)$) to[in=30, out=-90, looseness=1] (m23-3);
            \draw[mod] (m23-3) to[out=30, in=-90, looseness=1] ($(m12-1)+(-\shix,-\shiy)$);
            \draw[mod] (m23-2) to[out=30, in=-90, looseness=1] ($(m12-2)+(-\shix,-\shiy)$);
        }{}
	\end{tikzpicture}
}

\newcommand{\hamDerivation}[1]{
	\begin{tikzpicture}[scale=0.8,baseline=-3.5em]
    	\def\r{8pt};
    	\def\l{3.5};
    	\def\red{0.4};
    	\def\sp{0.16};
    	\coordinate (b1) at (\l/2,{sqrt(3)*\l/6-\red});
    	\coordinate (b2) at (\l/2,{-sqrt(3)*\l/6+\red});
    	\coordinate (a1) at (\l/4,{sqrt(3)*\l/4-\red});
    	\coordinate (a2) at (3*\l/4,{sqrt(3)*\l/4-\red});
    	\coordinate (a3) at (\l/4,{-sqrt(3)*\l/4+\red});
    	\coordinate (a4) at (3*\l/4,{-sqrt(3)*\l/4+\red});
    	\coordinate (a5) at ($(a3)+(0,-0.4)$);
    	\coordinate (a6) at ($(a4)+(0,-0.4)$);
                    
        \ifthenelse{#1 = 1}{
            \draw[line width = 0.7pt,->1-=0.7] (b1) --node[pos=0.6,yshift=5pt,sloped] {${\sss \alpha}$} (a1);
            \draw[line width = 0.7pt,->1-=0.7] (b1) --node[pos=0.6,yshift=5pt,sloped] {${\sss \beta}$} (a2);
            \draw[line width = 0.7pt,->1-=0.5] (a3) --node[pos=0.5,yshift=5pt, sloped] {${\sss \tilde \alpha}$} (b2);
            \draw[line width = 0.7pt,->1-=0.5] (a4) --node[pos=0.5,yshift=5pt, sloped] {${\sss \tilde \beta}$} (b2);
            \draw[line width = 0.7pt, ->1-=0.65] (b2) -- (b1);
            \node[xshift=-4pt, yshift=-2pt] at (\l/2,0) {${\sss \gamma}$};
            
            \node[mor, circle, inner sep=4pt] at (b1) {};
            \node at (b1) {${\sss j}$};
            \node[mor, circle, inner sep=4pt] at (b2) {};
            \node at (b2) {${\sss \tilde j}$};
        }{}
        \ifthenelse{#1 = 2 \OR #1 = 0}{
            \draw[line width = 0.7pt, ->1-=0.8] (a5) -- ($(a3)+(0,0.3)$);
            \draw[line width = 0.7pt, ->1-=0.8] (a6) -- ($(a4)+(0,0.3)$);
            \draw[mod] ($(a5)+(-0.8,0)$) -- ($(a6)+(0.8,0)$);
            \node[mor, circle, inner sep=4pt] at (a5) {};
            \node at (a5) {${\sss \tilde i}$};
            \node[mor, circle, inner sep=4pt] at (a6) {};
            \node at (a6) {${\sss \tilde l}$};
            \node[yshift=5pt] at ($(a5)+(-0.5,0)$) {${\sss A}$};
            \node[yshift=5pt] at ($(a6)+(0.5,0)$) {${\sss B}$};
            \node[yshift=5pt] at ($(a5)+0.5*(a6)-0.5*(a5)$) {${\sss \tilde C}$};
            \node[yshift=5pt] at ($(a3)+(0,0.3)$) {${\sss \tilde \alpha}$};
            \node[yshift=5pt] at ($(a4)+(0,0.3)$) {${\sss \tilde \beta}$};
            
            \ifthenelse{#1 = 0}{
                \draw[mod, dash pattern= on 0.2pt off 2.3pt] 
                    ($(a5)+(-0.9,0)$) -- ($(a5)+(-1.2,0)$)
                    ($(a6)+(0.9,0)$) -- ($(a6)+(1.2,0)$);
            }{}
        }{}
        \ifthenelse{#1 = 3}{
            \draw[line width = 0.7pt,->1-=0.7] (b1) --node[pos=0.6,yshift=5pt,sloped] {${\sss \alpha}$} (a1);
            \draw[line width = 0.7pt,->1-=0.7] (b1) --node[pos=0.6,yshift=5pt,sloped] {${\sss \beta}$} (a2);
            \draw[line width = 0.7pt, ->1-=0.65] (b2) -- (b1);
            \node[xshift=-4pt, yshift=-2pt] at (\l/2,0) {${\sss \gamma}$};
            
            \draw[line width = 0.7pt, ->1-=0.6, rounded corners=3pt] (a5) -- (a3) -- (b2) node[pos=0.45, yshift=5pt, sloped] {${\sss \tilde \alpha}$};
            \draw[line width = 0.7pt, ->1-=0.6, rounded corners=3pt] (a6) -- (a4) -- (b2) node[pos=0.45, yshift=5pt, sloped] {${\sss \tilde \beta}$};
            \draw[mod] ($(a5)+(-0.8,0)$) -- ($(a6)+(0.8,0)$);
            \node[yshift=5pt] at ($(a5)+(-0.5,0)$) {${\sss A}$};
            \node[yshift=5pt] at ($(a6)+(0.5,0)$) {${\sss B}$};
            \node[yshift=5pt] at ($(a5)+0.5*(a6)-0.5*(a5)$) {${\sss \tilde C}$};
            \node[mor, circle, inner sep=4pt] at (a5) {};
            \node at (a5) {${\sss \tilde i}$};
            \node[mor, circle, inner sep=4pt] at (a6) {};
            \node at (a6) {${\sss \tilde l}$};
            
            \node[mor, circle, inner sep=4pt] at (b1) {};
            \node at (b1) {${\sss j}$};
            \node[mor, circle, inner sep=4pt] at (b2) {};
            \node at (b2) {${\sss \tilde j}$};
        }{}
        \ifthenelse{#1 = 4}{
            \draw[line width = 0.7pt,->1-=0.7] ($(b1)+(0,{-sqrt(3)*\l/12-0.4})$) --node[pos=0.6,yshift=5pt,sloped] {${\sss \alpha}$} ($(a1)+(0,{-sqrt(3)*\l/12-0.4})$);
            \draw[line width = 0.7pt,->1-=0.7] ($(b1)+(0,{-sqrt(3)*\l/12-0.4})$) --node[pos=0.6,yshift=5pt,sloped] {${\sss \beta}$} ($(a2)+(0,{-sqrt(3)*\l/12-0.4})$);
            \draw[line width = 0.7pt, ->1-=0.65] ($(b2)+(0,{-sqrt(3)*\l/12-0.4})$) -- ($(b1)+(0,{-sqrt(3)*\l/12-0.4})$);
            \node[xshift=-4pt, yshift=-2pt] at ($(\l/2,0)+(0,{-sqrt(3)*\l/12-0.4})$) {${\sss \gamma}$};

            \draw[mod] ($(a5)+(-0.8,0)$) -- ($(a6)+(0.8,0)$);
            \node[yshift=5pt] at ($(a5)+(0,0)$) {${\sss A}$};
            \node[yshift=5pt] at ($(a6)+(0,0)$) {${\sss B}$};
            \node[mor, circle, inner sep=4pt] at ($(a5)+0.5*(a6)-0.5*(a5)$) {};
            \node at ($(a5)+0.5*(a6)-0.5*(a5)$) {${\sss k}$};
            
            \node[mor, circle, inner sep=4pt] at ($(b1)+(0,{-sqrt(3)*\l/12-0.4})$) {};
            \node at ($(b1)+(0,{-sqrt(3)*\l/12-0.4})$) {${\sss j}$};

        }{}
        \ifthenelse{#1 = 5}{
            \draw[line width = 0.7pt, ->1-=0.8] (a5) -- ($(a3)+(0,0.3)$);
            \draw[line width = 0.7pt, ->1-=0.8] (a6) -- ($(a4)+(0,0.3)$);
            \draw[mod] ($(a5)+(-0.8,0)$) -- ($(a6)+(0.8,0)$);
            \node[mor, circle, inner sep=4pt] at (a5) {};
            \node at (a5) {${\sss i}$};
            \node[mor, circle, inner sep=4pt] at (a6) {};
            \node at (a6) {${\sss l}$};
            \node[yshift=5pt] at ($(a5)+(-0.5,0)$) {${\sss A}$};
            \node[yshift=5pt] at ($(a6)+(0.5,0)$) {${\sss B}$};
            \node[yshift=5pt] at ($(a5)+0.5*(a6)-0.5*(a5)$) {${\sss C}$};
            \node[yshift=5pt] at ($(a3)+(0,0.3)$) {${\sss \alpha}$};
            \node[yshift=5pt] at ($(a4)+(0,0.3)$) {${\sss \beta}$};
        }{}
        \ifthenelse{#1 = 6}{
            \draw[line width = 0.7pt, ->1-=0.8] (a5) -- ($(a3)+(0,0.3)$);
            \draw[mod] ($(a5)+(-0.8,0)$) -- ($(a5)+(0.8,0)$);
            \node[mor, circle, inner sep=4pt] at (a5) {};
            \node at (a5) {${\sss \tilde i}$};
            \node[yshift=5pt] at ($(a5)+(-0.5,0)$) {${\sss A}$};
            \node[yshift=5pt] at ($(a5)+(0.5,0)$) {${\sss \tilde C}$};
            \node[yshift=5pt] at ($(a3)+(0,0.3)$) {${\sss \tilde \alpha}$};
        }{}
	\end{tikzpicture}
}

\newcommand{\object}[2]{
	\begin{tikzpicture}[scale=1,baseline=-2.5pt]
	    \clip (-0.05,-0.1) -- (1.05,-0.1) -- (1.05,0.26) -- (-0.05,0.26) --cycle;
    	\draw[#1] (0,0) -- (1,0) node[text=black, pos=0.5, yshift=4pt](){${\sss #2}$};
	\end{tikzpicture}
}

\newcommand{\morphism}[3]{
	\begin{tikzpicture}[scale=1,baseline=-2.8pt]
	    \clip (-0.05,-0.2) -- (1.05,-0.2) -- (1.05,0.26) -- (-0.05,0.26) --cycle;
    	\draw[#1] (0,0) -- (1,0) node[text=black, pos=0.1, yshift=4pt](){${\sss #2}$} node[text=black, pos=0.9, yshift=4pt](){${\sss #2}$} ;
    	\node[mor, circle, inner sep=4pt] at (0.5,0) {};
    	\node[] at (0.5,0) {${\sss #3}$};
	\end{tikzpicture}
}

\newcommand{\doubleMorphism}[3]{
	\begin{tikzpicture}[scale=1,baseline=-2.8pt]
	    \clip (-0.05,-0.2) -- (1.55,-0.2) -- (1.55,0.26) -- (-0.05,0.26) --cycle;
    	\draw[#1] (0,0) -- (1.5,0) node[text=black, pos=0.1, yshift=4pt](){${\sss #2}$} node[text=black, pos=0.9, yshift=4pt](){${\sss #2}$} ;
    	\node[mor, circle, inner sep=4pt] at (0.5,0) {};
    	\node[mor, circle, inner sep=4pt] at (1,0) {};
    	\node[] at (0.5,0) {${\sss #3}$};
    	\node[] at (1,0) {${\sss #3}$};
	\end{tikzpicture}
}

\newcommand{\splitSpace}[5]{
	\begin{tikzpicture}[scale=1,baseline=-13pt]
    	\def\l{3.5};
    	\def\sp{0.16};
    	\ifthenelse{#5=1}{
        	\draw[obj, ->1-=0.5] (0,-0.7) --node[pos=0.5,xshift=-4pt](){${\sss #2}$} (0,-0.1);
        	\draw[mor] ({-\sp*\l/2},0) to[out=-90, in=-90, looseness=1.5] ({\sp*\l/2},0);
        	\draw[mod] ({-\sp*\l/2},-0.7) --node[text=black,pos=0.5, xshift=-5pt](){${\sss #1}$} ({-\sp*\l/2},0);
        	\draw[mod] ({\sp*\l/2},-0.7) --node[text=black,pos=0.5, xshift=5pt](){${\sss #3}$} ({\sp*\l/2},0);
        	\node[yshift=-3pt] at (0,0) {${\sss #4}$};
    	}{}
    	\ifthenelse{#5=2}{
        	\draw[obj, ->1-=0.7] (0,-0.6) --node[pos=0.7,xshift=-4pt](){${\sss #2}$} (0,0);
        	\draw[mor] ({-\sp*\l/2},-0.7) to[out=90, in=90, looseness=1.5] ({\sp*\l/2},-0.7);
        	\draw[mod] ({-\sp*\l/2},-0.7) --node[text=black,pos=0.5, xshift=-5pt](){${\sss #1}$} ({-\sp*\l/2},0);
        	\draw[mod] ({\sp*\l/2},-0.7) --node[text=black,pos=0.5, xshift=5pt](){${\sss #3}$} ({\sp*\l/2},0);
        	\node[yshift=3pt] at (0,-0.7) {${\sss #4}$};
    	}{}
	\end{tikzpicture}
}

\newcommand{\chain}[3]{
    \def\morI{#1}
    \def\morII{#2}
    \def\morIII{#3}
    \chainCont
}

\newcommand{\chainCont}[7]{
	\begin{tikzpicture}[scale=1,baseline={([yshift=-1ex]current bounding box.center)}]
    	\def\r{8pt};
    	\def\l{3.5};
    	\def\sp{0.16};
    	\coordinate (a1) at (0,0);
        \coordinate (a2) at (\l/2,0);
    	\coordinate (a3) at (0,\l/2);
    	\coordinate (a4) at (\l/2,\l/2);

        \clip (-0.55,\l/4-0.2) rectangle (3*\l/2+0.55,\l/2+0.1);
        
    	\middleCoo{a1}{a2}{m12}{m12-1}{m12-2}{\sp};
    	\middleCoo{a1}{a3}{m13}{m13-1}{m13-3}{\sp};
    	\middleCoo{a2}{a4}{m24}{m24-2}{m24-4}{\sp};
    	\middleCoo{a3}{a4}{m34}{m34-3}{m34-4}{\sp};
    	\barycentreSq{a1}{a2}{a3}{a4}{b1234}{b1234-1}{b1234-2}{b1234-3}{b1234-4};
    	
    	\foreach \x in {0,\l/2,\l}{
        	\draw[obj, ->1-=0.6, shorten <= 7pt] ($(b1234)+(\x,0)$) -- ($(m34)+(\x,0)$);
        	\draw[mor] ($(m34-3)+(\x,0)$) to[out=-90, in=-90, looseness=1.5] ($(m34-4)+(\x,0)$);
            \draw[mod, rounded corners=\r] 
        	    ($(m13-3)+(\x,0)$) -- ($(b1234-3)+(\x,0)$) -- ($(m34-3)+(\x,0)$)
        	    ($(m24-4)+(\x,0)$) -- ($(b1234-4)+(\x,0)$) -- ($(m34-4)+(\x,0)$);
        }
        \draw[mod, dash pattern= on 0.2pt off 2.3pt] 
            (m13-3) -- ($(m13-3)+(-0.3,0)$)
            ($(m24-4)+(\l,0)$) -- ($(m24-4)+(\l+0.3,0)$);
        \node[yshift=1pt] at (\l/4,\l/4) {${\sss #5}$};
        \node[yshift=1pt] at (\l/4+\l/2,\l/4) {${\sss #6}$};
        \node[yshift=1pt] at (\l/4+\l,\l/4) {${\sss #7}$};
        \ifthenelse{\equal{#1}{\beta}}{
            \node[mor, rounded corners=2pt, inner sep=4pt] at (m13-3) {$\;\;\;$};
            \node[] at (m13-3) {${\sss {\rm e} \oplus {\rm o}}$};
            \node[above, xshift=11pt] at (m13-3) {${\sss #1}$};
        }{
            \node[above] at (m13-3) {${\sss #1}$};
        }
        \ifthenelse{\equal{#2}{\beta}}{
            \node[mor, rounded corners=2pt, inner sep=4pt] at ($(m13-3)+(\l/2,0)$) {$\;\;\;$};
            \node[] at ($(m13-3)+(\l/2,0)$) {${\sss {\rm e} \oplus {\rm o}}$};
            \node[above, xshift=-11pt] at ($(m13-3)+(\l/2,0)$) {${\sss #2}$};
            \node[above, xshift=11pt] at ($(m13-3)+(\l/2,0)$) {${\sss #2}$};
        }{
            \node[above] at ($(m13-3)+(\l/2,0)$) {${\sss #2}$};
        }
        \ifthenelse{\equal{#3}{\beta}}{
            \node[mor, rounded corners=2pt, inner sep=4pt] at ($(m13-3)+(\l,0)$) {$\;\;\;$};
            \node[] at ($(m13-3)+(\l,0)$) {${\sss {\rm e} \oplus {\rm o}}$};
            \node[above, xshift=-11pt] at ($(m13-3)+(\l,0)$) {${\sss #3}$};
            \node[above, xshift=11pt] at ($(m13-3)+(\l,0)$) {${\sss #3}$};
        }{
            \node[above] at ($(m13-3)+(\l,0)$) {${\sss #3}$};
        }
        \ifthenelse{\equal{#4}{\beta}}{
            \node[mor, rounded corners=2pt, inner sep=4pt] at ($(m13-3)+(3*\l/2,0)$) {$\;\;\;$};
            \node[] at ($(m13-3)+(3*\l/2,0)$) {${\sss {\rm e} \oplus {\rm o}}$};
            \node[above, xshift=-11pt] at ($(m13-3)+(3*\l/2,0)$) {${\sss #4}$};
        }{
            \node[above] at ($(m13-3)+(3*\l/2,0)$) {${\sss #4}$};
        }
        
        \node[yshift=-3.5pt] at ($(m34)+(0,0)$) {${\sss \morI}$};
        \node[yshift=-3.5pt] at ($(m34)+(\l/2,0)$) {${\sss \morII}$};
        \node[yshift=-3.5pt] at ($(m34)+(\l,0)$) {${\sss \morIII}$};
	\end{tikzpicture}
}

\newcommand{\PEPS}[5]{
    \def\sty{#1}
    \def\morI{#2}
    \def\morII{#3}
    \def\morIII{#4}
    \def\morIV{#5}
    \PEPScont
}

\newcommand{\PEPScont}[7]{
	\begin{tikzpicture}[scale=1,baseline={([yshift=-.5ex]current bounding box.center)}]
    	\def\r{8pt};
    	\def\l{3.5};
    	\def\sp{0.08};
    	
    	\ifthenelse{#7 = 1}{\def\sgn{1}}{}
        \ifthenelse{#7 = 2}{\def\sgn{-1}}{}
    	
        \coordinate (a1) at (0,0);
        \coordinate (a2) at (\l,0);
        \coordinate (a3) at (\l/2,{\sgn*sqrt(3)*\l/2});
    	    	    	
    	\clip ({(0.5-\sp)*\l*cos(60)-0.05},-\sgn*0.05) rectangle ({\l-(0.5-\sp)*\l*cos(60)+0.05},{\sgn*(\l-(0.5+\sp)*\l*sin(60))+\sgn*0.05});
    	
    	\middleCoo{a1}{a2}{m12}{m12-1}{m12-2}{\sp};
    	\middleCoo{a2}{a3}{m23}{m23-2}{m23-3}{\sp};
    	\middleCoo{a1}{a3}{m13}{m13-1}{m13-3}{\sp};
    	\barycentre{a1}{a2}{a3}{b123}{b123-1}{b123-2}{b123-3}{\sp};
    	
    	\ifthenelse{#7 = 1}{
            \draw[obj,->1-=0.6] (m12) --node[pos=0.5,xshift=-5pt] {${\sss #6}$} (b123);
            \draw[obj,->1-=0.6] (b123) --node[pos=0.35,yshift=\sgn*4pt, sloped] {${\sss #4}$} (m13);
            \draw[obj,->1-=0.6] (b123) --node[pos=0.35,yshift=\sgn*4pt, sloped] {${\sss #5}$} (m23);
        }{}
        \ifthenelse{#7 = 2}{
            \draw[obj,->1-=0.6] (b123) --node[pos=0.5,xshift=-5pt] {${\sss #6}$} (m12);
            \draw[obj,->1-=0.6] (m13) --node[pos=0.65,yshift=-4.5pt, sloped] {${\sss #4}$} (b123);
            \draw[obj,->1-=0.6] (m23) --node[pos=0.65,yshift=-4.5pt, sloped] {${\sss #5}$} (b123);
        }{}
        
        \draw[mor] 
            (m12-1) to[out=\sgn*90, in=\sgn*90, looseness=1.5] (m12-2)
            (m13-1) to[out=-\sgn*30, in=-\sgn*30, looseness=1.5] (m13-3)
            (m23-2) to[out=-\sgn*150, in=-\sgn*150, looseness=1.5] (m23-3);
        \node[mor, circle, inner sep=4pt] at (b123) {};
        
        \draw[mod, rounded corners=\r, \sty] 
            (m12-1) -- (b123-1) -- (m13-1)
            (m12-2) -- (b123-2) -- (m23-2)
            (m23-3) -- (b123-3) -- (m13-3);
        \node[yshift=-\sgn*5pt, xshift=-8pt] at (b123-1) {${\sss #1}$};
        \node[yshift=-\sgn*5pt, xshift=8pt] at (b123-2) {${\sss #2}$};
        \node[yshift=\sgn*9pt] at (b123-3) {${\sss #3}$}; 
        \node[] at (b123) {${\sss \morII}$};
        \node[yshift=\sgn*3.5pt] at (m12) {${\sss \morIII}$};
        \node[yshift=-\sgn*1.5pt, xshift=2.5pt] at (m13) {${\sss \morI}$};
        \node[yshift=-\sgn*1.5pt, xshift=-2.5pt] at (m23) {${\sss \morIV}$};
	\end{tikzpicture}
}

\newcommand{\HAM}[7]{
    \def\sty{#1}
    \def\morI{#2}
    \def\morII{#3}
    \def\morIII{#4}
    \def\morIV{#5}
    \def\morV{#6}
    \def\morVI{#7}
    \HAMcont
}

\newcommand{\HAMcont}[9]{
	\begin{tikzpicture}[scale=1,baseline={([yshift=-.5ex]current bounding box.center)}]
    	\def\r{8pt};
    	\def\l{3.5};
    	\def\sp{0.08}; 
    	\def\red{0.5};
    	
    	\clip ($({(0.5-\sp)*\l*cos(60)},-{(\l-(0.5+\sp)*\l*sin(60))})+(0,0.95*\red)$) rectangle ($({\l-(0.5-\sp)*\l*cos(60)},{(\l-(0.5+\sp)*\l*sin(60))})+(0,-0.95*\red)$);
    	        	
        \foreach \sgn in {1,-1}{
            \coordinate (a1) at ($(0,0)+(0,-\sgn*\red)$);
            \coordinate (a2) at ($(\l,0)+(0,-\sgn*\red)$);
            \coordinate (a3) at ($(\l/2,{\sgn*sqrt(3)*\l/2})+(0,-\sgn*\red)$);
                
        	\middleCoo{a1}{a2}{m12}{m12-1}{m12-2}{\sp};
        	\middleCoo{a2}{a3}{m23}{m23-2}{m23-3}{\sp};
        	\middleCoo{a1}{a3}{m13}{m13-1}{m13-3}{\sp};
        	\barycentre{a1}{a2}{a3}{b123}{b123-1}{b123-2}{b123-3}{\sp};
        	    
            \draw[mor] 
                (m13-1) to[out=-\sgn*30, in=-\sgn*30, looseness=1.5] (m13-3)
                (m23-2) to[out=-\sgn*150, in=-\sgn*150, looseness=1.5] (m23-3);
            \node[mor, circle, inner sep=4pt] at (b123) {};
            
            \draw[mod, rounded corners=\r, \sty] 
                (m13-1) -- (b123-1) -- ($(m12-1)+(0,\sgn*\red)$)
                (m23-2) -- (b123-2) -- ($(m12-2)+(0,\sgn*\red)$)
                (m23-3) -- (b123-3) -- (m13-3);
            
            \ifthenelse{\sgn = 1}{
                \begin{pgfonlayer}{backC}
                    \draw[obj,->1-=0.6] (b123) --node[pos=0.35,yshift=\sgn*4pt,sloped] {${\sss #6}$} (m13);
                    \draw[obj,->1-=0.6] (b123) --node[pos=0.35,yshift=\sgn*4pt,sloped] {${\sss #8}$} (m23);
                \end{pgfonlayer}
                \node[yshift=\sgn*9pt] at (b123-3) {${\sss #3}$}; 
                \node[] at (b123) {${\sss \morVI}$};
                \node[yshift=-\sgn*1.5pt, xshift=2.5pt] at (m13) {${\sss \morIV}$};
                \node[yshift=-\sgn*1.5pt, xshift=-2.5pt] at (m23) {${\sss \morV}$};
                \node[xshift=-8pt] at ($(m12-1)+(0,\sgn*\red)$) {${\sss #1}$};
                \node[xshift=8pt] at ($(m12-2)+(0,\sgn*\red)$) {${\sss #2}$};
                \node[xshift=-4pt, yshift=-2pt] at ($(m12)+(0,\sgn*\red)$) {${\sss #9}$};
            }{
                \begin{pgfonlayer}{backC}
                    \draw[obj, ->1-=0.65] (b123) -- (m12);
                    \draw[obj,->1-=0.6] (m13) --node[pos=0.65,yshift=\sgn*4.5pt, sloped] {${\sss #5}$} (b123);
                    \draw[obj,->1-=0.6] (m23) --node[pos=0.65,yshift=\sgn*4.5pt, sloped] {${\sss #7}$} (b123);
                \end{pgfonlayer}
                \node[yshift=\sgn*9pt] at (b123-3) {${\sss #4}$}; 
                \node[] at (b123) {${\sss \morIII}$};
                \node[yshift=-\sgn*1.5pt, xshift=2.5pt] at (m13) {${\sss \morI}$};
                \node[yshift=-\sgn*1.5pt, xshift=-2.5pt] at (m23) {${\sss \morII}$};
            }
        }
	\end{tikzpicture}
}

\newcommand{\MPO}[6]{
    \def\styA{#1}
    \def\styB{#2}
    \def\morI{#3}
    \def\morII{#4}
    \def\morIII{#5}
    \def\morIV{#6}
    \MPOcont
}

\newcommand{\MPOcont}[7]{
	\begin{tikzpicture}[scale=1,baseline={([yshift=-.5ex]current bounding box.center)}]
        \def\r{8pt};
        \def\l{3.5};
        \def\sp{0.16};
        \coordinate (a1) at (0,0);
        \coordinate (a2) at (\l/2,0);
        \coordinate (a3) at (0,\l/2);
        \coordinate (a4) at (\l/2,\l/2);
        
        \middleCoo{a1}{a2}{m12}{m12-1}{m12-2}{\sp};
        \middleCoo{a1}{a3}{m13}{m13-1}{m13-3}{\sp};
        \middleCoo{a2}{a4}{m24}{m24-2}{m24-4}{\sp};
        \middleCoo{a3}{a4}{m34}{m34-3}{m34-4}{\sp};
        \barycentreSq{a1}{a2}{a3}{a4}{b1234}{b1234-1}{b1234-2}{b1234-3}{b1234-4};
        
        \ifthenelse{#7 = 0}{
            \draw[mod,->3-=0.7,shorten <= 1pt, shorten >= 1pt] (m13) --node[pos=0.35,yshift=5pt,text=black] {${\sss #6}$} node[mask point, pos=0.5](mk1){}  (m24);
        }{}
        \ifthenelse{#7 = 1}{
            \draw[speObj,->2-=0.7] (m13) --node[pos=0.35,yshift=5pt] {${\sss #6}$} node[mask point, pos=0.5](mk1){}  (m24);
        }{}
        \ifthenelse{#7 = 2}{
            \draw[obj, dot] (m24) --node[pos=0.35,yshift=5pt] {${\sss #6}$} node[mask point, pos=0.5](mk1){}  (m13);
            \clipAroundNode{mk1}{
                \draw[obj,->1-=0.35] (m12) --node[pos=0.6,xshift=9pt] {${\sss #5}$} (m34);
            }  
        }{
            \clipAroundNode{mk1}{
                \draw[obj,->1-=0.35] (m12) --node[pos=0.3,xshift=5pt] {${\sss #5}$} (m34);
            }  
        }
        \draw[mor] 
        (m12-1) to[out=90, in=90, looseness=1.5] (m12-2)
        (m13-1) to[out=0, in=0, looseness=1.5] (m13-3)
        (m24-2) to[out=-180, in=-180, looseness=1.5] (m24-4)
        (m34-3) to[out=-90, in=-90, looseness=1.5] (m34-4);
        
        \draw[obj, rounded corners=\r, \styA] 
        (m12-1) -- (b1234-1) -- (m13-1)
        (m12-2) -- (b1234-2) -- (m24-2);
        \draw[obj, rounded corners=\r, \styB] 
        (m13-3) -- (b1234-3) -- (m34-3)
        (m34-4) -- (b1234-4) -- (m24-4);
        
        \node[yshift=3pt] at (m12) {${\sss \morIV}$};
        \node[yshift=-3.5pt] at (m34) {${\sss \morIII}$};
        \node[xshift=3pt] at (m13) {${\sss \morI}$};   
        \node[xshift=-3pt] at (m24) {${\sss \morII}$};   
        \node[xshift=-8pt,yshift=8pt] at (b1234-3) {${\sss #1}$};
        \node[xshift=8pt,yshift=8pt] at (b1234-4) {${\sss #2}$};
        \node[xshift=-8pt,yshift=-8pt] at (b1234-1) {${\sss #4}$};
        \node[xshift=8pt,yshift=-8pt] at (b1234-2) {${\sss #3}$}; 
	\end{tikzpicture}
}

\newcommand{\pulling}[1]{
	\begin{tikzpicture}[scale=1,baseline={([yshift=-.5ex]current bounding box.center)}]
        \def\r{8pt};
        \def\l{3.5};
        \def\sp{0.08};
        \def\red{0.5};
        
        \ifthenelse{#1=1}{
            \clip (-0.4,-0.05) rectangle (3.9,3.5);
            \coordinate (a1) at (0,0);
            \coordinate (a2) at (\l,0);
            \coordinate (a3) at (\l/2,{sqrt(3)*\l/2});	    	    	
        	\middleCoo{a1}{a2}{m12}{m12-1}{m12-2}{\sp};
        	\middleCoo{a2}{a3}{m23}{m23-2}{m23-3}{\sp};
        	\middleCoo{a1}{a3}{m13}{m13-1}{m13-3}{\sp};
        	\barycentre{a1}{a2}{a3}{b123}{b123-1}{b123-2}{b123-3}{\sp};
        	\foreach \x in {m13,m13-1,m13-3}{
        	    \coordinate (\x) at ($(\x)+({cos(30)*\red},{-sin(30)*\red})$);
        	}
        	\foreach \x in {m23,m23-2,m23-3}{
        	    \coordinate (\x) at ($(\x)+({-cos(30)*\red},{-sin(30)*\red})$);
        	}
            \draw[obj,->1-=0.6] (m12) --node[pos=0.5,xshift=-5pt] {${\sss \gamma}$} (b123);       
            \draw[mor] (m12-1) to[out=90, in=90, looseness=1.5] (m12-2);
            \draw[mod, rounded corners=\r] 
                (m12-1) -- (b123-1) -- (m13-1)
                (m12-2) -- (b123-2) -- (m23-2)
                (m23-3) -- (b123-3) -- (m13-3);
            
            \def\sp{0.16};
            \coordinate (b1) at ($({\l*cos(60)/4},{\l*sin(60)/4})+({cos(30)*\red},{-sin(30)*\red})$);
            \coordinate (b2) at ($({3*\l*cos(60)/4},{3*\l*sin(60)/4})+({cos(30)*\red},{-sin(30)*\red})$);
            \coordinate (b3) at ($(b1)+({-\l*cos(30)/2},{\l*sin(30)/2})$);
            \coordinate (b4) at ($(b2)+({-\l*cos(30)/2},{\l*sin(30)/2})$);
            \middleCoo{b1}{b3}{n13}{n13-1}{n13-3}{\sp};
            \middleCoo{b2}{b4}{n24}{n24-2}{n24-4}{\sp};
            \middleCoo{b3}{b4}{n34}{n34-3}{n34-4}{\sp};
            \barycentreSq{b1}{b2}{b3}{b4}{b1234}{b1234-1}{b1234-2}{b1234-3}{b1234-4};
            \draw[speObj, shorten >=-1pt] (n13) -- node[mask point, pos=0.5](mk1){} (n24);
            \clipAroundNode{mk1}{
                \draw[obj,->1-=0.5] (b123) --node[pos=0.4,yshift=4pt,sloped] {${\sss \alpha}$} (n34);
            } 
            \draw[mor] 
                (n34-3) to[out=-30, in=-30, looseness=1.5] (n34-4)
                (n13-1) to[out=60, in=60, looseness=1.5] (n13-3);
            \draw[mod, rounded corners=\r]
                (m13-1) -- (b1234-1) -- (n13-1)
                (m13-3) -- (b1234-2) -- (n24-2)
                (n13-3) -- (b1234-3) -- (n34-3)
                (n34-4) -- (b1234-4) -- (n24-4);
                
            \coordinate (c1) at ($(a2)+({-3*\l*cos(60)/4},{3*\l*sin(60)/4})+({-cos(30)*\red},{-sin(30)*\red})$);
            \coordinate (c2) at ($(a2)+({-\l*cos(60)/4},{\l*sin(60)/4})+({-cos(30)*\red},{-sin(30)*\red})$);
            \coordinate (c3) at ($(c1)+({\l*cos(30)/2},{\l*sin(30)/2})$);
            \coordinate (c4) at ($(c2)+({\l*cos(30)/2},{\l*sin(30)/2})$);
            \middleCoo{c1}{c3}{o13}{o13-1}{o13-3}{\sp};
            \middleCoo{c2}{c4}{o24}{o24-2}{o24-4}{\sp};
            \middleCoo{c3}{c4}{o34}{o34-3}{o34-4}{\sp};
            \barycentreSq{c1}{c2}{c3}{c4}{c1234}{c1234-1}{c1234-2}{c1234-3}{c1234-4};
            \draw[speObj,shorten <= -1pt] (o13) -- node[mask point, pos=0.5](mk1){} (o24);
            \clipAroundNode{mk1}{
                \draw[obj,->1-=0.5] (b123) --node[pos=0.4, yshift=4pt,sloped] {${\sss \beta}$} (o34);
            } 
            \draw[mor] 
                (o34-3) to[out=-150, in=-150, looseness=1.5] (o34-4)
                (o24-2) to[out=120, in=120, looseness=1.5] (o24-4);
            \draw[mod, rounded corners=\r]
                (m23-2) -- (c1234-2) -- (o24-2)
                (m23-3) -- (c1234-1) -- (o13-1)
                (o13-3) -- (c1234-3) -- (o34-3)
                (o34-4) -- (c1234-4) -- (o24-4);
            
            \draw[line width = 0.7pt, line join=round, line cap=join, color=BrickRed,->2-=0.52] (n24) to[out=60,in=120,looseness=1] node[pos=0.5,yshift=-4pt](){${\sss a}$} (o13);
            \draw[mod] (n24-2) to[out=60,in=120,looseness=1] (o13-1);
            \draw[mod] (n24-4) to[out=60,in=120,looseness=1] node[pos=0.5,yshift=6pt,text=black](){${\sss C}$} (o13-3);
            \node[mor, circle, inner sep=4pt] at (b123) {};    
            \node at (b123) {${\sss m}$};
            \node[yshift=9pt] at (b123-3) {${\sss F}$};
            \node[yshift=-6pt,xshift=-4pt] at (m13-1) {${\sss A}$};
            \node[yshift=-6pt,xshift=4pt] at (m23-2) {${\sss E}$};
            \node[xshift=-10pt,yshift=-3pt] at (b1234-3) {${\sss B}$};
            \node[xshift=10pt,yshift=-3pt] at (c1234-4) {${\sss D}$};
            \node[yshift=3pt] at (m12) {${\sss n}$};
            \node[xshift=1.5pt,yshift=2.6pt] at (n13) {${\sss i}$};
            \node[xshift=2.5pt,yshift=-1.5pt] at (n34) {${\sss k}$};
            \node[xshift=-1.5pt,yshift=2.6pt] at (o24) {${\sss j}$};
            \node[xshift=-2.5pt,yshift=-1.5pt] at (o34) {${\sss l}$};
        }{}
        
        \ifthenelse{#1=2}{
            \clip (0.65,-1.35) rectangle (2.85,1.85);
            \coordinate (a1) at (0,0);
            \coordinate (a2) at (\l,0);
            \coordinate (a3) at (\l/2,{sqrt(3)*\l/2});	    	    	
        	\middleCoo{a1}{a2}{m12}{m12-1}{m12-2}{\sp};
        	\middleCoo{a2}{a3}{m23}{m23-2}{m23-3}{\sp};
        	\middleCoo{a1}{a3}{m13}{m13-1}{m13-3}{\sp};
        	\barycentre{a1}{a2}{a3}{b123}{b123-1}{b123-2}{b123-3}{\sp};
        	\foreach \x in {m12,m12-1,m12-2}{
        	    \coordinate (\x) at ($(\x)+(0,\red)$);
        	}
            \draw[obj,->1-=0.6] (b123) --node[pos=0.35,yshift=4pt, sloped] {${\sss \alpha}$} (m13);
            \draw[obj,->1-=0.6] (b123) --node[pos=0.35,yshift=4pt, sloped] {${\sss \beta}$} (m23);     
            \draw[mor] 
                (m13-1) to[out=-30, in=-30, looseness=1.5] (m13-3)
                (m23-2) to[out=-150, in=-150, looseness=1.5] (m23-3);
            \draw[mod, rounded corners=\r] 
                (m12-1) -- (b123-1) -- (m13-1)
                (m12-2) -- (b123-2) -- (m23-2)
                (m23-3) -- (b123-3) -- (m13-3);
                
            \def\sp{0.16};
            \coordinate (b1) at (\l/4,-\l/2+\red);
            \coordinate (b2) at (3*\l/4,-\l/2+\red);
            \coordinate (b3) at (\l/4,\red);
            \coordinate (b4) at (3*\l/4,\red);
            \middleCoo{b1}{b3}{n13}{n13-1}{n13-3}{\sp};
            \middleCoo{b2}{b4}{n24}{n24-2}{n24-4}{\sp};
            \middleCoo{b1}{b2}{n12}{n12-1}{n12-2}{\sp};
            \barycentreSq{b1}{b2}{b3}{b4}{b1234}{b1234-1}{b1234-2}{b1234-3}{b1234-4};
            \draw[speObj,->2-=0.7] (n13) --node[pos=0.35,yshift=-4pt] {${\sss a}$} node[mask point, pos=0.5](mk1){} (n24);
            \clipAroundNode{mk1}{
                \draw[obj,->1-=0.6] (n12) --node[pos=0.6,xshift=-5pt] {${\sss \gamma}$} (b123);  
            } 
            \draw[mor] 
                (n12-1) to[out=90, in=90, looseness=1.5] (n12-2)
                (n13-1) to[out=0, in=0, looseness=1.5] (n13-3)
                (n24-2) to[out=-180, in=-180, looseness=1.5] (n24-4);
            \draw[mod, rounded corners=\r]
                (m12-1) -- (b1234-3) -- (n13-3)
                (m12-2) -- (b1234-4) -- (n24-4)
                (n12-1) -- (b1234-1) -- (n13-1)
                (n12-2) -- (b1234-2) -- (n24-2);
            \node[mor, circle, inner sep=4pt] at (b123) {};    
            \node at (b123) {${\sss m}$};
            \node[xshift=-8pt,yshift=-8pt] at (b1234-1) {${\sss A}$};
            \node[xshift=8pt,yshift=-8pt] at (b1234-2) {${\sss E}$};
            \node[xshift=-8pt,yshift=14pt] at (b1234-3) {${\sss B}$};
            \node[xshift=8pt,yshift=14pt] at (b1234-4) {${\sss D}$};
            \node[yshift=9pt] at (b123-3) {${\sss C}$};
            \node[yshift=3pt] at (n12) {${\sss n}$};
            \node[xshift=2pt,yshift=-1.5pt] at (m13) {${\sss k}$};
            \node[xshift=-2pt,yshift=-1.5pt] at (m23) {${\sss l}$};
            \node[xshift=3pt] at (n13) {${\sss i}$};
            \node[xshift=-3pt] at (n24) {${\sss j}$};
        }{}
	\end{tikzpicture}
}

\newcommand{\intertwinerPulling}[1]{
	\begin{tikzpicture}[scale=1,baseline={([yshift=-.5ex]current bounding box.center)}]
        \def\r{8pt};
        \def\l{3.5};
        \def\sp{0.08};
        \def\red{0.5};
        
        \ifthenelse{#1=1}{
            \clip (-0.4,-0.05) rectangle (3.9,3.5);
            \coordinate (a1) at (0,0);
            \coordinate (a2) at (\l,0);
            \coordinate (a3) at (\l/2,{sqrt(3)*\l/2});	    	    	
        	\middleCoo{a1}{a2}{m12}{m12-1}{m12-2}{\sp};
        	\middleCoo{a2}{a3}{m23}{m23-2}{m23-3}{\sp};
        	\middleCoo{a1}{a3}{m13}{m13-1}{m13-3}{\sp};
        	\barycentre{a1}{a2}{a3}{b123}{b123-1}{b123-2}{b123-3}{\sp};
        	\foreach \x in {m13,m13-1,m13-3}{
        	    \coordinate (\x) at ($(\x)+({cos(30)*\red},{-sin(30)*\red})$);
        	}
        	\foreach \x in {m23,m23-2,m23-3}{
        	    \coordinate (\x) at ($(\x)+({-cos(30)*\red},{-sin(30)*\red})$);
        	}
            \draw[obj,->1-=0.6] (m12) --node[pos=0.5,xshift=-5pt] {${\sss \nu}$} (b123);       
            \draw[mor] (m12-1) to[out=90, in=90, looseness=1.5] (m12-2);
            \draw[rounded corners=\r] 
                (m12-1) -- (b123-1) -- (m13-1)
                (m12-2) -- (b123-2) -- (m23-2)
                (m23-3) -- (b123-3) -- (m13-3);
            
            \def\sp{0.16};
            \coordinate (b1) at ($({\l*cos(60)/4},{\l*sin(60)/4})+({cos(30)*\red},{-sin(30)*\red})$);
            \coordinate (b2) at ($({3*\l*cos(60)/4},{3*\l*sin(60)/4})+({cos(30)*\red},{-sin(30)*\red})$);
            \coordinate (b3) at ($(b1)+({-\l*cos(30)/2},{\l*sin(30)/2})$);
            \coordinate (b4) at ($(b2)+({-\l*cos(30)/2},{\l*sin(30)/2})$);
            \middleCoo{b1}{b3}{n13}{n13-1}{n13-3}{\sp};
            \middleCoo{b2}{b4}{n24}{n24-2}{n24-4}{\sp};
            \middleCoo{b3}{b4}{n34}{n34-3}{n34-4}{\sp};
            \barycentreSq{b1}{b2}{b3}{b4}{b1234}{b1234-1}{b1234-2}{b1234-3}{b1234-4};
            \draw[mod, shorten <=1pt] (n13) -- node[mask point, pos=0.5](mk1){} (n24);
            \clipAroundNode{mk1}{
                \draw[obj,->1-=0.5] (b123) --node[pos=0.4,yshift=4pt,sloped] {${\sss \beta}$} (n34);
            } 
            \draw[mor] 
                (n34-3) to[out=-30, in=-30, looseness=1.5] (n34-4)
                (n13-1) to[out=60, in=60, looseness=1.5] (n13-3);
            \draw[rounded corners=\r]
                (m13-1) -- (b1234-1) -- (n13-1)
                (m13-3) -- (b1234-2) -- (n24-2);
            \draw[mod, rounded corners=\r]
                (n13-3) -- (b1234-3) -- (n34-3)
                (n34-4) -- (b1234-4) -- (n24-4);
                
            \coordinate (c1) at ($(a2)+({-3*\l*cos(60)/4},{3*\l*sin(60)/4})+({-cos(30)*\red},{-sin(30)*\red})$);
            \coordinate (c2) at ($(a2)+({-\l*cos(60)/4},{\l*sin(60)/4})+({-cos(30)*\red},{-sin(30)*\red})$);
            \coordinate (c3) at ($(c1)+({\l*cos(30)/2},{\l*sin(30)/2})$);
            \coordinate (c4) at ($(c2)+({\l*cos(30)/2},{\l*sin(30)/2})$);
            \middleCoo{c1}{c3}{o13}{o13-1}{o13-3}{\sp};
            \middleCoo{c2}{c4}{o24}{o24-2}{o24-4}{\sp};
            \middleCoo{c3}{c4}{o34}{o34-3}{o34-4}{\sp};
            \barycentreSq{c1}{c2}{c3}{c4}{c1234}{c1234-1}{c1234-2}{c1234-3}{c1234-4};
            \draw[mod,shorten >= 1pt] (o13) -- node[mask point, pos=0.5](mk1){} (o24);
            \clipAroundNode{mk1}{
                \draw[obj,->1-=0.5] (b123) --node[pos=0.4, yshift=4pt,sloped] {${\sss \gamma}$} (o34);
            } 
            \draw[mor] 
                (o34-3) to[out=-150, in=-150, looseness=1.5] (o34-4)
                (o24-2) to[out=120, in=120, looseness=1.5] (o24-4);
            \draw[rounded corners=\r]
                (m23-2) -- (c1234-2) -- (o24-2)
                (m23-3) -- (c1234-1) -- (o13-1);
            \draw[mod, rounded corners=\r]
                (o13-3) -- (c1234-3) -- (o34-3)
                (o34-4) -- (c1234-4) -- (o24-4);
            
            \draw[mod,->3-=0.52] (n24) to[out=60,in=120,looseness=1] node[pos=0.5,yshift=-4pt,text=black](){${\sss D}$} (o13);
            \draw[] (n24-2) to[out=60,in=120,looseness=1] (o13-1);
            \draw[mod] (n24-4) to[out=60,in=120,looseness=1] node[pos=0.5,yshift=6pt,text=black](){${\sss B}$} (o13-3);
            \node[mor, circle, inner sep=4pt] at (b123) {};    
            \node at (b123) {${\sss m}$};
            \node[yshift=9pt] at (b123-3) {${\sss \mu}$};
            \node[yshift=-6pt,xshift=-4pt] at (m13-1) {${\sss \alpha}$};
            \node[yshift=-6pt,xshift=4pt] at (m23-2) {${\sss \delta}$};
            \node[xshift=-10pt,yshift=-3pt] at (b1234-3) {${\sss A}$};
            \node[xshift=10pt,yshift=-3pt] at (c1234-4) {${\sss C}$};
            \node[yshift=3pt] at (m12) {${\sss n}$};
            \node[xshift=1.5pt,yshift=2.6pt] at (n13) {${\sss i}$};
            \node[xshift=2.5pt,yshift=-1.5pt] at (n34) {${\sss k}$};
            \node[xshift=-1.5pt,yshift=2.6pt] at (o24) {${\sss j}$};
            \node[xshift=-2.5pt,yshift=-1.5pt] at (o34) {${\sss l}$};
        }{}
        
        \ifthenelse{#1=2}{
            \clip (0.65,-1.35) rectangle (2.85,1.85);
            \coordinate (a1) at (0,0);
            \coordinate (a2) at (\l,0);
            \coordinate (a3) at (\l/2,{sqrt(3)*\l/2});	    	    	
        	\middleCoo{a1}{a2}{m12}{m12-1}{m12-2}{\sp};
        	\middleCoo{a2}{a3}{m23}{m23-2}{m23-3}{\sp};
        	\middleCoo{a1}{a3}{m13}{m13-1}{m13-3}{\sp};
        	\barycentre{a1}{a2}{a3}{b123}{b123-1}{b123-2}{b123-3}{\sp};
        	\foreach \x in {m12,m12-1,m12-2}{
        	    \coordinate (\x) at ($(\x)+(0,\red)$);
        	}
            \draw[obj,->1-=0.6] (b123) --node[pos=0.35,yshift=4pt, sloped] {${\sss \beta}$} (m13);
            \draw[obj,->1-=0.6] (b123) --node[pos=0.35,yshift=4pt, sloped] {${\sss \gamma}$} (m23);     
            \draw[mor] 
                (m13-1) to[out=-30, in=-30, looseness=1.5] (m13-3)
                (m23-2) to[out=-150, in=-150, looseness=1.5] (m23-3);
            \draw[mod, rounded corners=\r] 
                (m12-1) -- (b123-1) -- (m13-1)
                (m12-2) -- (b123-2) -- (m23-2)
                (m23-3) -- (b123-3) -- (m13-3);
                
            \def\sp{0.16};
            \coordinate (b1) at (\l/4,-\l/2+\red);
            \coordinate (b2) at (3*\l/4,-\l/2+\red);
            \coordinate (b3) at (\l/4,\red);
            \coordinate (b4) at (3*\l/4,\red);
            \middleCoo{b1}{b3}{n13}{n13-1}{n13-3}{\sp};
            \middleCoo{b2}{b4}{n24}{n24-2}{n24-4}{\sp};
            \middleCoo{b1}{b2}{n12}{n12-1}{n12-2}{\sp};
            \barycentreSq{b1}{b2}{b3}{b4}{b1234}{b1234-1}{b1234-2}{b1234-3}{b1234-4};
            \draw[mod,->3-=0.7, shorten <= 1pt, shorten >= 1pt] (n13) --node[pos=0.35,yshift=-4pt,text=black] {${\sss D}$} node[mask point, pos=0.5](mk1){} (n24);
            \clipAroundNode{mk1}{
                \draw[obj,->1-=0.6] (n12) --node[pos=0.6,xshift=-5pt] {${\sss \nu}$} (b123);  
            } 
            \draw[mor] 
                (n12-1) to[out=90, in=90, looseness=1.5] (n12-2)
                (n13-1) to[out=0, in=0, looseness=1.5] (n13-3)
                (n24-2) to[out=-180, in=-180, looseness=1.5] (n24-4);
            \draw[mod,rounded corners=\r]
                (m12-1) -- (b1234-3) -- (n13-3)
                (m12-2) -- (b1234-4) -- (n24-4);
              \draw[rounded corners=\r]  
                (n12-1) -- (b1234-1) -- (n13-1)
                (n12-2) -- (b1234-2) -- (n24-2);
            \node[mor, circle, inner sep=4pt] at (b123) {};    
            \node at (b123) {${\sss m}$};
            \node[xshift=-8pt,yshift=-8pt] at (b1234-1) {${\sss \alpha}$};
            \node[xshift=8pt,yshift=-8pt] at (b1234-2) {${\sss \delta}$};
            \node[xshift=-8pt,yshift=14pt] at (b1234-3) {${\sss A}$};
            \node[xshift=8pt,yshift=14pt] at (b1234-4) {${\sss C}$};
            \node[yshift=9pt] at (b123-3) {${\sss B}$};
            \node[yshift=3pt] at (n12) {${\sss n}$};
            \node[xshift=2pt,yshift=-1.5pt] at (m13) {${\sss k}$};
            \node[xshift=-2pt,yshift=-1.5pt] at (m23) {${\sss l}$};
            \node[xshift=3pt] at (n13) {${\sss i}$};
            \node[xshift=-3pt] at (n24) {${\sss j}$};
        }{}
	\end{tikzpicture}
}

\newcommand{\adjacency}[7]{
	\begin{tikzpicture}[scale=1,baseline={([yshift=-.5ex]current bounding box.center)}]
	    \ifthenelse{#1=1}{
    	    \coordinate (a) at (0,0);
    	    \coordinate (b) at (1,0);
    	    \coordinate (c) at ({1+cos(30)},0.5);
    	    \coordinate (d) at ({1+cos(30)},-0.5);
    	    \coordinate (e) at ($(c)+(1,0)$);
    	    \coordinate (f) at ($(d)+(1,0)$);
    	    \draw (a) -- (b) -- (c) -- (e);
    	    \draw (b) -- (b) -- (d) -- (f);
    	    \draw (c) -- (d);
    	    \draw (b) to[out=45, in=135, looseness=1, min distance=15pt] (b);
    	    \draw (c) to[out=45, in=135, looseness=1, min distance=15pt] (c);
    	    \draw (d) to[out=-45, in=-135, looseness=1, min distance=15pt] (d);
    	    \foreach \x in {a,b,c,d,e,f}{
    	        \node[draw=none, fill=black, circle, inner sep=1.2pt] at (\x) {};
    	    }
    	    \node[left] at (a) {${\sss #2}$};
    	    \node[right] at (e) {${\sss #3}$};
    	    \node[right, yshift=1pt] at (f) {${\sss #3^2}$};
    	    \node[below] at (b) {${\sss #4}$};
    	    \node[above right, yshift=-2pt] at (d) {${\sss #3^2 #4}$};
    	    \node[below right] at (c) {${\sss #3 #4}$};
	    }{}
	    \ifthenelse{#1=2}{
    	    \coordinate (a) at (0,0);
    	    \coordinate (b) at (1,0);
    	    \coordinate (c) at ({1+cos(30)},0.5);
    	    \coordinate (d) at ({1+cos(30)},-0.5);
    	    \draw (a) -- (b) -- (c);
    	    \draw (b) -- (d);
    	    \draw (b) to[out=45, in=135, looseness=1, min distance=15pt] (b);
    	    \draw (c) to[out=45, in=135, looseness=1, min distance=15pt] (c);
    	    \draw (d) to[out=45, in=135, looseness=1, min distance=15pt] (d);
    	    \draw (a) to[out=60, in=160, looseness=1] (c);
    	    \draw (a) to[out=-60, in=-160, looseness=1] (d);
    	    \foreach \x in {a,b,c,d}{
    	        \node[draw=none, fill=black, circle, inner sep=1.2pt] at (\x) {};
    	    }
    	    \node[left] at (a) {${\sss #5}$};
    	    \node[below] at (b) {${\sss #6}$};
    	    \node[right] at (d) {${\sss #5 #3^2}$};
    	    \node[right] at (c) {${\sss #5 #3}$};
	    }{}
	    \ifthenelse{#1=3}{
    	    \coordinate (a) at (0,0);
    	    \coordinate (b) at (1,0);
    	    \draw (a) -- (b);
    	    \draw (b) to[out=45, in=135, looseness=1, min distance=15pt] (b);
    	    \draw (b) to[out=45, in=135, looseness=1, min distance=22pt] (b);
    	    \draw (b) to[out=45, in=135, looseness=1, min distance=29pt] (b);
    	    \foreach \x in {a,b}{
    	        \node[draw=none, fill=black, circle, inner sep=1.2pt] at (\x) {};
    	    }
    	    \node[left] at (a) {${\sss #7}$};
    	    \node[below] at (b) {${\sss #4 #7}$};
	    }{}
	    \ifthenelse{#1=4}{
    	    \coordinate (a) at (0,0);
    	    \coordinate (b) at (1,0);
    	    \coordinate (c) at (2,0);
    	    \draw (a) -- (c);
    	    \draw (b) to[out=45, in=135, looseness=1, min distance=15pt] (b);
    	    \foreach \x in {a,b,c}{
    	        \node[draw=none, fill=black, circle, inner sep=1.2pt] at (\x) {};
    	    }
    	    \node[left] at (a) {${\sss #2}$};
    	    \node[below] at (b) {${\sss #3}$};
    	    \node[right] at (c) {${\sss #4}$};
	    }{}
	\end{tikzpicture}
}

%% file: _Introduction.tex
\section{Introduction}

\noindent
Since the dawn of mathematics, scholars have been fascinated by the existence of dualities. An early example is the idea of dual polyhedra obtained by interchanging faces and vertices, as described in volume XV of \emph{Euclid’s Elements}  by Isidore of Miletus, whereby the isocahedron is dual to the dodecahedron and the tetrahedron is self-dual \cite{CORFIELD201755}. Crucially, dual polyhedra share the same \emph{symmetry group}.  This notion of symmetry, and other abstract generalizations, have remained at the heart of the concept of duality in modern mathematics and physics. Dualities effectively express different ways in which abstract symmetries can establish themselves and their representations can be transmuted into one another.

Dualities play a particularly important role in the field of statistical mechanics and quantum phase transitions, and an essential part of the canon of quantum spin physics consists of constructions such as the Jordan-Wigner transformation \cite{1928ZPhy...47..631J,1961AnPhy..16..407L,RevModPhys.36.856,PhysRevLett.63.322,Batista:2001}, the Kramers-Wannier duality \cite{PhysRev.60.252} and generalizations thereof involving \emph{gauging} procedures \cite{weyl1929elektron,PhysRevD.11.395,RevModPhys.51.659, Fisher2004, RevModPhys.52.453,doi:10.1063/1.1665530}. Symmetries, in a generic sense as defined below, are again center stage here and the corresponding dualities relate theories that implement those symmetries in a different way. In the aforementioned examples, the duality transformation maps local Hamiltonians to local Hamiltonians. More generally, any symmetric local operator is mapped to a dual symmetric local operator. What makes the duality non-trivial is that local operators that are not symmetric in one theory are mapped to \emph{non-local} non-symmetric operators in the dual theory. We require these properties to hold for any (non-trivial) duality transformation.

An especially rich source of dualities has been the field of exactly solvable lattice models \cite{baxter}. Famously, the Kramers-Wannier duality enabled the exact determination of the critical temperature of the 2d Ising model \cite{PhysRev.60.252}, while the Jordan-Wigner transformation trivializes the computation of its partition function by mapping it to a theory of free fermions \cite{PhysRev.76.1232,PhysRev.76.1244,NishimoriOrtiz2011}. Typically, there is some algebraic structure underlying these integrable lattice models that ensures the existence of an exact solution. It is the identification of such an algebra that allowed Onsager to compute exactly the thermodynamics of the 2d Ising model \cite{PhysRev.65.117}. This so-called Onsager algebra, and generalizations thereof such as the Temperley-Lieb \cite{temperley1971relations} and Birman-Murakami-Wenzl \cite{birman1989braids,ojm/1200780357} algebras, have since been used to construct a wealth of integrable lattice models. It was realized early on that different representations of the same algebra allow one to construct dual integrable lattice models \cite{PhysRev.65.117,temperley1971relations,baxter,1988CMaPh.118..355P,roche1990ocneanu}. In this work, we reconsider this observation purely from the point of view of symmetries, and define duality without necessarily reference to a particular model.

Traditionally, symmetries are effected by group transformations and, typically in quantum systems, they are realized as unitary representations of the group. In this work we consider a more general notion of symmetry that involves transformations whose composition law agrees with that of a \emph{fusion ring}. This includes the traditional group symmetries, as well as \emph{non-invertible} transformations that do not admit a unitary representation. These generalized symmetries and their multiplication rules are encoded into higher mathematical structures known as \emph{fusion categories}, and as such they are most accurately referred to as \emph{categorical symmetries}. In quantum lattice systems, they are in general \emph{non-local}, in the sense that they cannot be realized as tensor products of local operators. Instead, they are realized as \emph{matrix product operators} (MPOs) \cite{Buerschaper:2013nga,bultinckMPOs,bultinckAnyons,hauruDefects,aasenDefects,williamsonSET,Aasen:2020jwb}, a tensor network parametrization which captures the non-trivial entanglement structure present in these operators \cite{pirvu2010matrix,haegeman2017diagonalizing,cirac2020matrix}. Mathematically, an MPO realization of a fusion category symmetry is described by a choice of \emph{module category} over this fusion category symmetry \cite{Lootens:2020mso}. The data contained in this module category allows one to construct the local tensors that build up the MPO, as well as all possible \emph{symmetric operators} that commute with these MPO symmetries.

A well known class of models exhibiting MPO symmetries are the \emph{anyonic chains} \cite{PhysRevLett.98.160409,PhysRevB.87.235120,PhysRevLett.101.050401,PhysRevLett.103.070401,ardonneAnyonChain,Buican:2017rxc}. These can be thought of as generalizations of the Heisenberg model, whereby the spin degrees of freedom are promoted to objects in a fusion category interpreted as topological charges of quasi-particles, and the tensor product of SU(2) representations is replaced by the fusion rules of the category \cite{etingof2016tensor}. It is well known that these models satisfy symmetry relations with respect to operators labelled by objects in a fusion category. Such categorical symmetries have received widespread attention in recent years  \cite{Kapustin:2009av,Kapustin:2010if,Bhardwaj:2017xup,Tachikawa:2017gyf,Thorngren:2019iar,Huang:2021ytb,Thorngren:2021yso,Kaidi:2021xfk,Choi:2021kmx,tantivasadakarn2021longrange}, including their applications in the much older classical statistical mechanics counterparts of the anyonic chains \cite{aasenDefects,Aasen:2020jwb}. In the case that these models are critical, their low-energy physics is described by a conformal field theory, in which these non-local symmetries are identified as lattice regularizations of the topological defects \cite{PhysRevB.86.155111,PhysRevB.101.134111,PhysRevLett.121.177203,changDefects} of the continuum field theory. These quantum spin Hamiltonians can then be understood as the gapless edge theories of a (2+1)d system with topological order \cite{hauruDefects,PhysRevLett.121.177203,Buican:2017rxc,lootens2019cardy,Kong:2019byq,Kong:2019cuu}, realizing a lattice version of the well known holographic relation between topological and conformal field theories (CFTs) \cite{cmp/1104178138,MOORE1989422,Fuchs:2007tx,Frohlich:2006ch,Fuchs:2010hk}.

It is then natural to exploit the powerful formalism of category theory to establish protocols realizing arbitrary duality maps, between seemingly unrelated quantum Hamiltonians, in a methodical fashion. The main contribution of this paper is to demonstrate that, at least for the case of one-dimensional quantum lattice systems, there is a systematic approach to constructing dualities between local Hamiltonians based on fusion and module categories. A key consequence of our categorical approach is that it provides a \emph{classification of all possible dualities of a given model Hamiltonian} based on the classification of fusion and module categories. Furthermore, our formulation explicitly provides \emph{MPO intertwiners} that implement the duality at the level of the Hilbert space. Those MPO connect local symmetric operators to dual local operators symmetric with respect to a different realization of the symmetry. Explicit dual Hamiltonians can then be constructed by taking linear combinations of such symmetric operators. While here we focus on the case of quantum chains, a completely equivalent exposition is possible for the case of two-dimensional models of classical statistical mechanics by replacing the Hamiltonian with a transfer matrix \cite{PhysRevLett.121.177203,Vanhove_2022} constructed from symmetric operators.

\begin{figure}[tb]
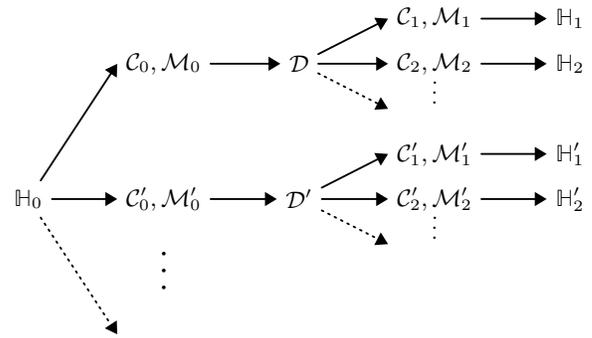

    \includeTikz{0}{flowChart}{\flowChart}
    \caption{Flowchart for generation of dualities. Given a Hamiltonian $\mathbb{H}_0$, we choose the MPO realization of the categorical symmetry $\mc{C}_0$, providing a module category ${\cal M}_0$ over $\mc C_0$. This data in turn stipulates an input category ${\cal D}$ corresponding to a bond algebra. One can then generate \emph{all} possible dual Hamiltonians $\mathbb{H}_i$ by choosing different module categories $\mc M_i$ over $\mc D$. Had one chosen a different MPO symmetry ${\cal C}'_0$ and module category ${\cal M}'_0$, one would have generated a different input category ${\cal D}'$ leading to a different set of dual maps.}
    \label{fig:Flowchart}
\end{figure}

More concretely, our recipe for generating duality maps is summarized in Fig. \ref{fig:Flowchart}. Given a Hamiltonian $\mathbb{H}_0$, we choose the symmetry that will be used to perform the duality. It does not need to be the full set of symmetries of $\mathbb{H}_0$. As mentioned above, in general this symmetry is described by a fusion category, that we call $\mc{C}_0$, and is realized as an MPO whose explicit representation is determined by the module category $\mc{M}_0$ over $\mc{C}_0$. From this data, one can construct all local symmetric operators that commute with these MPOs. These local symmetric operators are built from generalized Clebsch-Gordan coefficients, and they generate an algebra called the \emph{bond algebra} \cite{cobanera2010unified,bondAlgebra} to which $\mathbb{H}_0$ belongs. These generalized Clebsch-Gordan coefficients are obtained from the module category $\mc{M}_0$ thought of as a module category over another fusion category $\mc{D}$, that we refer to as the \emph{input category}. It is the input category $\mc{D}$ that governs the bond algebra, and it is defined by generalized 6j-symbols providing the \emph{recoupling theory} of the generalized Clebsch-Gordan coefficients. The crux of our approach is that one can find distinct realizations of the symmetric operators so that they satisfy the same algebraic relations by choosing different module categories $\mc{M}_i$ over the input category $\mc{D}$ \cite{Lootens:2020mso}. These local symmetric operators define dual models with distinct realizations of the same symmetry, while the bond algebra they generate remains the same. The corresponding dual symmetry MPOs are encoded into fusion categories $\mc{C}_i$ that satisfy the so-called \emph{Morita equivalence} \cite{etingof2016tensor}. Crucially, the categorical data of these different module categories $\mc{M}_i$ allows us to explicitly construct MPO intertwiners that implement the duality transformation between symmetric operators at the level of the states. 

Our construction makes the role of symmetries in dualities very explicit, as the MPO intertwiner only depends on the different choices of module categories; once this MPO is constructed, it will serve as the intertwiner between any two Hamiltonians constructed out of linear combinations of dual bond algebra elements. The resulting dual models can be thought of as generalizations of anyonic chains where the microscopic degrees of freedom are now determined by the module categories. Generally, the Hilbert spaces of these models are not necessarily tensor product spaces. Dual models may have very different local degrees of freedom, and may only be defined on a subspace where allowed configurations are subject to local constraints. This means dualities may be realized between a theory and a subspace of a different dual theory; this is known as {\it emergent duality} \cite{bondAlgebra}. Whenever the chosen duality map reveals itself as a (otherwise hidden) symmetry  of the original model Hamiltonian, as in the Kramers-Wannier example, the transformation is known as a {\it self-duality} \cite{cobanera2010unified}. A necessary (but not sufficient) condition for a given duality to be a self-duality is that  dual and original symmetries are the same. A key advantage of the categorical approach is that these dual symmetries can be simply determined, allowing us to rule out the possibility that certain dualities are self-dualities.

The central merit of the tensor network representation, and more specifically of the MPO construction, is the fact that global non-local duality transformations can be implemented in a local way at the cost of introducing additional entanglement degrees of freedom. This allows us to explicitly construct the \emph{isometries} relating dual Hamiltonians, ensuring their spectra are related, something extremely hard to achieve by other means given the non-local character of the transformation. Although we shall not discuss it here, let us remark that the intertwining MPO completely determines the {\it dual variables}, including those relating local order parameters of a system to {\it string order parameters} of its dual theory. Finally, note that the procedure of \emph{gauging} is incorporated in our framework of duality transformations, and our construction recovers the well-known fact that group-like symmetries represented with non-trivial 3-cocycles cannot be gauged as the required module categories do not exist.

The implementation of a duality at the level of the Hilbert space as an MPO intertwiner has direct physical implications. Generically, a duality transformation may map a gapped phase of a system to another different gapped phase. Recent work \cite{tantivasadakarn2021longrange} has exploited this property as a means to prepare states associated to a specified phase from a product (un-entangled) state. They showed, on a case-by-case basis, that certain dualities can be implemented using constant depth circuits supplemented with measurement and classical communication \cite{tantivasadakarn2021longrange}, and thus provide an operational recipe to implement them in experimental setups. Our exact MPO intertwiners provide the general theory for these explicit duality implementations, and translating them into the quantum circuit language will provide a way to realize a general duality transformation in experimental setups.

Another far-reaching mathematical and physical consequence of our category theoretical approach is the elucidation of the long-sought  {\it non-abelian duality} problem \cite{COBANERA2013574}. In the categorical formulation, we define a non-abelian duality as a duality whose fusion category $\mc{D}$ governing the bond algebra is derived from a non-abelian group, and at least one of the module categories involved in the duality is also based on a non-abelian group. Within our construction, we circumvent the difficulties associated with traditional approaches, which partly stem from these non-abelian dualities typically giving rise to dual symmetries that can no longer be described by a group. These findings could lead to rigorous derivations of generalized particle-vortex dualities \cite{PESKIN1978122}, some of them relating fermionic to bosonic theories, in (2+1)d quantum and topological field theories.

\medskip \noindent

An outline of this paper is as follows: First, we give a heuristic exposition of our construction in Sec. \ref{SectionII}. After reviewing relevant key concepts of category theory, we present our systematic recipe for Hamiltonian models with non-trivial dual theories based on the formalism of MPO symmetries, and relate this construction to the bond-algebra approach to dualities. Finally, in Sec. \ref{Section III}, we illustrate our approach with dualities between distinct realizations of symmetries encoded into various fusion categories. Using our formalism, we present a new emergent duality of the $t$-$J_z$ chain model, as well as other emergent dualities in sections \ref{sec:RepS3}, \ref{Vertexmodel} and \ref{Haagerup}. Non-abelian dualities, extensions, bulk-boundary correspondence and future work on higher spatial dimensions are discussed in Sec. \ref{Section IV}.

%% file: _Hamiltonian.tex
\section{Hamiltonians with categorical symmetries}
\label{SectionII}
\noindent
\emph{We present in this section a systematic construction of categorically symmetric local operators that yield dual {\rm (1+1)d} quantum Hamiltonians.}
\subsection{Heuristics}

\noindent
Before going into technical details, let us motivate our formalism. We consider (1+1)d quantum Hamiltonians---a priori defined on infinite chains---of the form $\mathbb H_{ A} = \sum_{\msf i} \mathbb h_{ A,\msf i}$, where for simplicity we restrict the local terms $\mathbb h_{A, \msf i}$ to include up to two-site interactions. In general, $\mathbb h_{ A,\msf i}$ is defined in terms of a three-valent tensor $A$ as
\begin{equation}
    \label{eq:localTerms}
    \mathbb h_{ A,\msf i} = \sum_k A_{i'l'}^k \tilde{A}_{il}^k \, | i' ,l'\ra \la i , l | \equiv 
    \includeTikz{0}{heuristics0A}{\heuristics{0}{A}{}} \, ,
\end{equation}
and from now on we will use the usual graphical notation for describing tensors and their contractions. We are interested in Hamiltonians that satisfy specific symmetry relations. At the level of the tensors $A$, these require the existence of MPO \emph{symmetries} satisfying so-called \emph{pulling-through} conditions:
\begin{equation}
    \label{eq:heuristics}
    \includeTikz{0}{heuristics1AObj}{\heuristics{1}{A}{speObj}} = 
    \includeTikz{0}{heuristics2AObj}{\heuristics{2}{A}{speObj}} \, ,
\end{equation}
where red curved lines depict indices along which MPO tensors get contracted to one another. Elementary examples of such tensors $A$ are built from Clebsch-Gordan coefficients of a finite group $G$, which are invariant under the action of any of its generators. The MPO symmetries are thus labeled by group variables and can be fused together via group multiplication. More generally, we are interested in MPO symmetries with operators that are not necessarily invertible and whose properties are encoded into a fusion category. The tensors $A$ can then be thought of as generalized Clebsch-Gordan coefficients invariant under the action of such operators.

A central insight of this paper is the following: Given a tensor $A$, we can often define another tensor $B$ related to $A$ via the existence of an MPO \emph{intertwiner}, which is {\it distinct} from an MPO symmetry, encoding the action of a duality:
\begin{equation}
    \label{eq:heuristics_intertwiner}
    \includeTikz{0}{heuristics1AMod}{\heuristics{1}{A}{mod}} = 
    \includeTikz{0}{heuristics2BMod}{\heuristics{2}{B}{mod}} \, .
\end{equation}
Crucially, tensors $B$ obtained in this way satisfy symmetry conditions of the form \eqref{eq:heuristics} with respect to MPOs whose properties are encoded into a fusion category that is similar---in a sense to be specified below---to that of the symmetry operators of $A$. The Hamiltonian $\mathbb H_B$ built out of tensors $B$ is then dual to $\mathbb H_A$. In particular, the MPO intertwiner can be used to construct an isometry connecting $\mathbb H_A$ and $\mathbb H_B$, thus implying a relation between their spectra.

More concretely, given MPO symmetries encoded into a given fusion category $\mc D$, the tensors $A$ can be constructed from a piece of data known as the $F$-symbols of $\mc D$. The existence of distinct tensors $B$ satisfying the properties outlined above is then guaranteed if there are distinct so-called \emph{module categories} over $\mc D$. These module categories $\mc M$ contain data that can be used to construct the aforementioned generalized Clebsch-Gordan coefficients. Analogously to the way ordinary Clebsch-Gordan coefficients can be recoupled via \emph{Wigner $6j$-symbols}, the generalized Clebsch-Gordan coefficients can be recoupled using the $F$-symbols of $\mc D$ regardless of the choice of $\mc M$. The fact that the tensors $A$ and $B$ share the same recoupling theory confirms that they are strongly related to one another. 

A particular manifestation of this relationship can be observed when considering the algebra generated by all local symmetric operators constructed from these generalized Clebsch-Gordan coefficients, the so-called \emph{bond algebra} \cite{bondAlgebra}. The bond algebras constructed from different generalized Clebsch-Gordan coefficients are isomorphic, since their structure constants are derived from the same recoupling theory. In \cite{bondAlgebra}, it was argued that this implies the existence of an isometry relating these bond algebras and by extension the two Hamiltonians, motivating the notion of isomorphic bond algebras as a formalization of duality. Our formulation of these concepts allows us to go beyond in that we are able to {\it explicitly construct this isometry map from the MPO intertwiners}. It turns out that the precise nature of this transformation requires a thorough understanding of the different symmetry sectors of the models, information that is naturally provided in the categorical formulation \cite{aasenDefects,PhysRevLett.121.177203}.

\bigskip
\noindent
To gain some intuition, consider the simple case of $\mathbb Z_2$ symmetry. For concreteness, we examine the transverse field Ising chain model, however, we note that the MPO intertwiners implementing the relevant dualities are the same for any $\mathbb Z_2$ symmetric model. We contemplate an infinite chain with $\mathbb Z_2$-valued ``matter'' degrees of freedom at sites labeled by half-integers, governed by the Hamiltonian $\mathbb H_A = \sum_{\msf i}\mathbb h_{A,\msf i}$, 
\begin{equation}
    \mathbb h_{A,\msf i} = -J( X_{\msf i-\frac{1}{2}}X_{\msf i+\frac{1}{2}} + gZ_{\msf i+\frac{1}{2}}) \, .
\end{equation}
This Hamiltonian has a (global) $\mathbb Z_2$ symmetry realized by the tensor product of Pauli $Z$ operators acting on half-integer sites. The local terms $\mathbb h_{A,\msf i}$ can be conveniently constructed from a tensor $A$ as in eq.~\eqref{eq:localTerms} such that the $\mathbb Z_2$ symmetry descends from the pulling-through condition
\begin{equation}
    \includeTikz{0}{heuristicsIsing1AObjZ}{\heuristicsIsing{1}{A}{speObj}{Z}} = \includeTikz{0}{heuristicsIsing2BObjZ}{\heuristicsIsing{2}{A}{speObj}{Z}} \, .
\end{equation}
A duality can be obtained by gauging the global $\mathbb Z_2$ symmetry, which in this setting is achieved by following the procedure outlined in \cite{PhysRevX.5.011024}. At sites labeled by integers---in-between the matter degrees of freedom---we introduce additional $\mathbb Z_2$ ``gauge'' degrees of freedom together with local constraints
\begin{equation}
    \mc G_{i+\frac{1}{2}} := Z_{\msf i} Z_{\msf i+\frac{1}{2}} Z_{\msf i+1} \stackrel{!}{=} \mathbb 1 \, .
\end{equation}
Local projectors onto the $\mc G = \mathbb 1$ subspaces are then given by $\mathbb P_\mc G = (\mathbb 1 + \mc G)/2$. Since projectors on different sites commute, states in the constrained subspace can be obtained by applying the following MPO intertwiner on the matter degrees of freedom
\begin{equation}
    \includeTikz{0}{heuristicsProjector}{\heuristicsProjector} \, ,
\end{equation}
at the expense of doubling the gauge degrees of freedom. This doubling is done to implement the constraints in a local way, with projectors centered on the matter degrees of freedom. While it is not strictly required in this case and we will even undo the doubling in a later step, it is indicative of the way the categorical framework implements these kinds of constraints. The matter degrees of freedom at sites ${\msf i + \frac{1}{2}}$ can be disentangled from the gauge degrees of freedom at sites ${\msf i}$ and ${\msf i+1}$ by acting with a local unitary and can therefore be discarded. A more conventional notation for the building block of this MPO is then provided by
\begin{align}
    \sum_{a,b=0,1} \!\!\!\! 
    \includeTikz{-1}{heuristicsKWMPO1}{\heuristicsKWMPO{a}{a+b}{b}{a}{b}{}{}{}}
    \equiv \!
    \sum_{a,b=0,1} |a \ra |a,b \ra \la b| \la a+b| \, ,
    \label{eq:interKW}
\end{align}
where the addition $a+b$ is modulo 2. The action of the MPO intertwiner on $\mathbb h_{A,\msf i}$ can be understood as follows. Acting on the $Z_{\msf i + \frac{1}{2}}$ term, we find the dual term
\begin{equation}
    \includeTikz{0}{heuristicsKWMPO2}{\heuristicsKWMPO{}{}{}{}{}{Z}{}{}} \!\! = \!\! \includeTikz{0}{heuristicsKWMPO3}{\heuristicsKWMPO{}{}{}{}{}{}{Z}{Z}} \!\! ,
\end{equation}
while for $X_{\msf i-\frac{1}{2}}X_{\msf i+\frac{1}{2}}$, the dual term is
\begin{equation}
    \includeTikz{0}{heuristicsKWMPO4}{\heuristicsKWMPOBis{X}{}} \; = \; \includeTikz{0}{heuristicsKWMPO5}{\heuristicsKWMPOBis{}{X}} \; .
\end{equation}
These dual terms can be written in terms of a tensor $B$, which now satisfies the pulling-through condition
\begin{equation}
    \includeTikz{0}{heuristics1BObjX}{\heuristicsIsing{1}{B}{speObj}{XX}} = \includeTikz{0}{heuristics2BObjX}{\heuristicsIsing{2}{B}{speObj}{XX}} \, .
\end{equation}
Notice that due to the doubling, we have two gauge degrees of freedom per site $\msf i + \frac{1}{2}$, and as such the MPO symmetry locally acts as $X_{\msf i} X_{\msf i+1}$. Removing this redundancy leads to tensors symmetric with respect to tensor products of Pauli $X$ operators acting on integer sites, which is an equivalent but distinct implementation of the global $\mathbb Z_2$ symmetry. Putting everything together, we obtain the dual Hamiltonian $\mathbb H_B = \sum_{\msf i}\mathbb h_{B,\msf i}$ such that 
\begin{equation}
    \mathbb h_{B,\msf i} = -J(X_{\msf i}+gZ_{\msf i} Z_{\msf i+1}) \, .
\end{equation}
This Hamiltonian has a global $\mathbb Z_2$ symmetry with respect to a tensor product of Pauli $X$ operators acting on the integer sites. Up to a local change of basis, we recognize the Kramers-Wannier dual of the initial Hamiltonian. Despite having derived the MPO intertwiner in the context of the Ising model, it is general in the sense that it performs the Kramers-Wannier duality for a generic $\mathbb Z_2$ symmetric model.

A second duality of the Ising model---perhaps not always thought of as such---is the mapping to free fermions via the Jordan-Wigner transformation \cite{bondAlgebra}. In essence, this transformation reinterprets a bosonic spin up/down degree of freedom at some site as the presence/absence of a fermion at that site. For a single spin, this is achieved through the following substitution:
\begin{equation}
    S^+_{\msf i} = \frac{1}{2}(X_{\msf i} + i Y_{\msf i}) \mapsto c^\dag_{\msf i} \, , \q
    S^-_{\msf i} = \frac{1}{2}(X_{\msf i} - i Y_{\msf i}) \mapsto c^{\;}_{\msf i} \, ,
\end{equation}
where the fermionic creation operator $c_{\msf i}^\dagger$ satisfies canonical anticommutation relations, $\{c_{\msf i}^{\;},c_{\msf j}^\dagger\}=\delta_{\msf i \msf j}$, whereas independent spin operators commute. For several spins, one is thus required to consider the transformation \cite{1928ZPhy...47..631J}
\begin{equation}
    \label{eq:JWtransfo}
    S^+_{\msf i} \mapsto K_{\msf i}^{\;} c^\dag_{\msf i} \, , \q
    S^-_{\msf i} \mapsto K_{\msf i}^{\;} c^{\;}_{\msf i} \, , \q
    Z_{\msf i} \mapsto 1-2c^\dag_{\msf i}c^{\;}_{\msf i} \, ,
\end{equation}
with $K^{\;}_{\msf i} = \exp{\big(i\pi\sum_{\msf j=-\infty}^{\msf i - 1} c^\dag_{\msf j}c^{\;}_{\msf j}\big)}$ an operator that defines the total fermionic parity at sites $\msf j < \msf i$, ensuring that the spin operators satisfy the expected commutative algebra. Applying these relations to $\mathbb H_A$ yields the dual Hamiltonian\footnote{When applying the transformations given in eq.~\eqref{eq:JWtransfo} to $\mathbb H_A$, we are implicitly performing the shift $\msf i \mapsto \msf i+\frac{1}{2}$.}
\begin{align}
    \nn
    \mathbb H_C = - J \sum_{\msf i} \big(
    &c^\dagger_{\msf i-\frac{1}{2}} c^{\phantom \dagger}_{\msf i+\frac{1}{2}} + c^\dagger_{\msf i-\frac{1}{2}} c^\dagger_{\msf i+\frac{1}{2}} + \text{h.c.}
    \\[-0.5em]
    &- g(2c^\dagger_{\msf i+\frac{1}{2}} c^{\phantom \dagger}_{\msf i + \frac{1}{2}} - 1) \big) \, ,
\end{align}
a model of non-interacting fermions that can be readily diagonalized. 
The Jordan-Wigner transformation can be interpreted as an MPO intertwiner, with the MPO represented in the form of a fermionic tensor network \cite{bultinck2017fermionic}. Denote by $n_{\msf i}(a) = 0,1$ the fermionic occupation number at site $\msf i$ with $n_{\msf i}(0)=0$ and $n_{\msf i}(1)=1$ such that $|n_{\msf i}(a)\ra \equiv (c_{\msf i}^\dagger)^{n_{\msf i}(a)}| \varnothing \ra$. It follows from the anticommutation relations of fermionic operators that $|n_{\msf i}(a)\ra | n_{\msf j}(b) \ra = (-1)^{ab}|n_{\msf j}(b)\ra | n_{\msf i}(a) \ra$. Given these notations, the MPO intertwiner implementing the Jordan-Wigner transformation has building block
\begin{align}
    \nn
    &\sum_{a,b=0,1} \!\!\!\! 
    \includeTikz{0}{heuristicsJWMPO}{\heuristicsJWMPO{n_{\msf i-\frac{1}{2}}(a)}{b}{n_{\msf i+\frac{1}{2}}(a+b)}{n_{\msf i}(b)}}
    \\
    &\q \equiv 
    \sum_{a,b=0,1} |n_{\msf i-\frac{1}{2}}(a) \ra |n_{\msf i}(b) \ra \la n_{\msf i+\frac{1}{2}}(a+b)| \la b| \, ,
    \label{eq:interJW}
\end{align}
where again additions are modulo 2. Notice that the total parity of fermionic occupation numbers of this tensor is even. This implies that when constructing the MPO intertwiner by contracting copies of this building block along virtual degrees of freedom, the parity of the virtual occupation number $n_{\msf i}(a)$ at site $\msf i$ is equal to the total parity associated with fermions at sites $\msf j < \msf i$. Acting with $X_{\msf i} = S^+_{\msf i} + S^-_{\msf i}$ on the spin degree of freedom of the MPO tensor at site $\msf i$, it follows from the previous remark that
\begin{align}
    &\sum_{a,b} |n_{\msf i-\frac{1}{2}}(a) \ra |n_{\msf i}(b) \ra \la n_{\msf i+\frac{1}{2}}(a+b)| \la b+1| 
    \\
    \nn 
    & \q = \sum_{a,b} |n_{\msf i-\frac{1}{2}}(a) \ra |n_{\msf i}(b+1) \ra \la n_{\msf i+\frac{1}{2}}(a+b+1)| \la b|
    \\
    \nn
    & \q =  \sum_{a,b} |n_{\msf i-\frac{1}{2}}(a) \ra (c^\dag_{\msf i} + c^{\;}_{\msf i})|n_{\msf i}(b) \ra \la n_{\msf i+\frac{1}{2}}(a+b+1)| \la b|
    \\
    \nn
    & \q =  \sum_{a,b}  K^{\;}_{\msf i} (c^\dag_{\msf i} + c^{\;}_{\msf i}) |n_{\msf i-\frac{1}{2}}(a) \ra |n_{\msf i}(b) \ra \la n_{\msf i+\frac{1}{2}}(a+b+1)| \la b| \, .
\end{align}
Denoting the MPO intertwiner as $O_{\text JW}$ and noting that $c^\dag_{\msf i} + c^{\;}_{\msf i}$ commutes with the even MPO tensors, one can write
\begin{equation}
    O_{\text JW} X_{\msf i} = K_{\msf i}(c^\dag_{\msf i} + c^{\;}_{\msf i}) O_{\text JW}',
\end{equation}
showing that this MPO intertwiner indeed implements the Jordan-Wigner transformation. Here, $O_{\text JW}'$ has an additional operator on the virtual leg between sites $\msf{i}$ and $\msf{i+1}$ that flips the parity. As in the Kramers-Wannier duality example, the MPO intertwiner performs the Jordan-Wigner transformation for any $\mathbb Z_2$ symmetric model, and can be obtained from the categorical formulation.

\subsection{Technical preliminaries}

\noindent
As alluded to above, our construction requires two pieces of category theoretical data, namely a \emph{fusion category} and a \emph{module category} over it (see \cite{etingof2016tensor} for precise definitions). Let $\mc D$ be a (spherical) fusion category. Succinctly, it is a collection of \emph{simple objects} interpreted as topological charges of various quasi-particles that can fuse with one another. Throughout this manuscript, isomorphism classes of simple objects in $\mc D$  are notated via $\alpha, \beta, \ldots \in \mc I_\mc D$ (Greek lowercase letter) and their respective quantum dimensions via $d_\alpha, d_\beta, \ldots \in \mathbb C$. The fusion of objects is defined by the so-called \emph{monoidal} structure $(\otimes,\mathbb 1,\F{})$ of $\mc D$, where $\otimes$ is an \emph{associative} product rule, $\mathbb 1$ is a distinguished object interpreted as the \emph{trivial} charge and $F$ is an isomorphism $F : \alpha \otimes (\beta \otimes \gamma) \xrightarrow{\sim} (\alpha \otimes \beta) \otimes \gamma$ referred to as the \emph{monoidal associator}. Denoting by $\mc H^{\gamma}_{\alpha,\beta} := \Hom_{\mc D}(\alpha \otimes \beta, \gamma) \ni |\alpha \beta \gamma ,i \ra$ the vector space of maps from $\alpha \otimes \beta$ to $\gamma$, we have $\alpha \otimes \beta \simeq \bigoplus_\gamma N_{\alpha \beta}^\gamma \, \gamma$, where $N_{\alpha \beta}^\gamma := {\rm dim}_\mathbb C \, \mc H^{\gamma}_{\alpha,\beta} \in \mathbb N$ are the fusion \emph{multiplicities}. The monoidal associator boils down to a collection of isomorphisms
\begin{equation}
    \F{}^{\alpha \beta \gamma}_\delta: 
    \bigoplus_{\nu}\mc H_{\beta,\gamma}^\nu \otimes \mc H_{\alpha,\nu}^\delta
    \xrightarrow{\sim} 
    \bigoplus_{\mu}\mc H_{\alpha, \beta}^\mu \otimes \mc H_{\mu, \gamma}^\delta  
    \label{eq:Fdef}
\end{equation}
that can be graphically depicted as
\begin{equation}
    \includeTikz{0}{monoidalAssociator2}{\monoidalAssociator{2}} \!\!\!\!\!\! 
    = \sum_{\mu} \sum_{i,l}
    \big( \F{}^{\alpha \beta \gamma}_\delta\big)^{\nu,jk}_{\mu,il}
    \includeTikz{0}{monoidalAssociator1}{\monoidalAssociator{1}} \, ,
\end{equation}
where $i,j,k,l$ label basis vectors in $\mc H_{\alpha, \beta}^\mu$, $\mc H_{\beta,\gamma}^\nu$, $\mc H_{\alpha,\nu}^\delta$ and $\mc H_{\mu,\gamma}^\delta$, respectively. In these diagrams, tensor products are ordered from top to bottom in agreement with eq.~\eqref{eq:Fdef}.

Let us now consider a (finite semi-simple $\mathbb C$-linear) right module category $\mc M$ over $\mc D$, which is broadly speaking a category with a right \emph{action} over $\mc D$. Henceforth, isomorphism classes of simple objects in $\mc M$ are denoted by $A,B,\ldots \in \mc I_\mc M$ (Roman capital letters). 
Simple objects of $\mc M$ act on the topological charges of $\mc D$ via the module structure $(\cat,\F{\cat})$, where $- \cat -: \mc M \times \mc D \to \mc M$ is the action and $\F{\cat}$ is an isomorphism $\F{\cat}: A \cat (\alpha \otimes \beta) \xrightarrow{\sim} (A \cat \alpha) \cat \beta$ referred to as the \emph{module associator}. For instance, every fusion category $\mc D$ defines a module category over itself, which we refer to as the \emph{regular} module category. Introducing the notation $\mc V_{A,\alpha}^B := \Hom_{\mc M}(A \cat \alpha,B) \ni |A \alpha B,i \ra$, the module associator $\F{\cat}$ boils down to a collection of isomorphisms
\begin{equation}
    \F{\cat}^{A \alpha \beta}_B: 
    \bigoplus_{\gamma}\mc H_{\alpha,\beta}^\gamma \otimes \mc V_{A,\gamma}^B
    \xrightarrow{\sim} 
    \bigoplus_{C}\mc V_{A,\alpha}^C \otimes \mc V_{C,\beta}^B  
\end{equation}
that can be depicted as
\begin{equation}
    \label{eq:FcatMove}
    \includeTikz{0}{moduleAssociator2}{\moduleAssociator{2}} \!\!\!\!\!\! 
    = \sum_{C} \sum_{i,l}
    \big( \F{\cat}^{A \alpha \beta}_B\big)^{\gamma,jk}_{C,il}
    \;\; 
    \includeTikz{0}{moduleAssociator1}{\moduleAssociator{1}} \, ,
\end{equation}
where $i,j,k,l$ label basis vectors in $\mc V_{A,\alpha}^C$, $\mc H_{\alpha,\beta}^\gamma$, $\mc V_{A,\gamma}^B$ and $\mc V_{C,\beta}^B$, respectively. Notice that we use a different colour (purple) for strands labelled by simple objects ($A,B,\ldots$) in the module category $\cal M$, and we shall refer to those as `module strands'.
The matrix entries of these isomorphisms, which are referred to as $\F{\cat}$-\emph{symbols}, can be graphically represented as\footnote{Up to aesthetic considerations, these diagrams can be thought as 2d projections of decorated tetrahedra, where the gray patches are identified with the vertices.}
\begin{equation}
    \begin{split}
        \label{eq:FSymbols}
        \includeTikz{0}{PEPSF}{\PEPS{}{i}{j}{k}{l}{A}{B}{C}{\alpha}{\beta}{\gamma}{1}} :=
        &\big( \F{\cat}^{A \alpha \beta}_B\big)^{\gamma,jk}_{C,il} \, ,
        \\[-3em]
        &\big( \Fbar{\cat}^{A \alpha \beta}_B\big)^{\gamma,jk}_{C,il} =:
        \includeTikz{0}{PEPSFBar}{\PEPS{}{i}{j}{k}{l}{A}{B}{C}{\alpha}{\beta}{\gamma}{2}} \, ,
    \end{split}
\end{equation}
where, for completeness, we also included the matrix entries associated with the inverse $\Fbar{\cat}$ of the module associator. Unless otherwise stated, we work with $\F{\cat}$ symbols that are unitary and real, which in our convention implies $\big( \F{\cat}^{A \alpha \beta}_B\big)^{\gamma,jk}_{C,il} = \big(\Fbar{\cat}^{A \alpha \beta}_B\big)^{\gamma,jk}_{C,il}$. Note that we kept the module strands unoriented as the corresponding labels will be summed over in practice. By convention, $\F{\cat}$-symbols for which the fusion rules are not everywhere satisfied vanish. Crucially, the module associator $\F{\cat}$ must satisfy a consistency condition, known as the \emph{pentagon axiom}, involving the monoidal associator $\F{}$ and which ensures that the equation
\begin{equation}
    \label{eq:pentagon}
    \sum_q \!\!\! \includeTikz{0}{pent2}{\pent{2}} \hspace{-2em} = \sum_\mu \! \sum_{i,l,p} \! \big( \F{}^{\alpha \beta \gamma}_\delta\big)^{\nu,jk}_{\mu,il} \hspace{-1.8em} \includeTikz{0}{pent1}{\pent{1}}
\end{equation}
holds for any choice of simple objects and basis vectors. Henceforth, we omit drawing gray patches associated with basis vectors that are being contracted (e.g., $p$ and $q$ in the previous equation). 

By interpreting the diagrams in eq.~\eqref{eq:FSymbols} as the non-vanishing components of four-valent tensors, it follows from eq.~\eqref{eq:pentagon} that these tensors can be used to define a tensor network representation \cite{Buerschaper:2013nga,PhysRevB.79.085119,PhysRevB.79.085118,bultinckMPOs,bultinckAnyons} of the ground state subspace of a \emph{string-net model} with input data $\mc D$ \cite{Turaev:1992hq,Barrett:1993ab,Levin:2004mi,kirillov2011stringnet}. Crucially, these tensors exhibit symmetry conditions with respect to non-trivial MPOs defined by tensors whose non-vanishing components are of the form
\begin{equation}
    \includeTikz{0}{MPO}{\MPO{mod}{mod}{i}{j}{k}{l}{A}{B}{D}{C}{\alpha}{a}{1}} \, .
\end{equation}
The symmetry conditions then ensure that these operators can be freely deformed throughout the tensor network away from their endpoints according to the pulling-through condition
\begin{equation}
    \label{eq:pulling}
    \sum_F \, 
    \includeTikz{0}{pulling1}{\pulling{1}} \; = \!\!
    \includeTikz{0}{pulling2}{\pulling{2}}  .
\end{equation} 
These symmetry operators, whose properties are encoded into another fusion category $\mc C \cong \mc D_{\mc M}^*$ known as the \emph{Morita dual} of $\mc D$ with respect to $\mc M$ \cite{etingof2016tensor}, can then be used to characterize degenerate ground states and create \emph{anyonic excitations} of the topological string-net model.  
In this context, different choices of module categories $\mc M$ yields different tensor network representations of the same topological model \cite{Lootens:2020mso}. Interestingly, given a string-net model, it is possible to define distinct tensor network representations across different regions of the underlying lattice via the introduction of MPO intertwiners. Crucially, these intertwining operators can be fused with the symmetry operators associated with either representation in an associative way. Analogously to the symmetry operators, these can be freely moved through the lattice ensuring that the tensor network representations are \emph{locally indistinguishable}.

\subsection{Categorically symmetric local operators}

\noindent
Before addressing our recipe for constructing duality maps, let us consider the following situation:
Let $\mc D$ be a fusion category and $\mc M$ a module category over it. Invoking the graphical calculus sketched above, every tree-like diagram of the form
\begin{equation}
    \includeTikz{5}{hamDerivation0}{\hamDerivation{0}} \, ,
\end{equation}
labelled by simple objects in $\mc I_D$ and $\mc I_\mc M$ such that the vector spaces $\mc V_{A, \tilde \alpha}^{\tilde C}$ and $\mc V_{\tilde C, \tilde \beta}^{B}$ are non-trivial, is interpreted as a state $\cdots \otimes | A \tilde \alpha \tilde C,\tilde i \ra \otimes | \tilde C \tilde \beta B, \tilde l \ra \otimes  \cdots $ in a Hilbert space. Diagrams of the form
\begin{equation}
    \includeTikz{0}{hamDerivation1}{\hamDerivation{1}}
\end{equation}
labelled by simple objects in $\mc I_D$ such that the vector spaces $\mc H_{\alpha,\beta}^\gamma$ and $(\mc H_{\tilde \alpha, \tilde \beta}^\gamma)^\star$ are non-trivial can then be interpreted as \emph{local} operators acting on such a Hilbert space of tree-like diagrams. The action of these operators can be readily computed via the change of basis provided by eq.~\eqref{eq:FcatMove} and its inverse:
\begin{equation*}
    \includeTikz{0}{hamDerivation1}{\hamDerivation{1}}
    \!\!\!\!\!\circ \; \includeTikz{9}{hamDerivation2}{\hamDerivation{2}}\, := \, \includeTikz{29}{hamDerivation3}{\hamDerivation{3}}
    \\[-2.4em]
\end{equation*}
\begin{align}
    \label{eq:hamDerivation}
    & \q =
    \sum_k \big( \Fbar{\cat}^{A \tilde \alpha \tilde \beta}_B\big)^{\gamma,\tilde j k}_{\tilde C,\tilde i \tilde l}
    \, \includeTikz{19}{hamDerivation4}{\hamDerivation{4}}
    \\ & \q =
    \sum_{k,i,l}\sum_{C}    \big( \F{\cat}^{A \alpha \beta}_B\big)^{\gamma,jk}_{C,il} \,
    \big( \Fbar{\cat}^{A \tilde \alpha \tilde \beta}_B\big)^{\gamma,\tilde j k}_{\tilde C,\tilde i \tilde l} \, 
    \includeTikz{9}{hamDerivation5}{\hamDerivation{5}} \, .
    \nn
\end{align}
This is a generalization of the usual anyonic chain construction \cite{PhysRevLett.98.160409,PhysRevB.87.235120,PhysRevLett.101.050401,PhysRevLett.103.070401,ardonneAnyonChain,Buican:2017rxc}.
Note that the definition of these operators only depends on $\mc D$, whereas the Hilbert space is specified by a choice of $\mc M$, suggesting that distinct choices of $\mc D$-module categories should yield dual models. 

Let us now formalize this construction.
Given a spherical fusion category $\mc D$ and a $\mc D$-module category $\mc M$, we are interested in local operators acting on (total) Hilbert spaces of the form 
\begin{align}
    \label{eq:Hilbert}
    \mc H &= \bigoplus_{\{A\}}\bigoplus_{\{\alpha\}} \bigotimes_{\msf i}
    \mc V_{\msf i + \frac{1}{2}}
    \\ \nn
    &\equiv \bigoplus_{\{A\}} \bigoplus_{\{\alpha\}}
    \includeTikz{1}{chainGen}{\chain{\mc V_{\msf i- \! \frac{3}{2}}}{\mc V_{\msf i- \! \frac{1}{2}}}{\mc V_{\msf i+ \! \frac{1}{2}}}{A_{\msf i-2}}{A_{\msf i-1}}{A_{\msf i}}{A_{\msf i+1}}{\alpha_{\msf i - \frac{3}{2}}}{\alpha_{\msf i - \frac{1}{2}}}{\alpha_{\msf i + \frac{1}{2}}}}
\end{align}
with $\mc V_{\msf i + \frac{1}{2}} := \Hom_{\mc M}(A_\msf i \cat \alpha_{\msf i + \frac{1}{2} }, A_{\msf i +1})$. Throughout this manuscript, we shall implicitly work with infinite chains, unless otherwise stated. Importantly, the Hilbert space \eqref{eq:Hilbert} is typically not a tensor product of local Hilbert spaces. We choose the convention that unlabelled module strands denote the morphism
\begin{equation}
    \label{eq:unlabelled}
    \includeTikz{2.4}{objectMod}{\object{mod}{}} \equiv \sum_{A \in \mc I_\mc M} \sum_{i} \includeTikz{1}{morphismMod}{\morphism{mod}{A}{i}} \, ,
\end{equation}
where the second sum is over basis vectors in the endomorphism spaces $\End_{\mc M}(A)$. As suggested by the definition, we shall also consider (fermionic) module categories whose simple objects may have non-trivial endomorphism algebras, but from now on we shall take all endomorphism spaces to be isomorphic to $\mathbb C$ unless otherwise stated. In the same spirit, unlabelled gray patches as depicted below are provided by
\begin{equation}
    \begin{split}
        \label{eq:unlabelledPatch}
        \includeTikz{0}{splitSpace1}{\splitSpace{A}{\alpha}{B}{}{1}} &\equiv
        \sum_{i}
        \includeTikz{0}{splitSpace2}{\splitSpace{A}{\alpha}{B}{i}{1}} 
        |A\alpha B, i \ra 
        \\
        \includeTikz{0}{splitSpace3}{\splitSpace{A}{\alpha}{B}{}{2}} &\equiv
        \sum_{i}
        \includeTikz{0}{splitSpace4}{\splitSpace{A}{\alpha}{B}{i}{2}} \la A\alpha B, i | 
    \end{split} \, ,
\end{equation}
where $|A \alpha B,i\ra \in \mc V^{B}_{A,\alpha}$ and $\la A \alpha B,i| \in (\mc V^B_{A,\alpha})^\star$ are basis vectors, for any $i = 1, \dots, {\rm dim}_\mathbb C \, \mc V^B_{A,\alpha}$. Given this notation, we consider local operators $\mathbb b_{a,\msf i}^{\mc M}$ of the form
\begin{equation}
    \label{eq:bondDef}
    \mathbb b_{a, {\msf i}}^{\mc M} \equiv
    \sum_{\substack{\alpha, \beta,\gamma \\ \tilde \alpha , \tilde \beta}} \sum_{j ,\tilde j}
    b_a(\alpha,\tilde \alpha, \beta, \tilde \beta, \gamma,j,\tilde j)
    \!\! 
    \includeTikz{0}{HamGen1}{\HAM{}{}{}{\tilde j}{}{}{j}{}{}{}{}{\tilde \alpha}{\alpha}{\tilde \beta}{\beta}{\gamma}}  ,
\end{equation}
where $b_a(\alpha,\tilde \alpha, \beta, \tilde \beta, \gamma, j, \tilde j) \in \mathbb C$. In virtue of eqs.~(\ref{eq:FSymbols},\ref{eq:unlabelled},\ref{eq:unlabelledPatch}), the operator on the r.h.s. is such that
\begin{align}
    &\big(\la A  \alpha  C,  i| \otimes \la C  \beta B ,  l | \big) 
    \, \mathbb b_{a,\msf i}^{\mc M} \, 
    \big(|\tilde C \tilde \beta B ,\tilde l\ra \otimes |A \tilde \alpha \tilde C,\tilde i \ra  \big)
    \\
    \nn
    & \;\; = 
    \sum_\gamma \sum_{k, j, \tilde j}\!
    b_a(\alpha, \tilde \alpha, \beta, \tilde \beta, \gamma, j , \tilde j)
    \big( \F{\cat}^{A \alpha \beta}_B\big)^{\gamma,jk}_{C,il} \,
    \big( \Fbar{\cat}^{A \tilde \alpha \tilde \beta}_B\big)^{\gamma,\tilde j k}_{\tilde C,\tilde i \tilde l} ,
\end{align}
where the contribution of the $\F{\cat}$-symbols
\begin{equation}
    \sum_k
    \big( \F{\cat}^{A \alpha \beta}_B\big)^{\gamma,jk}_{C,il} \,
    \big( \Fbar{\cat}^{A \tilde \alpha \tilde \beta}_B\big)^{\gamma,\tilde j k}_{\tilde C,\tilde i \tilde l} =
    \includeTikz{0}{HamGen2}{\HAM{}{\tilde i}{\tilde l}{\tilde j}{i}{l}{j}{A}{B}{C}{\tilde C}{\tilde \alpha}{\alpha}{\tilde \beta}{\beta}{\gamma}}
\end{equation}
matches that in eq.~\eqref{eq:hamDerivation}.

We commented earlier that the tensors whose non-vanishing components evaluate to the $\F{\cat}$-symbols satisfy symmetry conditions translating into pulling-through conditions of the form depicted in eq.~\eqref{eq:pulling}. It follows from the definition of $\mathbb b_{a,\msf i}^{\mc M}$ in terms of these tensors that they commute with the corresponding symmetry MPOs \cite{PhysRevLett.98.160409,PhysRevB.87.235120,Buican:2017rxc}. These non-local operators and their properties being encoded into (fusion) categories---in contrast to mere groups for instance---justifies why we refer to the $\mathbb b_{a,\msf i}^{\mc M}$ defined in this section as \emph{categorically symmetric} operators. We point out that we restrict ourselves to two-site operators for simplicity. Multiple-site operators can be constructed in the same way.

\subsection{Bond algebras and duality}

\noindent
Given an input category $\mc D$ and the symmetric local operators $\mathbb b_{a,\msf i}^{\mc M}$, we would like to argue that the different representations of the $\mathbb b_{a,\msf i}^{\mc M}$ associated with any choice of $\mc D$-module category $\mc M$ provide a way of constructing dual models. The concept of quantum duality was formalized in \cite{bondAlgebra} introducing the notion of bond algebras. This formulation turns out to be particularly suited to our construction. 

Consider the set of all symmetric local operators $\mathbb b_{a,\msf i}^{\mc M}$ as defined previously and let us refer to them as \emph{bonds}. These bonds define an algebra $\mc A\{\mathbb b_{a,{\msf i}}^{\mc M}\}$ known as the \emph{bond algebra}, generated by taking all possible finite products of all possible bonds, as well as the identity operator:
\begin{equation}
   \{\mathbb{id}, \mathbb b_{a,{\msf i}}^{\mc M}, \mathbb b_{b,{\msf j}}^{\mc M} \mathbb b_{c,{\msf k}}^{\mc M}, 
   \mathbb b_{a,\msf i}^{\mc M} \mathbb b_{b,\msf j}^{\mc M} \mathbb b_{c, \msf k}^{\mc M}, \ldots\} \, .
\end{equation}
In general, elements of the bond algebra as defined above are not all linearly independent but we can find a basis $\{\mc O_x^{\mc M}\}$ so that bonds and products of bonds can be decomposed into it. By definition, these basis elements satisfy \emph{operator product expansions}
\begin{equation}
    \mc O_x^{\mc M} \mc O_y^{\mc M} = \sum_z f_{xy}^{z,\mc M} \mc O_z^{\mc M} \, ,
\end{equation}
where $f_{xy}^{z,\mc M}$ are the \emph{structure constants} of the bond algebra so that bond algebras with the same structure constants are isomorphic.

Given our definition of the bonds in eq.~\eqref{eq:bondDef}, products of bonds are computed by invoking the recoupling theory encoded into eq.~\eqref{eq:pentagon} of tensors \eqref{eq:FSymbols} thought as some generalized Clebsch-Gordan coefficients. For a choice of basis of the bond algebra, repeated use of eq.~\eqref{eq:pentagon} can thus be used to explicitly compute the structure constants. Crucially, this recoupling theory is invariant under a change of $\mc D$-module category $\mc M$. Indeed, it is manifest from eq.~\eqref{eq:pentagon} that recoupling does not involve the objects in $\mc M$ and only depends on the monoidal structure of $\mc D$ via its $\F{}$-symbols. As such the structure constants of the bond algebras associated with our models only depend on the choice of $\mc D$:
\begin{equation}
    f_{xy}^{z,\mc M} = f_{xy}^z(F) \, .
\end{equation}
Consequently, the bonds $\{\mathbb b_{a,\msf i}^{\mc M}\}$ generate isomorphic bond algebras for any choice of $\mc D$-module category $\mc M$. It follows from the results in \cite{bondAlgebra} that categorically symmetric local operators that only differ by the choice of $\mc D$-module category are \emph{dual} to one another, formalizing the intuition that dualities are maps between local operators preserving their algebraic relations.

In our formalism, the existence of such an isomorphism between the two bond algebras is ensured by the fact that we have access to MPO intertwiners mapping bonds associated with distinct module categories onto one another. In the scenario where one of the modules is taken to be the regular module category, these intertwining operators admit particularly simple expressions in terms of tensors whose non-vanishing components evaluate to the $\F{\cat}$-symbols:
\begin{equation}
    \includeTikz{0}{MPOintertwiner}{\MPO{obj}{mod}{i}{j}{k}{l}{A}{B}{\beta}{\alpha}{\gamma}{C}{0}} 
    = \big(\F{\cat}^{C\alpha \gamma}_B\big)^{\beta,lj}_{A,ik} \, .
    \label{eq:interDef}
\end{equation}
Pulling such an intertwining operator through bonds associated with the regular module category yields bonds associated with $\mc M$ according to
\begin{equation}
    \sum_{\mu} \, 
    \includeTikz{0}{intertwinerPulling1}{\intertwinerPulling{1}} \; = \!\! \includeTikz{0}{intertwinerPulling2}{\intertwinerPulling{2}}  .
    \label{eq:interPull}
\end{equation} 
These operators implement the bond algebra isomorphism and therefore the duality at the level of the local tensors, where the non-locality is captured by the fact that the virtual bond dimension of this operator is non-trivial. Intertwiners between bond algebras obtained from two generic module categories can be obtained from the ones above via composition; they are described by \emph{module functors} \cite{Lootens:2020mso}.

To appreciate how the bond-algebraic formulation of duality works, let us consider two Hamiltonians $\mathbb H = \sum_{\msf i} \mathbb h_{\msf i}$ whose local terms are constructed by taking some linear combination of the bonds
\begin{equation}
    \mathbb h_{\msf i} = \sum_a J_a \mathbb b_{a,\msf i}^{\mc M} \, .
\end{equation}
This defines two dual Hamiltonians $\mathbb H_A$ and $\mathbb H_B$, acting on Hilbert spaces $\mc H_A^{\phantom{}}$ and $\mc H_B^{\phantom{}}$ respectively, where the Hamiltonian $\mathbb H_A$ is constructed by taking the regular $\mc D$-module category and $\mathbb H_B$ is built from an arbitrary $\mc D$-module category $\mc M$. The MPO symmetries of these models are then given by the Morita duals $\mc D_{\mc D}^* \cong \mc D$ and $\mc D_{\mc M}^* \cong \mc C$, respectively, and the Hamiltonians are transformed into one another by action of the MPO intertwiner. To understand the action of the duality map at the level of the Hilbert spaces, we consider the models on finite size rings. The presence of MPO symmetries indicates that the Hilbert spaces $\mc H_A^{\phantom{}}$ and $\mc H_B^{\phantom{}}$ can be decomposed into direct sums of $n$ sectors:
\begin{equation}
    \mc H_A^{\phantom{}} = \bigoplus_i^n \mc H_{A,i}^{\phantom{}} \q {\rm and} \q \mc H_B^{\phantom{}} = \bigoplus_i^n \mc H_{B,i}^{\phantom{}} \, ,
\end{equation}
where $i$ roughly labels all possible charges under the MPO symmetry, as well as all symmetry twisted boundary conditions. Consequently, the Hamiltonians are block diagonal and decompose as
\begin{equation}
    \mathbb H_A = \bigoplus_i^n \mathbb H_{A,i} \q {\rm and} \q \mathbb H_B = \bigoplus_i^n \mathbb H_{B,i} \, .
\end{equation}
The Hilbert spaces $\mc H_A^{\phantom{}}$ and $\mc H_B^{\phantom{}}$ need not be the same dimension, as the fusion categories describing the MPO symmetries $\mc D$ and $\mc C$ are typically not (monoidally) equivalent, but the number of sectors is the same for both. Mathematically, this is guaranteed by the fact that the fusion categories $\mc D$ and $\mc C$ are Morita equivalent; the sectors are given by the monoidal centers of these fusion categories, which are equivalent for Morita equivalent fusion categories \cite{etingof2016tensor}. At the level of the MPO symmetries, the monoidal center can be constructed from the \emph{tube algebra} \cite{bultinckAnyons,williamsonSET}, the central idempotents of which correspond to the different sectors in the model. The dimension of these central idempotents is in general different for the two models, which is reflected in the difference in Hilbert space dimension.

The fact that these models have isomorphic bond algebras implies the existence of a set of unitary transformations:
\begin{equation}
    \begin{split}
        \mathbb U_i: \mc H_{A,i}^{\phantom{}} \times \mc H_{A,i}^{\rm aux} &\rightarrow \mc H_{B,i}^{\phantom{}} \times \mc H_{B,i}^{\rm aux}
        \\
        {\rm s.t.} \q 
        \mathbb U_i (\mathbb H_{A,i} \otimes \mathbb 1_{A,i})\mathbb U_i^\dag &= \mathbb H_{B,i} \otimes \mathbb 1_{B,i} \, ,
    \end{split}
    \label{eq:unitaryTF}
\end{equation}
where the auxiliary Hilbert spaces $\mc H^{\rm aux}$ are chosen to account for the potential mismatch in Hilbert space dimension between $\mc H_{A,i}^{\phantom{}}$ and $\mc H_{B,i}^{\phantom{}}$; a prototypical instance where this occurs is in dualities obtained by gauging a non-abelian symmetry. The existence of such unitary transformations has been discussed in \cite{bondAlgebra} and implies that up to degeneracies, the Hamiltonians $\mathbb H_A$ and $\mathbb H_B$ have the same spectrum, but its explicit construction for generic models has not been obtained. In our formalism however, these unitary transformations can be explicitly constructed from the MPO intertwiners together with the knowledge of their interaction with the MPO symmetries of the two models. A detailed description requires an analysis of the sectors of these two models in terms of the idempotents of tube algebras and generalizations thereof involving MPO intertwiners. 

Finally, the local operators in these theories admit a similar characterization in terms of tube algebras, and are able to change the sector by acting on a state in a process that is equivalent to the fusion of anyons. A detailed exposition of these aspects will be presented elsewhere.

%% file: _Examples.tex
\section{Examples}
\label{Section III}
\noindent
\emph{In this section we illustrate the previous construction for various choices of input category-theoretic data and bonds building up a Hamiltonian. We consider examples realizing familiar dualities, which can be studied without invoking category theory, as well as more exotic examples  to showcase the general applicability of our approach.}

\subsection{$\mathbb Z_2$: transverse field Ising model\label{sec:Ising}}

\noindent
Let $\mc D = \Vect_{\mathbb Z_2}$ be the category of $\mathbb Z_2$-graded vector spaces, where we write the simple objects of $\Vect_{\mathbb Z_2}$ as $\{\mathbb 1,m\}$. The non-trivial fusion rules read $\mathbb 1 \otimes \mathbb 1 \simeq \mathbb 1 \simeq m \otimes m$ and $\mathbb 1 \otimes m \simeq m \simeq m \otimes \mathbb 1$. As an example, we consider the model 
\begin{equation}
    \label{eq:Ising}
    \mathbb H = -J \sum_{\msf i} \mathbb b_{1,\msf i}^{\mc M} -Jg \sum_{\msf i} \mathbb b_{2,\msf i}^{\mc M} \, .   
\end{equation}
defined by the bonds
\begin{align}
    \nn
    \mathbb b_{1,\msf i}^{\mc M} 
    &= 
    \includeTikz{0}{HamIsing1}{\HAM{}{}{}{1}{}{}{1}{}{}{}{}{m}{\mathbb 1}{m}{\mathbb 1}{\gamma}} 
    +
    \includeTikz{0}{HamIsing2}{\HAM{}{}{}{1}{}{}{1}{}{}{}{}{\mathbb 1}{m}{m}{\mathbb 1}{\gamma}}
    + \; (\mathbb 1 \leftrightarrow m) 
    \\
    \label{eq:bondIsing}
    {\rm and} \q \mathbb b_{2,\msf i}^{\mc M} &= 
    \includeTikz{0}{HamIsing3}{\HAM{}{}{}{1}{}{}{1}{}{}{}{}{\mathbb 1}{\mathbb 1}{\mathbb 1}{\mathbb 1}{\gamma}} - 
    \includeTikz{0}{HamIsing4}{\HAM{}{}{}{1}{}{}{1}{}{}{}{}{m}{m}{m}{m}{\gamma}}  ,
\end{align}
where $\gamma \in \mc I_\mc D$ is uniquely specified by the fusion rules. The bond $b_{2,\msf i}^{\mc M}$ is in fact a one-site operator in disguise, which we write as a two-site operator for consistency with the other bonds in this work. For this example, the module associator evaluates to the identity morphism for any choice of module category so that $\F{\cat}$-symbols equals 1 when the fusion rules are satisfied, and 0 otherwise.

\medskip
\noindent
$\bul$ Let $\mc M = \Vect_{\mathbb Z_2}$ be the regular $\Vect_{\mathbb Z_2}$-module category. As per eq.~\eqref{eq:unlabelled}, we have $\includeTikz{2.2}{objectMod}{} = \includeTikz{2.2}{objectMod1}{\object{mod}{\mathbb 1}} \oplus \includeTikz{2.2}{objectModm}{\object{mod}{m}}$. Imposing hom-spaces to be non-trivial in the definition eq.~\eqref{eq:Hilbert}  of the total Hilbert space $\mc H$ constrains objects $\{\alpha\}$ in $\mc I_\mc D$ to be determined by a choice of objects $\{A\}$ in $\mc I_\mc M$ via the fusion rules. Since hom-spaces in $\mc M$ are all one-dimensional, it follows that $\mc H$ is isomorphic to $\bigotimes_{\msf i} (\mathbb C \oplus \mathbb C)$, where $\mathbb C \oplus \mathbb C \simeq \mathbb C[ \includeTikz{2.2}{objectMod}{} ] \simeq \mathbb C^2$, such that the physical spins are located in the `middles' of the module strands. The operator $\mathbb b_{1,\msf i}^{\mc M}$ acts on the strand $\msf i$ as $|\mathbb 1 / m \ra \mapsto | m / \mathbb 1\ra$, whereas the operator $\mathbb b_{2,\msf i}^{\mc M}$ projects out states whose strands $\msf i$$-$1 and $\msf i$$+$1 have distinct labelling objects, and acts as the identity operator otherwise.
It follows that in the Pauli $Z$ basis, the Hamiltonian \eqref{eq:Ising} reads
\begin{align}
    \nn
    \mathbb H &= - J \sum_{\msf i} \big(
    \mathbb 1_{\msf i -1}X_\msf i \mathbb 1_{\msf i+1} + \frac{g}{2}(Z_{\msf i-1}Z_{\msf i}\mathbb 1_{\msf i+1} + \mathbb 1_{\msf i-1}Z_{\msf i}Z_{\msf i+1}) \big)
    \\
    \label{eq:IsingVecZ2}
    &= -J\sum_{\msf i} \big( X_\msf i + g Z_\msf i Z_{\msf i+1} \big) \, ,
\end{align}
which we recognize as the \emph{transverse field Ising} model, seemingly first introduced in  \cite{DEGENNES1963132}. 

\medskip
\noindent
$\bul$ Let us now consider the category $\mc M = \Vect$, which is a $\Vect_{\mathbb Z_2}$-module category via the forgetful functor $\Vect_{\mathbb Z_2} \to \Vect$. We notate the unique simple object in $\Vect$ via $\mathbb 1 \simeq \mathbb C$ such that $\mathbb 1 \cat \alpha \simeq \mathbb 1$ for any $\alpha \in \mc I_\mc D$. According to eq.~\eqref{eq:Hilbert} the total Hilbert space is 
\begin{equation}
    \mc H = \bigotimes_{\msf i}\bigoplus_{\alpha_{\msf i+\frac{1}{2}}}\Hom_{\mc M}(\mathbb 1 \cat \alpha_{\msf i + \frac{1}{2}}, \mathbb 1) 
    \simeq \bigotimes_{\msf i}\mathbb C^2  \, .
\end{equation}
It follows immediately from the definitions of $\mathbb b^{\cal M}_{1,\msf i}$ and $\mathbb b^{\cal M}_{2,\msf i}$ that in the Pauli $Z$ basis the Hamiltonian \eqref{eq:Ising} now reads
\begin{equation}
    \label{eq:IsingVec}
    \mathbb H = - J \sum_{\msf i} 
    \big( X_{\msf i-\frac{1}{2}}X_{\msf i+\frac{1}{2}} + g Z_{\msf i + \frac{1}{2}} \big) \, , 
\end{equation}
which we recognize as the \emph{Kramers-Wannier} dual of the Hamiltonian given in eq.~\eqref{eq:IsingVecZ2}. The MPO intertwiner corresponding to this duality is obtained from the module associator via eq.~\eqref{eq:interDef} and in this simple case reduces to eq.~\eqref{eq:interKW}.

Denoting the model associated with the regular $\Vect_{\mathbb Z_2}$-module category as $\mathbb H_A$ and the one 
with $\Vect$ by $\mathbb H_B$, their duality implies the existence of a unitary transformation mapping one to the other, as discussed above. On closed boundary conditions, these Hamiltonians are block diagonal in the four symmetry sectors, corresponding to even/odd charge under the global $\mathbb Z_2$ symmetry and periodic/antiperiodic boundary conditions. These sectors are all of the same dimension and as such no auxiliary Hilbert spaces are required to construct a unitary transformation relating the individual blocks of the Hamiltonians. Interestingly, this unitary transformation, as provided by the MPO intertwiners, interchanges the even (odd), periodic (antiperiodic) sector of $\mathbb H_A$ with the periodic (antiperiodic), even (odd) sector of $\mathbb H_B$, illustrating the subtle interplay between duality transformations and symmetry properties of the Hamiltonians involved.

\medskip
\noindent
$\bul$ To investigate another duality of the Ising model, it is convenient to think of the input category $\mc D$ as the category $\msf{sVec}$ of super vector spaces, which is equivalent, as a fusion category, to $\Vect_{\mathbb Z_2}$. We denote the simple objects of $\msf{sVec}$ as $\{\mathbb 1, \psi\}$. Let $\mc M = \mathsf{sVec} / {\la \psi \simeq \mathbb 1 \ra}$ be the (fermionic) $\msf{sVec}$-module category whose unique simple object $\mathbb 1$  satisfies $\mathbb 1 \cat \mathbb 1 \simeq \mathbb C^{1|0} \cdot \mathbb 1$ and $\mathbb 1 \cat \psi \simeq \mathbb C^{0|1} \cdot \mathbb 1$, where $\mathbb C^{1|0} \simeq \mathbb C$ is usually omitted, and $\mathbb C^{0|1}$ is the purely \emph{odd} one-dimensional \emph{super vector space}. It follows that the local Hilbert spaces $\mc V_{\msf i+ \frac{1}{2}}$ are now given by
\begin{align}
    \nn
    \mc V_{\msf i+\frac{1}{2}} 
    &= \Hom_{\mc M}(\mathbb 1 \cat \mathbb 1, \mathbb 1) \oplus \Hom_\mc M(\mathbb 1 \cat \psi , \mathbb 1)
    \\
    &\simeq \mathbb C^{1|0} \oplus \mathbb C^{0|1} = \mathbb C^{1|1} \, ,
\end{align}
where $\mathbb C^{1|1}$ is a super vector space with one even and one odd basis vectors. We identify the basis vector in $\mathbb C^{1|0}$ with the empty state $| \varnothing \ra$ and the basis vector in the odd vector space $\mathbb C^{0|1}$ with $c_\msf i^\dagger | \varnothing \ra$ where $c_\msf i$ is a fermionic operator satisfying $\{c^{\phantom \dagger}_\msf i, c^\dagger_\msf j\} = \delta_{\msf i \msf j}$ and $\{c_\msf i, c_\msf j\} =0$. Replacing $m$ by $\psi$ in eq.~\eqref{eq:bondIsing}, we notice that the first two terms in the definition of $\mathbb b^{\cal M}_{1,\msf i}$ act on states
\begin{equation}
    (c^\dagger_{\msf i-\frac{1}{2}})^{n_{\msf i - \frac{1}{2}}}
    (c^\dagger_{\msf i+\frac{1}{2}})^{n_{\msf i + \frac{1}{2}}}
    | \varnothing \ra \in \mathbb C^{1|1} \otimes \mathbb C^{1|1}
\end{equation}
via $c^\dagger_{\msf i-\frac{1}{2}} c^{\phantom \dagger}_{\msf i+\frac{1}{2}}$ and $c^\dagger_{\msf i-\frac{1}{2}} c^\dagger_{\msf i+\frac{1}{2}}$, respectively, and similarly for the last two terms, whereas the operator $\mathbb b^{\cal M}_{2,\msf i}$ acts as $\frac{1}{2}((-1)^{n_{\msf i -  \frac{1}{2}}}+ (-1)^{n_{\msf i + \frac{1}{2}}})$.
It follows that the Hamiltonian \eqref{eq:Ising} now reads
\begin{align}
    \nn
    \mathbb H = - J \sum_{\msf i} \big(
    &c^\dagger_{\msf i-\frac{1}{2}} c^{\phantom \dagger}_{\msf i+\frac{1}{2}} + c^\dagger_{\msf i-\frac{1}{2}} c^\dagger_{\msf i+\frac{1}{2}} + \text{h.c.}
    \\[-0.5em]
    &- g(2c^\dagger_{\msf i+\frac{1}{2}} c^{\phantom \dagger}_{\msf i + \frac{1}{2}} - 1) \big) \, ,
\end{align}
which we recognize as the \emph{Jordan-Wigner} dual of the Hamiltonian given in eq.~\eqref{eq:IsingVec}. The MPO intertwiner corresponding to this duality is obtained from the module associator via eq.~\eqref{eq:interDef} and reduces to eq.~\eqref{eq:interJW} when combined with the Kramers-Wannier MPO intertwiner.

\bigskip
\noindent
There exists a fairly straightforward generalization of this class of examples obtained by replacing the spherical fusion category $\mc D$ by the category $\Vect_G$ of $G$-graded vector spaces, with $G$ an arbitrary finite group. This is the input data of topological gauge theories known as (untwisted) Dijkgraaf-Witten theories \cite{cmp/1104180750}, whose lattice Hamiltonian incarnations were studied by Kitaev in \cite{KITAEV20032}. Indecomposable module categories over $\Vect_G$ are well understood and are labelled by pairs $(L,[\psi])$ with $L \subset G$ a subgroup of $G$ and $[\psi]$ a cohomology class in $H^2(L,\mathbb C^\times)$ \cite{2002math......2130O}. Applying our construction for various choices of module categories, we expect to recover dual versions of the Ising-like Hamiltonians investigated in \cite{Albert:2021vts,PhysRevB.98.245135}. As with $\Vect_{\mathbb Z_2}$, two particularly interesting choices of $\Vect_G$-module categories are provided by $\mc M=\Vect$ and $\mc M = \Vect_G$.
The MPOs intertwining the corresponding tensor network representations of the $\Vect_G$ string-net model are particularly simple:
\begin{equation*}
    \includeTikz{0}{MPOintertwinerVecG}{\MPO{obj}{dot}{1}{1}{1}{1}{}{}{g_2}{g_1^{-1}}{g_1g_2}{}{2}} 
    \equiv \includeTikz{-1}{gauging1}{\gauging{1}} \! =1
    \, , \q \forall \, g_1,g_2 \in G \, ,
\end{equation*}
where the black dots represent group delta functions.
As mentioned earlier, these intertwining operators map the bonds of the Hamiltonian associated with $\mc M= \Vect_G$ to those of the Hamiltonian associated with $\mc M = \Vect$. Invoking the property
\begin{equation}
    \label{eq:gauging}
    \includeTikz{0}{gauging2}{\gauging{2}} =  \includeTikz{0}{gauging3}{\gauging{3}} \, ,
\end{equation}
where $R_h: |g \ra \mapsto |gh^{-1}\ra$ and $L_h: |g \ra \mapsto |hg \ra$,
we can readily confirm that this operator implements the duality transformation between the Ising-like Hamiltonians. This tensor network can also be thought of as implementing a gauging map. In particular, applying such an operator to product states yield GHZ-like states thus providing a non-local map relating short- and long-range order. A further generalization is obtained by considering the fusion category $\Vect_G^\alpha$ that only differs from $\Vect_G$ by its monoidal associator, which now evaluates to some normalized representative of a cohomology class $[\alpha]$ in $H^3(G,{\rm U}(1))$. Module categories over this category are also well understood \cite{2002math......2130O}. Interestingly, as long as $[\alpha]$ is not the trivial class, it is not possible to choose $\Vect$ as a module category. This corresponds to the ``anomaly'' that presents an obstruction to gauging the corresponding quantum models.


\subsection{$\mathbb Z_2$: Heisenberg XXZ to $t$-$J_z$ model}

\noindent
In the previous section, we illustrated how our approach reproduces the well-known Kramers-Wannier and Jordan-Wigner dualities. We will now show how our approach leads to new dualities that were previously not known. Taking again $\mc D = \Vect_{\mathbb Z_2}$, we consider the model
\begin{equation}
    \mathbb H = 2 t\sum_{\msf i} \mathbb b_{1,\msf i}^{\mc M} + J_z \sum_{\msf i} \mathbb b_{2,\msf i}^{\mc M}
\end{equation}
defined by the bonds
\begin{equation}
    \mathbb b_{1,\msf i}^{\mc M} 
    =  \!\!
    \includeTikz{0}{HamIsing2}{\HAM{}{}{}{1}{}{}{1}{}{}{}{}{\mathbb 1}{m}{m}{\mathbb 1}{\gamma}}
    \!\!\! +
    (\mathbb 1 \leftrightarrow m) \, , \q
    \mathbb b_{2,\msf i}^{\mc M} \!\!
    = 
    \includeTikz{0}{HamIsing4}{\HAM{}{}{}{1}{}{}{1}{}{}{}{}{m}{m}{m}{m}{\gamma}} \hspace{-2em}
\end{equation}
where $\gamma$ is again uniquely specified by the fusion rules.

\medskip
\noindent
$\bul$ Taking the module category $\mc M = \Vect$ and using the same arguments as before, it is straightforward to see that this leads to the Hamiltonian
\begin{align}
    \nn
    \mathbb H = t &\sum_{\msf i} \big( X_{\msf i-\frac{1}{2}}X_{\msf i+\frac{1}{2}} + Y_{\msf i-\frac{1}{2}}Y_{\msf i+\frac{1}{2}}\big) 
    \\ 
    + \frac{J_z}{4} &\sum_{\msf i}\big(Z_{\msf i-\frac{1}{2}} + \mathbb 1_{\msf i-\frac{1}{2}}\big)\big(Z_{\msf i+\frac{1}{2}} + \mathbb 1_{\msf i+\frac{1}{2}}\big) \, .
\end{align}
Up to an irrelevant constant, this is the spin-$1/2$ Heisenberg XXZ model in the presence of a magnetic field.

\medskip
\noindent
$\bul$ We now consider $\mathsf{sVec}$ as a module category over $\Vect_{\mathbb Z_2}$. As before, we denote the simple objects of $\mathsf{sVec}$ as $\{\mathbb 1,\psi\}$; the fusion rules are $A \cat \mathbb 1 \simeq \mathbb C^{1|0} \cdot A$, $\mathbb 1 \cat m \simeq \mathbb C^{0|1} \cdot \psi$ and $\psi \cat m \simeq \mathbb C^{0|1} \cdot \mathbb 1$. The local Hilbert spaces are given by
\begin{equation}
    \mc V_{\msf i + \frac{1}{2}} = \bigoplus_{A,B,\alpha} \Hom_{\mc M}(A \cat \alpha,B) \simeq \mathbb C^{2|2} \, ,
\end{equation}
where $\mathbb C^{2|2}$ is the super vector space with two even and two odd basis vectors. We choose basis vectors for the two-dimensional even component given by $|\varnothing_{\pm} \ra = \frac{1}{\sqrt{2}}(|\mathbb 1 \mathbb 1 \mathbb1 \ra \pm |\psi \mathbb 1 \psi \ra)$, while for the two-dimensional odd component we define the basis vectors $| \! \uparrow \, \ra = |\mathbb 1 m \psi \ra$ and $| \! \downarrow \, \ra = |\psi m \mathbb 1 \ra$. We note that the identification of the degrees of freedom here differs from the $\mc M = \Vect_{\mathbb Z_2}$ considered above, where we interpreted the module strands as the effective degrees of freedom. In this basis, we denote via $\bar{c}_{\msf i,\sigma}$, with $\sigma \in \{\uparrow,\downarrow\}$, the \emph{constrained} fermion operators defined by their non-zero components $\bar{c}_{\msf i,\sigma}^{\dag} |\varnothing_+ \ra = |\sigma \ra$ and $\bar{c}_{\msf i,\sigma} |\sigma \ra = |\varnothing_+ \ra$. These operators correspond to the creation/annihilation of fermions with the constraint that no two fermions, regardless of their spin, can occupy the same site \cite{Batista:2001}. The full Hilbert space is a subspace of $\bigotimes_{\msf i} \mathbb C^{2|2}$. This subspace is characterized by the fact that two fermions separated by any number of holes $|\varnothing_+\ra$ (including none) must have opposite spins, i.e., it displays anti-phase domains related to string order. As we will see, configurations involving $|\varnothing_-\ra$ are projected out by the Hamiltonian, and as such we can effectively restrict the local degrees of freedom to $\{|\varnothing_+\ra,| \! \uparrow \, \ra,| \! \downarrow \, \ra\}$. Acting on this subspace, and using that the $\F{\cat}$-symbols of the module category $\msf{sVec}$ are all trivial, the Hamiltonian can be written in terms of constrained fermion operators as
\begin{equation}
    \mathbb H = t\sum_{\msf i,\sigma} \big( \bar{c}_{\msf i-\frac{1}{2},\sigma}^\dag \bar{c}_{\msf i+\frac{1}{2},\sigma} + \text{h.c.}\big) + J_z \sum_{\msf i}S^z_{\msf i-\frac{1}{2}} S^z_{\msf i+\frac{1}{2}} \, ,
\end{equation}
which is the Hamiltonian of the well-known $t$-$J_z$ model. Here, the spin operators are defined as $S^z_{\msf i} = (\bar{n}_{\msf i,\uparrow} - \bar{n}_{\msf i,\downarrow})$, with $\bar{n}_{\msf i,\sigma} = \bar{c}_{\msf i,\sigma}^{\dag}\bar{c}_{\msf i,\sigma}$ the constrained fermion number operators. This model is particularly relevant in the context of cuprate superconductivity \cite{doi:10.1142/2945}. 

Therefore, we established a new {\it emergent} duality between the $t$-$J_z$ model in a subspace and the XXZ model in a magnetic field. In  \cite{PhysRevLett.85.4755} it was shown that the ground state of the $t$-$J_z$ model belongs to that subspace and this {\it quasi-exact solvability} was used to determine its quantum phase diagram. By recasting this insight into our formalism we are able to demonstrate that the relation between these two models is in fact a duality, which was previously unknown.

Importantly, the duality defined above can be extended to the full Hilbert space of the $t$-$J_z$ model by introducing \emph{symmetry twists} into the definition of the Hilbert space. These effectively act as anti-phase domain walls for the $\{\mathbb 1,\psi\}$ degrees of freedom labeling the module strands, and when inserted between fermions (separated by any number of holes) forces them to have the same spin. In this way, one can build the full Hilbert space of the $t$-$J_z$ model; the Hamiltonian is block diagonal in the number of symmetry twists. Through duality maps each of these blocks can be related to an XXZ-type model that will depend on the number of symmetry twists. The precise nature of this mapping requires a detailed understanding of symmetry twists and their relation to charge sectors under duality \cite{Lootens:2022avn}; we will elaborate on this in future work. The $t$-$J_z$ has recently been studied in the context of Hilbert space fragmentation \cite{PhysRevB.101.125126,PhysRevX.12.011050}, and we expect this new duality to contribute to our understanding of the thermalization properties of these systems.


\subsection{$\mathbb Z_2 \times \mathbb Z_2$: Kennedy-Tasaki transformation}

\noindent
As mentioned above, indecomposable module categories over $\Vect_G$ are labelled by pairs $(L,[\psi])$, with $L \subseteq G$ and $[\psi] \in H^2(L,\mathbb C^\times)$. In order to showcase the effect of the 2-cocycle $\psi$, we shall consider the simplest example where the cohomology class $[\psi]$ is non-trivial. Let $\mc D = \Vect_{\mathbb Z_2 \times \mathbb Z_2}$ be the category of $(\mathbb Z_2 \times \mathbb Z_2)$-graded vector spaces. We write the simple objects as $\{\mathbb 1\mathbb 1,\mathbb 1 m, m \mathbb 1, mm\}$. The associator of $\Vect_{\mathbb Z_2 \times \mathbb Z_2}$ evaluates to the identity morphism so that the $F$-symbols equal $1$ as long as all the fusion rules are satisfied, and $0$ otherwise. As an example, we consider the model
\begin{equation}
    \mathbb H = J_1 \sum_{\msf i} \mathbb b_{1,\msf i}^{\mc M} + J_2 \sum_{\msf i} \mathbb b_{2,\msf i}^{\mc M} + J_3 \sum_{\msf i} \mathbb b_{3,\msf i}^{\mc M}
\end{equation}
defined by the bonds
\begin{align}
    \nn
    \mathbb b_{1,\msf i}^{\mc M} 
    &= 
    \includeTikz{0}{HamKT1}{\HAM{}{}{}{1}{}{}{1}{}{}{}{}{\mathbb 1 m \;}{m m \;\,}{\;\, m m}{\; \mathbb 1 m}{\gamma}} 
    -
    \includeTikz{0}{HamKT2}{\HAM{}{}{}{1}{}{}{1}{}{}{}{}{\mathbb 1 m \;}{m m \;\,}{\; \mathbb 1 m}{\;\, m m}{\gamma}}
    + \; (\mathbb 1 m \leftrightarrow m m) \, ,
\end{align}
\begin{align}
    \nn
    \mathbb b_{2,\msf i}^{\mc M} 
    &= 
    \includeTikz{0}{HamKT3}{\HAM{}{}{}{1}{}{}{1}{}{}{}{}{\mathbb 1 m \;}{m \mathbb 1 \;}{\; \mathbb 1 m}{\; m \mathbb 1}{\gamma}} 
    +
    \includeTikz{0}{HamKT4}{\HAM{}{}{}{1}{}{}{1}{}{}{}{}{\mathbb 1 m \;}{m \mathbb 1 \;}{\; m \mathbb 1}{\; \mathbb 1 m}{\gamma}}
    + \; (\mathbb 1 m \leftrightarrow m \mathbb 1) \, ,
\end{align}
and
\begin{align}
    \nn
    \mathbb b_{3,\msf i}^{\mc M} 
    &= 
    \includeTikz{0}{HamKT5}{\HAM{}{}{}{1}{}{}{1}{}{}{}{}{m \mathbb 1 \;}{m m \;\,}{\; m \mathbb 1}{\;\, m m}{\gamma}}
    +
    \includeTikz{0}{HamKT6}{\HAM{}{}{}{1}{}{}{1}{}{}{}{}{m \mathbb 1 \;}{m m \;\,}{\;\, m m}{\; m \mathbb 1}{\gamma}}
    + \; (m \mathbb 1 \leftrightarrow m m) \, ,
\end{align}
where $\gamma$ is again uniquely specified by the fusion rules.

\medskip
\noindent
$\bul$ Let $\mc M = \Vect$, which is obtained by choosing $L = \mathbb Z_2 \times \mathbb Z_2$ and taking $[\psi]$ to be the trivial cohomology class in $H^2(\mathbb Z_2 \times \mathbb Z_2,\mathbb C^\times) \simeq \mathbb Z_2$. We denote the unique simple object of $\Vect$ as $\mathbb 1$, such that $\mathbb 1 \cat \alpha \simeq \mathbb 1$ for any $\alpha \in \mc I_\mc D$. The non-vanishing $\F{\cat}$-symbols are all equal to 1. Given the definition of the bonds $\mathbb b_{a,\msf i}^{\mc M}$ and the fusion rules in $\Vect_{\mathbb Z_2 \times \mathbb Z_2}$, the Hamiltonian acts on the effective total Hilbert space
\begin{equation}
    \mc H \stackrel{\rm eff.}{=} \bigotimes_{\msf i}\bigoplus_{\alpha_{\msf i+\frac{1}{2}} \neq \mathbb 1 \mathbb 1}\Hom_{\mc M}(\mathbb 1 \cat \alpha_{\msf i + \frac{1}{2}}, \mathbb 1) 
    \simeq \bigotimes_{\msf i}\mathbb C^3  \, .
\end{equation}
 With this choice of module category, the Hamiltonian becomes
\begin{equation}
    \label{eq:ktHeisenberg}
    \mathbb H = J_1 \sum_{\msf i} S^x_{\msf i}S^x_{\msf i+1} + J_2 \sum_{\msf i} S^y_{\msf i}S^y_{\msf i+1} + J_3 \sum_{\msf i} S^z_{\msf i}S^z_{\msf i+1}
\end{equation}
where we work in the spin-$1$ basis provided by
\begin{equation*}
    S^x = \begin{pmatrix}
        0 & 0 & -i\\
        0 & 0 & 0\\
        i & 0 & 0
    \end{pmatrix} \! ,\q
    S^y = \begin{pmatrix}
        0 & 1 & 0\\
        1 & 0 & 0\\
        0 & 0 & 0
    \end{pmatrix} \! ,\q
    S^z = \begin{pmatrix}
        0 & 0 & 0\\
        0 & 0 & 1\\
        0 & 1 & 0
    \end{pmatrix} \! .
\end{equation*}
This is the Hamiltonian of the spin-$1$ Heisenberg XYZ model \cite{1928ZPhy...49..619H}.

\medskip
\noindent
$\bul$ Let $\mc M = \Vect^\psi$, which is obtained by choosing $L = \mathbb Z_2 \times \mathbb Z_2$ and taking $[\psi]$ to be the non-trivial cohomology class in $H^2(\mathbb Z_2 \times \mathbb Z_2,\mathbb C^\times) \simeq \mathbb Z_2$. Simple object and fusion rules are the same as for $\Vect$, so that the effective total Hilbert space is unchanged. The difference lies in the $\F{\cat}$-symbols, the relevant entries of which are now given by
\begin{equation}
    \big( \F{\cat}^{\mathbb 1 \alpha \beta}_{\mathbb 1}\big)^{\alpha \otimes \beta,11}_{\mathbb 1,11} = 
    \psi(\alpha,\beta) =
    \begin{pmatrix}
        1 & 1 & 1\\
        -1 & 1 & -1\\
        -1 & 1 & -1
    \end{pmatrix}_{\!\!\! \alpha\beta} \!\!\! ,
\end{equation}
where $\{\alpha,\beta\} \in \{\mathbb 1 m, m \mathbb 1, mm\}$. With this choice of module category, the Hamiltonian can be readily checked to be
\begin{align}
    \nn
    \mathbb H = -J_1 \sum_{\msf i} \; &S^x_{\msf i}S^x_{\msf i+1} + J_2 \sum_{\msf i} S^y_{\msf i}e^{i\pi (S^z_{\msf i} + S^x_{\msf i+1})}S^y_{\msf i+1} \\
    &- J_3 \sum_{\msf i} S^z_{\msf i}S^z_{\msf i+1} \, .
\end{align}
The duality relating this Hamiltonian to the one given in eq.~\eqref{eq:ktHeisenberg} is known as the \emph{Kennedy-Tasaki} transformation \cite{kennedy1992hidden}. It provides a map between a model with symmetry protected topological order in the ground state, such as the Haldane phase of the spin-$1$ Heisenberg model, to a model without symmetry protected topological order. This transformation was generalized in \cite{else2013hidden,duivenvoorden2013symmetry} to the case of abelian groups $G$ of the form $G \simeq H \times H$, where $H$ is some group. In a sense, our approach further generalizes this transformation to arbitrary $G$.


\subsection{$\msf{Ising}$: critical Ising model}

\noindent
Let $\mc D = \mathsf{Ising}$ be the fusion category with simple objects $\{\mathbb 1 , \psi\} \oplus \{\sigma\}$. The non-trivial fusion rules read $\psi \otimes \psi \simeq \mathbb 1$, $\sigma \otimes \psi \simeq \sigma$ and $\sigma \otimes \sigma \simeq \mathbb 1 \oplus \psi$ such that $d_\mathbb 1 =  1 = d_\psi$ and $d_\sigma = \sqrt{2}$. The non-trivial $F$-symbols are then provided by
\begin{equation}
    \F{}^{\sigma \sigma \sigma}_{\sigma} 
    = \frac{1}{\sqrt{2}}
    \begin{pmatrix}
        1 & 1 \\
        1 & -1
    \end{pmatrix}
    \, , \q
    \F{}^{\sigma \psi \sigma}_{\psi}
    =
    \F{}^{\psi \sigma \psi}_{\sigma}
    = (-1) \, .
\label{eq:Fising}
\end{equation}
As an example, we consider the model
\begin{equation}
    \mathbb H = - \sum_{\msf i} \mathbb b_{\msf i}^{\mc M}
\end{equation}
defined by the bond
\begin{equation}
    \label{eq:bondIsing2}
    \mathbb b_{\msf i}^{\mc M} =
    \includeTikz{0}{HamCritIsing1}{\HAM{}{}{}{1}{}{}{1}{}{}{}{}{\sigma}{\sigma}{\sigma}{\sigma}{\mathbb 1}}
    -  \includeTikz{0}{HamCritIsing2}{\HAM{}{}{}{1}{}{}{1}{}{}{}{}{\sigma}{\sigma}{\sigma}{\sigma}{\psi}}
    \equiv     
    \includeTikz{0}{HamCritIsing3}{\HAM{}{}{}{1}{}{}{1}{}{}{}{}{\sigma}{\sigma}{\sigma}{\sigma}{\gamma}} ,
\end{equation}
where $\gamma = \mathbb 1 \ominus \psi$ is a shorthand defined via the diagrams above.

\medskip
\noindent
$\bul$ Let $\mc M = \msf{Ising}$ be the regular $\msf{Ising}$-category. As per eq.~\eqref{eq:unlabelled}, we have
\begin{equation}
    \label{eq:modStrandIsing}
    \includeTikz{2.2}{objectMod}{\object{mod}{}} = 
    \includeTikz{2.2}{objectMod1}{\object{mod}{\mathbb 1}} \oplus 
    \includeTikz{2.2}{objectModPsi}{\object{mod}{\psi}} \oplus 
    \includeTikz{2.2}{objectModSigma}{\object{mod}{\sigma}} \, .
\end{equation}
Given the definition of $\mathbb b_{\msf i}^{\mc M}$ and the fusion rules in $\msf{Ising}$, the Hamiltonian acts on the effective total Hilbert space
\begin{equation}
    \begin{split}
        \label{eq:decompIsing}
        \mc H
        \stackrel{\rm eff.}{=} \; & \mathbb C \Big[ \!\!\!
        \includeTikz{0}{chainCritIsing1}{\chain{1}{1}{1}{\mathbb 1 \oplus \psi}{\sigma}{\mathbb 1 \oplus \psi}{\sigma}{\sigma}{\sigma}{\sigma}} \!\!\! \Big]
        \\ 
        \oplus \; &\mathbb C \Big[ \!\!\! 
        \includeTikz{0}{chainCritIsing2}{\chain{1}{1}{1}{\sigma}{\mathbb 1 \oplus \psi}{\sigma}{\mathbb 1 \oplus \psi}{\sigma}{\sigma}{\sigma}} \!\!\! \Big]  .
    \end{split}
\end{equation}
Let us focus on either one of the two Hilbert spaces appearing in this decomposition. First, notice that all the hom-spaces are one-dimensional. We then identify a module strand labelled by $\sigma$ as the presence of a \emph{defect} and $\mathbb C[\includeTikz{2.2}{objectMod1}{}, \includeTikz{2.2}{objectModPsi}{}] \simeq \mathbb C^2$ as the local Hilbert space of a physical spin, which we locate as before in the middle of the corresponding strand. The operator $\mathbb b_{\msf i}^{\mc M}$ acts differently depending on whether the module strand at the site $\msf i$ is labelled by $\mathbb 1 / \psi$ or $\sigma$. In the former case, it follows from the definition of the $F$-symbols that $\mathbb b_{\msf i}^{\mc M}$ acts on the strand $\msf i$ as $| \mathbb 1 / \psi \ra \mapsto \frac{1}{2}(| \mathbb 1 / \psi \ra + | \psi / \mathbb 1 \ra - | \mathbb 1 / \psi \ra + | \psi / \mathbb 1 \ra) = |\psi / \mathbb 1 \ra$ and identically on the strands $\msf i$$-$1 and $\msf i$+1 labelled by $\sigma$. In the latter case, we notice that the first term in the definition of $\mathbb b_{\msf i}^{\mc M}$ acts as the identity or the zero operator depending on whether the strands $\msf i$$-$1 and $\msf i$+1 have matching labelling objects, and vice versa for the second term. Putting everything together, we obtain in the Pauli $X$ basis the Hamiltonian
\begin{align}
    \nn
    \mathbb H 
    = &- \sum_\msf i 
    \big( | \sigma \ra \la \sigma| \otimes Z_{2\msf i} \otimes | \sigma \ra \la \sigma| +  X_{2\msf i} \otimes |\sigma \ra \la \sigma | \otimes X_{2 \msf i+2}\big)
    \\
    \label{eq:IsingIsing}
    \stackrel{{\rm eff.}}{=} \! &- \sum_\msf i \big( Z_\msf i + X_{\msf i}X_{\msf i +1} \big) \, ,
\end{align}
which we recognize as the critical transverse field Ising model. 
Taking into account the second Hilbert space in the decomposition \eqref{eq:decompIsing} of $\mc H$, we obtain a direct sum of two copies of the model.

The critical Ising model can be used to illustrate the distinction between non-invertible symmetries and duality transformations. At criticality, the symmetries of the Ising model are encoded into the fusion category $\mc D_{\mc M}^* = \msf{Ising}$ such that the MPO labelled by $\psi$ performs the global spin-flip $\mathbb Z_2$ symmetry, and the MPO labelled by $\sigma$ essentially swaps the two terms in the decomposition \eqref{eq:decompIsing} of the total Hilbert space. 
Note that it is crucial to define the model on the total Hilbert space \eqref{eq:decompIsing} for $\sigma$ to correspond to a symmetry, otherwise it is interpreted as performing the Kramers-Wannier duality between the two terms \cite{aasenDefects,Aasen:2020jwb}. Promoting Kramers-Wannier duality to a symmetry amounts to constructing the $\msf{Ising}$ fusion category by ``gauging'' the duality defect labelled by the $\Vect_{\mathbb Z_2}$-module category $\Vect$ \cite{PhysRevB.101.134111}.\footnote{Mathematically, the $\msf{Ising}$ fusion category corresponds to a $\mathbb Z_2$-extension of $\Vect_{\mathbb Z_2}$ \cite{etingof2010fusion}.} 
Although it can be generalized, note that this approach is limited to dualities where the different realizations of the symmetries are encoded into monoidally equivalent fusion categories, whereas in general Morita equivalence is sufficient to ensure duality. Furthermore, a generic model with $\msf{Ising}$ symmetry does not necessarily split into a direct sum; therefore, the $\sigma$ MPO cannot be interpreted as a duality transformation in general. 

\medskip
\noindent
$\bul$ Let $\mc M = \msf{Ising} / {\la \psi \simeq \mathbb 1 \ra}$ be the (fermionic) $\msf{Ising}$-module category obtained via fermion condensation of $\psi$ (see \cite{Aasen:2017ubm,Bultinck_2017} for detailed constructions). This module category has two simple objects denoted by $\{\mathbb 1, \beta\}$ such that $\End_{\mc M}(\mathbb 1) \simeq \mathbb C$ and $\End_{\mc M}(\beta) \simeq \mathbb C^{1|1}$. The non-trivial fusion rules are indicated in the table below
\begin{align}
	\setlength{\tabcolsep}{0.5em}
	\begin{tabularx}{0.45\columnwidth}{c|ccc}
	    $\cat$ & $\mathbb 1$ & $\psi$ & $\sigma$
	    \\
	    \midrule
	    $\mathbb 1$ & $\mathbb 1$ & $\mathbb C^{0|1}\cdot \mathbb 1$ & $\beta$ 
	    \\
	    $\beta$ & $\beta$ & $\beta$ & $\mathbb C^{1|1} \cdot \mathbb 1$ 
	\end{tabularx} \, .
\end{align}
Components of the  module associator $\F{\cat}$ are \emph{even} isomorphisms and a non-exhaustive list of non-vanishing $\F{\cat}$-symbols is given by
\begin{align*}
    \big( \F{\cat}^{\beta \sigma \sigma}_{\beta} \big)^{\mathbb 1,\ee \oo}_{\mathbb 1,\oo \ee}
    = 
    \big( \F{\cat}^{\beta \sigma \sigma}_{\beta} \big)^{\mathbb 1,\ee \oo}_{\mathbb 1,\ee \oo}
    = 
    -i \big( \F{\cat}^{\beta \sigma \sigma}_{\beta} \big)^{\mathbb 1,\ee \ee}_{\mathbb 1,\oo \oo} 
    &=
    \frac{1}{d_\sigma} \, ,
    \\
    \big( \F{\cat}^{\beta \sigma \sigma}_{\beta} \big)^{\psi,\ee \oo}_{\mathbb 1,\oo \ee}
    =
    - \big( \F{\cat}^{\beta \sigma \sigma}_{\beta} \big)^{\psi,\ee \oo}_{\mathbb 1,\ee \oo}
    =
    i\big( \F{\cat}^{\beta \sigma \sigma}_{\beta} \big)^{\psi,\ee \ee}_{\mathbb 1,\oo \oo}
    &=
    \frac{1}{d_\sigma} \, ,
    \\
    \big( \F{\cat}^{\mathbb 1 \sigma \sigma}_{\mathbb 1} \big)^{\mathbb 1,\ee \ee}_{\mathbb 1,\ee \ee} 
    =
    \big( \F{\cat}^{\mathbb 1 \sigma \sigma}_{\mathbb 1} \big)^{\psi, \ee \oo}_{\mathbb 1,\ee \oo} 
    =
    \big( \F{\cat}^{\mathbb 1 \sigma \sigma}_{\mathbb 1} \big)^{\psi, \ee \oo}_{\mathbb 1,\oo \ee} 
    &= 1 \, ,\phantom{\frac{1}{d_\sigma}} 
\end{align*}
where the labels `o' and `e' refer to the oddness and evenness of the basis vectors. We note that these $\F{\cat}$ symbols are not real, and one has to take a complex conjugate to obtain $\Fbar{\cat}$.
As per eq.~\eqref{eq:unlabelled}, we have
\begin{equation}
    \includeTikz{2.2}{objectMod}{\object{mod}{}} = 
    \includeTikz{2.2}{objectMod1}{\object{mod}{\mathbb 1}} \oplus 
    \includeTikz{2.2}{objectModBeta}{\object{mod}{\beta}} \oplus 
    \includeTikz{0.8}{morphismModBeta}{\morphism{mod}{\beta}{{\rm o}}} \, .
\end{equation}
Henceforth, we assume that $\includeTikz{0.8}{doubleMorphism}{\doubleMorphism{mod}{\beta}{{\rm o}}} = i \includeTikz{2.2}{objectModBeta}{}$, where the odd basis vectors on the l.h.s. are ordered from left to right.
Noticing that the odd basis vector in $\mc V^{\mathbb 1}_{\beta ,\sigma} \simeq \mathbb C^{1|1}$ can be obtained from the even basis vector by acting with the odd element in $\End_\mc M(\beta)$, we can fix basis vectors in split spaces to be even. It follows that the Hamiltonian acts on the effective total Hilbert space
\begin{equation}
    \begin{split}
        \label{eq:decompFermion}
        \mc H
        \stackrel{\rm eff.}{=} \; &\mathbb C \Big[ \!\!\!
        \includeTikz{0}{chainCritIsingFermionic1}{\chain{{\rm e}}{{\rm e}}{{\rm e}}{\mathbb 1}{\beta}{\mathbb 1}{\beta}{\sigma}{\sigma}{\sigma}} \!\!\! \Big]
        \\ 
        \oplus \;  &\mathbb C \Big[ \!\!\!
        \includeTikz{0}{chainCritIsingFermionic2}{\chain{{\rm e}}{{\rm e}}{{\rm e}}{\beta}{\mathbb 1}{\beta}{\mathbb 1}{\sigma}{\sigma}{\sigma}} \!\!\! \Big] ,
    \end{split}
\end{equation}
at which point our derivation follows closely that of \cite{Aasen:2017ubm}.
Focusing on either one of the Hilbert spaces appearing in the decomposition above, we identify the module strand labelled by $\mathbb 1$ as a vacancy and $\mathbb C[\includeTikz{2.2}{objectModBeta}{}, \includeTikz{0.8}{morphismModBeta}{}] \simeq \mathbb C^{1|1}$ as the local Hilbert space of a physical fermion. Analogously to the previous choice of $\mc M$, we distinguish two actions for the operator $\mathbb b_{\msf i}^{\mc M}$ depending on the labelling of the module strand $\msf i$. If the labelling of the strand $\msf i$ takes value in $\End_{\mc M}(\beta)$, it follows from $\mathbb 1 \cat \psi \simeq \mathbb C^{0|1} \cdot \mathbb 1$ and the evenness of $\F{\cat}$ that $\mathbb b_{\msf i}^{\mc M}$ acts as the fermion parity operator. If the strand $\msf i$ is labelled by $\mathbb 1$, the definition of the $\F{\cat}$-symbols guarantees that the operator $\mathbb b_{\msf i}^{\mc M}$ acts on the strands {$\msf i$}{$-$}1 and $\msf i$+1 as the hopping or pairing operator, e.g. $|\ee,\oo \ra \mapsto \frac{1}{2}(| \ee,\oo \ra + |\oo,\ee \ra - | \ee, \oo \ra + |\oo,\ee \ra ) = |\oo,\ee \ra$. 
Putting everything together, we obtain the massless free fermion Hamiltonian
\begin{align}
    \mathbb H \stackrel{{\rm eff.}}{=} 
    - \sum_\msf i \big( c^\dagger_\msf i c^{\phantom{\dagger}}_{\msf i+1} + c^\dagger_\msf i c^\dagger_{\msf i+1} + {\rm h.c.} - (2c^\dagger_\msf i c^{\phantom{\dagger}}_\msf i - 1)\big)
\end{align}
which is the Jordan-Wigner dual of the Hamiltonian given in eq.~\eqref{eq:IsingIsing}.


\subsection{$\mathsf{Ising}^\msf{op} \boxtimes \mathsf{Ising}$: Heisenberg XXZ model} 

\noindent
Let $\mathcal{D}=\mathsf{Ising}^\msf{op} \boxtimes \mathsf{Ising}$ be the Deligne tensor product of two copies of $\msf{Ising}$,\footnote{As a category, $\msf{Ising}^{\rm op}$ is equivalent to $\msf{Ising}$, but its monoidal structure is such that $\alpha \otimes^{\rm op} \beta = \beta \otimes \alpha$ for any $\alpha,\beta \in \mc I_\mc D$.} such that the simple objects are of the form $(\alpha_1, \alpha_2) \equiv \alpha_1 \boxtimes \alpha_2$ with $\alpha_1, \alpha_2 \in \mc I_{\msf{Ising}}$. The fusion rules are obtained from those of $\msf{Ising}$ according to $(\alpha_1 \boxtimes \alpha_2) \otimes (\beta_1 \boxtimes \beta_2) = (\beta_1 \otimes \alpha_1)\boxtimes(\alpha_2 \otimes \beta_2)$ and the $F$-symbols are given by
\begin{equation}
    \F{}^{(\alpha_1 \boxtimes \alpha_2)(\beta_1 \boxtimes \beta_2)(\gamma_1 \boxtimes \gamma_2)}_{(\delta_1 \boxtimes \delta_2)} 
    = \Fbar{}^{\gamma_1\beta_1\alpha_1}_{\delta_1} \otimes \F{}^{\alpha_2\beta_2\gamma_2}_{\delta_2} \, ,
\end{equation}
where the $\msf{Ising}$ $F$-symbols on the r.h.s. were defined in eq.~\eqref{eq:Fising}. As an example, we consider the model
\begin{equation}
    \mathbb H = 
    -J\sum_{\msf i} \mathbb b_{1,\msf i}^{\mc M}
    -J \sum_{\msf i} \mathbb b_{2,\msf i}^{\mc M}
    + Jg \sum_{\msf i} \mathbb b_{3,\msf i}^{\mc M}
\end{equation}
defined by the bonds
\begin{equation}
    \mathbb b_{a,\msf i}^{\mc M} = \includeTikz{0}{HamXXZ}{\HAM{}{}{}{1}{}{}{1}{}{}{}{}{\sigma\sigma}{\sigma\sigma}{\sigma\sigma}{\sigma\sigma}{\gamma_{\!a}}} , \q
    \Bigg\{
    \begin{array}{l}
        \!\! \gamma_1 = (\mathbb 1 \oplus \psi)\boxtimes(\mathbb 1 \ominus \psi)\\
        \!\! \gamma_2 = (\mathbb 1 \ominus \psi)\boxtimes(\mathbb 1 \oplus \psi)\\
        \!\! \gamma_3 = (\mathbb 1 \ominus \psi)\boxtimes(\mathbb 1 \ominus \psi)
    \end{array} ,
\end{equation}
where $\mathbb 1 \ominus \psi$ is defined via eq.~\eqref{eq:bondIsing2} and we introduced the shorthand $\sigma \sigma \equiv \sigma \boxtimes \sigma$ that we shall use in diagrams.

\medskip
\noindent
$\bul$ Let $\mc M = \mathsf{Ising}^\msf{op} \boxtimes \mathsf{Ising}$ be the regular module category. We have
\begin{equation}
    \includeTikz{2.2}{objectMod}{} = \big(\includeTikz{2.2}{objectMod1}{} \oplus \includeTikz{2.2}{objectModPsi}{} \oplus \includeTikz{2.2}{objectModSigma} \big)^{\boxtimes\,2} \, ,
\end{equation}
so that, given the definition of $\mathbb b_{a,\msf i}^{\mc M}$, the Hamiltonian acts on an effective total Hilbert space that decomposes into four terms as
\begin{align}
        \nn
        \mc H
        \stackrel{\rm eff.}{=} \; & \mathbb C \Big[ 
        \includeTikz{0}{chainXXZ1}{\chain{1}{1}{1}{(\mathbb 1 \oplus \psi,\sigma)}{(\sigma,\mathbb 1 \oplus \psi)}{(\mathbb 1 \oplus \psi,\sigma)}{(\sigma,\mathbb 1 \oplus \psi)}{\sigma\sigma}{\sigma\sigma}{\sigma\sigma}} \Big]
        \\
        \nn
        \oplus \; &\mathbb C \Big[  
        \includeTikz{0}{chainXXZ2}{\chain{1}{1}{1}{\sigma\sigma}{\stackanchor[1pt]{${\sss (\mathbb 1 \oplus \psi,}$}{${\sss \;\, \mathbb 1 \oplus \psi)}$}}{\sigma \sigma}{\stackanchor[1pt]{${\sss (\mathbb 1 \oplus \psi,}$}{${\sss \;\, \mathbb 1 \oplus \psi)}$}}{\sigma\sigma}{\sigma\sigma}{\sigma\sigma}} \Big]  
        \\
        \label{eq:decompIsingIsing}
        \oplus \; &( \includeTikz{2.2}{objectModSigma}{} \leftrightarrow \includeTikz{2.2}{objectMod1Psi}{\object{mod}{\mathbb 1 \oplus \psi}} ) \, .
\end{align}
Given the obvious symmetry, we focus on the first two terms in this decomposition. Mimicking the previous study for $\mc D = \msf{Ising}$, we find that the operators $\mathbb b_{1,\msf i}^{\mc M}$ and $\mathbb b_{2,\msf i}^{\mc M}$ effectively act in the same way on these two vector spaces as $\mathbb 1_\msf i \otimes \widetilde{\mathbb h}_\msf i$ and $\mathbb h_\msf i \otimes \widetilde{\mathbb 1}_\msf i$, respectively, where $\mathbb h_\msf i = Z_\msf i Z_{\msf i+1}+X_\msf i$ and $\widetilde{\mathbb h}_{\msf i} = \widetilde{Z}_\msf i \widetilde{Z}_{\msf i+1}+\widetilde{X}_{\msf i+1}$ in the Pauli $Z$ basis. But we distinguish different actions for the operator $\mathbb b_{3,\msf i}^{\mc M}$. Indeed, it effectively acts on the first two terms in  eq.~\eqref{eq:decompIsingIsing} as $\widetilde{Z}_{\msf i -1}X_\msf i \widetilde{Z}_{\msf i} + Z_{\msf i-1}\widetilde{X}_{\msf i-1} Z_{\msf i}$ and $X_\msf i\widetilde{X}_{\msf i} + Z_{\msf i}\widetilde{Z}_{\msf i}Z_{\msf i+1}\widetilde{Z}_{\msf i+1}$, respectively. Putting everything together, we obtain a Hamiltonian $\mathbb H = \mathbb H_1 \oplus \mathbb H_2$ with
\begin{equation}
    \begin{split}
        \label{eq:coupledIsing1}
        \mathbb H_1 \stackrel{{\rm eff.}}{=} 
        -J \sum_{\msf i} \big( &\mathbb h_{\msf i} \otimes \widetilde{\mathbb 1}_{\msf i} 
        +\mathbb 1_{\msf i} \otimes \widetilde{\mathbb h}_{\msf i} 
        \\[-1em]
        &-g (\widetilde{Z}_{\msf i-1}X_\msf i \widetilde{Z}_{\msf i} +Z_{\msf i-1}\widetilde{X}_{\msf i-1} Z_{\msf i}) \big)
    \end{split}
\end{equation}
and
\begin{equation}
    \begin{split}
        \label{eq:coupledIsing2}
        \mathbb H_2 \stackrel{{\rm eff.}}{=} 
        -J \sum_{\msf i} \big( &\mathbb h_{\msf i} \otimes \widetilde{\mathbb 1}_{\msf i} 
        +\mathbb 1_{\msf i} \otimes \widetilde{\mathbb h}_{\msf i} 
        \\[-1em]
        &-g (X_{\msf i} \widetilde{X}_{\msf i} + Z_{\msf i}\widetilde{Z}_{\msf i}Z_{\msf i+1}\widetilde{Z}_{\msf i+1})\big)
    \end{split}
\end{equation}
which both describe two coupled \emph{critical} Ising models that decouple for $g=0$. 

\medskip
\noindent
$\bul$ Let $\mc M = \mathsf{Ising}$ be the module category over $\mathsf{Ising}^\msf{op} \boxtimes \mathsf{Ising}$ defined via $A \cat (\alpha_1 \boxtimes \alpha_2) \simeq (\alpha_1 \otimes A) \otimes \alpha_2$, for any $A \in \mc I_{\mc M}$. In particular one has $\sigma \cat (\sigma, \sigma) \simeq 2 \cdot \sigma$ and $\mathbb 1 \cat (\sigma, \sigma) \simeq \mathbb 1 \oplus \psi$.
A non-exhaustive list of non-vanishing $\F{\cat}$-symbols is given by
\begin{align*}
    &\big(
    \F{\cat}^{\sigma (\sigma ,\sigma)(\sigma ,\sigma)}_\sigma    
    \big)^{(\gamma_1,\gamma_2),11}_{\sigma,il}
    \! =
    \frac{1}{d_\sigma^2} \!
    \begin{pmatrix}
        1 & 1 & 1 & 1
        \\
        1 & -1 & 1 & -1
        \\
        1 & 1 & -1 & -1
        \\
        -1 & 1 & 1 & -1
    \end{pmatrix}_{\!\!\!\! (il)(\gamma_1,\gamma_2)}
    \\
    &\big(
    \F{\cat}^{A (\sigma, \sigma)(\sigma ,\sigma)}_B    
    \big)^{(\gamma_1 \otimes A,\gamma_2),11}_{C,11}
    \! =
    \frac{1}{d_\sigma}
    \begin{pmatrix}
        1 & 1 
        \\
        1 & -1
    \end{pmatrix}_{\!\! C\gamma_1} \,
\end{align*}
where $A,B,C,\gamma_1,\gamma_2 \in \{\mathbb 1, \psi\}$ and $i,l \in \{1,2\}$ such that the fusion spaces are all non-trivial. Unlabelled module strands are given by eq.~\eqref{eq:modStrandIsing} so that the effective total Hilbert space decomposes as
\begin{equation}
    \begin{split}
        \label{eq:decompXXZ}
        \mc H
        \stackrel{\rm eff.}{=} \;
        &\mathbb C \Big[ \!\!\!
        \includeTikz{0}{chainXXZIsing1}{\chain{1 \oplus 2}{1 \oplus 2}{1 \oplus 2}{\sigma}{\sigma}{\sigma}{\sigma}{\sigma\sigma}{\sigma\sigma}{\sigma\sigma}} \!\!\! \Big]
        \\ 
        \oplus \;
        &\mathbb C \Big[ \!\!\!
        \includeTikz{0}{chainXXZIsing2}{\chain{1}{1}{1}{\mathbb 1 \oplus \psi}{\mathbb 1 \oplus \psi}{\mathbb 1 \oplus \psi}{\mathbb 1 \oplus \psi}{\sigma\sigma}{\sigma\sigma}{\sigma\sigma}}
        \!\!\! \Big] ,
    \end{split}
\end{equation}
where $1 \oplus 2$ refer to the two basis vectors in $\mc V_{\sigma,(\sigma\sigma)}^\sigma$.
The operator $\mathbb b_{1,\msf i}^{\mc M}+ \mathbb b_{2,\msf i}^{\mc M} - g \mathbb b_{3,\msf i}^{\mc M}$ acts on the first vector space appearing in this decomposition in the Pauli $Z$ basis as $Z_{\msf i-\frac{1}{2}} X_{\msf i+\frac{1}{2}}
+ X_{\msf i-\frac{1}{2}} Z_{\msf i+\frac{1}{2}}- gY_{\msf i-\frac{1}{2}} Y_{\msf i+\frac{1}{2}}$, whereas it acts on the second term as $Z_{\msf i-1} X_{\msf i}\mathbb 1_{\msf i+1}
+ \mathbb 1_{\msf i -1} X_{\msf i} Z_{\msf i+1}- gZ_{\msf i-1} \mathbb 1_\msf i Z_{\msf i+1}$. Conjugating every other site of these spin chains via a \emph{Hadamard} matrix yields an effective Hamiltonian $\mathbb H = \mathbb H_1 \oplus \mathbb H_2$ with
\begin{equation}
    \label{eq:XXZA}
    \mathbb H_1 \stackrel{\rm eff.}{=} - J\sum_\msf i
    \big(
    X_{\msf i-\frac{1}{2}} X_{\msf i+\frac{1}{2}} + Z_{\msf i-\frac{1}{2}} Z_{\msf i+\frac{1}{2}} +g Y_{\msf i-\frac{1}{2}}Y_{\msf i+\frac{1}{2}} \big) \, , 
\end{equation}
which we recognize as the spin-1/2 Heisenberg XXZ model \cite{1928ZPhy...49..619H}, and
\begin{equation}
    \begin{split}
        \label{eq:XXZB}
        \mathbb H_2 \stackrel{\rm eff.}{=} - J\sum_\msf i
        \big(
        &X_\msf i X_{\msf i+1} + Z_\msf i Z_{\msf i+1}
        \\[-1em]
        &- g(X_{\msf i}X_{\msf i+2} + Z_{\msf i}Z_{\msf i+2})
        \big) \, .
    \end{split}
\end{equation}
We can readily check via repeated usage of Kramers-Wannier duality that these Hamiltonians are indeed dual to the coupled chains provided in eq.~\eqref{eq:coupledIsing1} and eq.~\eqref{eq:coupledIsing2}, respectively \cite{baxter,FENDLEY1989549,PhysRevD.25.1103,bondAlgebra}. Our construction complements the treatment of the $\msf{Ising}^{\rm op}\boxtimes \msf{Ising}$ CFT carried out in \cite{Thorngren:2019iar}.

\medskip \noindent
$\bul$ Let $\mc M = \msf{Ising} / {\la \psi \simeq \mathbb 1 \ra}$ be the (fermionic) $\msf{Ising}^{\rm op}\boxtimes \msf{Ising}$-module category obtained from the module category $\msf{Ising}$ via fermion condensation. The underlying category was described in the previous example and the module structure is such that $\mathbb 1 \cat (\sigma, \sigma) \simeq \mathbb C^{1|1} \cdot \mathbb 1$, $\mathbb 1 \cat (\psi,\mathbb 1) \simeq \mathbb C^{0|1}\cdot \mathbb 1$ and $\beta \cat (\sigma, \sigma) \simeq \mathbb C^{2|2} \cdot \beta$. Amongst others, we have the following $\F{\cat}$-symbols: 

\begin{align*}
    \big(\F{\cat}^{\mathbb 1(\sigma, \sigma)(\sigma, \sigma)}_{\mathbb 1} \big)_{\mathbb 1,\ee \ee}^{(\mathbb 1,\mathbb 1),1 \ee}
    =
    -\big(\F{\cat}^{\mathbb 1(\sigma, \sigma)(\sigma, \sigma)}_{\mathbb 1} \big)_{\mathbb 1,\oo \oo}^{(\mathbb 1,\mathbb 1),1 \ee} &= \frac{1}{d_\sigma} \, ,
    \\
    \big(\F{\cat}^{\mathbb 1(\sigma, \sigma)(\sigma, \sigma)}_{\mathbb 1} \big)_{\mathbb 1,\oo \ee}^{(\psi,\mathbb 1),1 \oo} 
    =
    i\big(\F{\cat}^{\mathbb 1(\sigma, \sigma)(\sigma, \sigma)}_{\mathbb 1} \big)_{\mathbb 1,\ee \oo}^{(\psi,\mathbb 1),1 \oo} &= \frac{1}{d_\sigma} \, ,
    \\
    \big(\F{\cat}^{\mathbb 1(\sigma, \sigma)(\sigma, \sigma)}_{\mathbb 1} \big)_{\mathbb 1,\oo \ee}^{(\mathbb 1,\psi),1 \oo} 
    =
    -i\big(\F{\cat}^{\mathbb 1(\sigma, \sigma)(\sigma, \sigma)}_{\mathbb 1} \big)_{\mathbb 1,\ee \oo}^{(\mathbb 1,\psi),1 \oo} &= \frac{1}{d_\sigma} \, ,
    \\
    \big(\F{\cat}^{\mathbb 1(\sigma, \sigma)(\sigma, \sigma)}_{\mathbb 1} \big)_{\mathbb 1,\ee \ee}^{(\psi,\psi),1 \ee} 
    =
    \big(\F{\cat}^{\mathbb 1(\sigma, \sigma)(\sigma, \sigma)}_{\mathbb 1} \big)_{\mathbb 1,\oo \oo}^{(\psi,\psi),1 \ee} &= \frac{1}{d_\sigma} \, .
\end{align*}
It follows from the fusion rules that the Hamiltonian acts on the total Hilbert space
\begin{equation}
    \begin{split}
        \label{eq:decompDFermio}
        \mc H
        \stackrel{\rm eff.}{=} \; &\mathbb C \Big[ \!\!\!
        \includeTikz{0}{chainXXZFermionic1}{\chain{{\rm e} \oplus {\rm o}}{{\rm e} \oplus {\rm o}}{{\rm e} \oplus {\rm o}}{\mathbb 1}{\mathbb 1}{\mathbb 1}{\mathbb 1}{\sigma\sigma}{\sigma\sigma}{\sigma\sigma}} \!\!\! \Big]
        \\ 
        \oplus \;  &\mathbb C \Big[ \!\!\!
        \includeTikz{0}{chainXXZFermionic2}{\chain{1}{1}{1}{\beta}{\beta}{\beta}{\beta}{\sigma\sigma}{\sigma\sigma}{\sigma\sigma}} \!\!\! \Big] ,
    \end{split}
\end{equation}
where as before ${\rm e} \oplus {\rm o}$ refer to the purely even and odd basis vectors in $\mathbb C^{1|1}$.
Applying the same techniques as previously, up to local unitaries, we find an effective Hamiltonian $\mathbb H = \mathbb H_1 \oplus \mathbb H_2$ with
\begin{equation}
    \begin{split}
        \mathbb H_1 \stackrel{\rm eff.}{=} - J\sum_\msf i
        \big(
        &2c^\dagger_{\msf i-\frac{1}{2}} c^{\phantom \dagger}_{\msf i+\frac{1}{2}} + 2c^\dagger_{\msf i+\frac{1}{2}} c^{\phantom \dagger}_{\msf i-\frac{1}{2}}
        \\[-0.2em]
        &+g(2c^\dagger_{\msf i-\frac{1}{2}} c^{\phantom \dagger}_{\msf i-\frac{1}{2}}-1)(2c^\dagger_{\msf i+\frac{1}{2}} c^{\phantom \dagger}_{\msf i+\frac{1}{2}}-1)
        \big)
    \end{split}
\end{equation}
\begin{equation}
    \begin{split}
        \mathbb H_2 \stackrel{\rm eff.}{=} &- J\sum_\msf i
        \big(
        2c^\dagger_{\msf i} c^{\phantom \dagger}_{\msf i+1} + 2c^\dagger_{\msf i+1} c^{\phantom \dagger}_{\msf i}
        \\[-0.2em]
        &+2g(c^\dagger_{\msf i}(2c^\dagger_{\msf i} c^{\phantom \dagger}_{\msf i}-1)c^{\phantom \dagger}_{\msf i+2} + c^\dagger_{\msf i+2} (2c^\dagger_{\msf i} c^{\phantom \dagger}_{\msf i}-1) c^{\phantom \dagger}_{\msf i})
        \big)
    \end{split}
\end{equation}
i.e. the Jordan-Wigner duals of the Hamiltonians \eqref{eq:XXZA} and \eqref{eq:XXZB}, respectively. We note that $\mathbb H_1$ corresponds to the standard \emph{Kogut-Susskind} prescription for discretizing a Dirac fermion \cite{PhysRevD.11.395}.


\subsection{$\mathsf{Rep}(\mc S_3)$: Heisenberg XXZ model}
\label{sec:RepS3}

\noindent
The category of finite-dimensional representations of a finite group $G$ form a fusion category, denoted as $\mathsf{Rep}(G)$. Similar to the $\mathsf{Vec}_G$ case, (indecomposable) module categories over $\mathsf{Rep}(G)$ are classified by pairs $(L,[\psi])$ with $L \subseteq G$ a subgroup of $G$ and $[\psi]$ a cohomology class in $H^2(L,\mathbb C^\times)$. The module categories are then denoted as $\mathsf{Rep}^{\psi}(L)$, the categories of (projective) representations of the subgroup $L$. Let $\mc{D} = \mathsf{Rep}(\mc S_3)$ be the category of finite dimensional representations of the symmetric group $\mc S_3$. There are three isomorphism classes of simple objects denoted by $\ub 0, \ub 1, \ub 2$, respectively, corresponding to the trivial, sign and two-dimensional irreducible representations. The non-trivial fusion rules are given by $\ub 1 \otimes \ub 1 \simeq \ub 0$, $\ub 1 \otimes \ub 2 \simeq \ub 2 \otimes \ub 1 \simeq \ub 2$ and $\ub 2 \otimes \ub 2 \simeq \ub 0 \oplus \ub 1 \oplus \ub 2$. The subset of relevant $F$-symbols for our construction is given by
\begin{align*}
    \big(\F{}^{\ub 0 \ub 2 \ub 2}_{\ub 0}\big)_{\ub 2,11}^{\ub 0,11} = 
    \big(\F{}^{\ub 1 \ub 2 \ub 2}_{\ub 1}\big)_{\ub 2,11}^{\ub 0,11} =
    \big(\F{}^{\ub 0 \ub 2 \ub 2}_{\ub 2}\big)_{\ub 2,11}^{\ub 2,11} = 
    - \big(\F{}^{\ub 1 \ub 2 \ub 2}_{\ub 2}\big)_{\ub 2,11}^{\ub 2,11} &= 1 \, ,
    \\
    \big(\F{}^{\ub 0 \ub 2 \ub 2}_{\ub 1}\big)_{\ub 2,11}^{\ub 1,11} = 
    \big(\F{}^{\ub 1 \ub 2 \ub 2}_{\ub 0}\big)_{\ub 2,11}^{\ub 1,11} = 
    \big(\F{}^{\ub 2 \ub 2 \ub 2}_{\ub 0}\big)_{\ub 2,11}^{\ub 2,11} = 
    - \big(\F{}^{\ub 2 \ub 2 \ub 2}_{\ub 1}\big)_{\ub 2,11}^{\ub 2,11} &= 1 \, ,
    \\
    \big(\F{}^{\ub 2 \ub 2 \ub 2}_{\ub 2}\big)_{\ub 0,11}^{\ub 0,11} = 
    \big(\F{}^{\ub 2 \ub 2 \ub 2}_{\ub 2}\big)_{\ub 1,11}^{\ub 0,11} = 
    \big(\F{}^{\ub 2 \ub 2 \ub 2}_{\ub 2}\big)_{\ub 0,11}^{\ub 1,11} = 
    \big(\F{}^{\ub 2 \ub 2 \ub 2}_{\ub 2}\big)_{\ub 1,11}^{\ub 1,11} &= \frac{1}{2}
\end{align*}
and
\begin{align*}
    \big(\F{}^{\ub 2 \ub 2 \ub 2}_{\ub 2}\big)_{\ub 2,11}^{\ub 0,11} 
    =
    \big(\F{}^{\ub 2 \ub 2 \ub 2}_{\ub 2}\big)_{\ub 0,11}^{\ub 2,11}  
    &= \frac{1}{\sqrt{2}} \, ,
    \\
    \big(\F{}^{\ub 2 \ub 2 \ub 2}_{\ub 2}\big)_{\ub 2,11}^{\ub 1,11} 
    = 
    \big(\F{}^{\ub 2 \ub 2 \ub 2}_{\ub 2}\big)_{\ub 1,11}^{\ub 2,11}
    &= - \frac{1}{\sqrt{2}} \, .
\end{align*}
As an example, we consider the model
\begin{equation}
    \mathbb H = 
    2J\sum_{\msf i} \mathbb b_{1,\msf i}^{\mc M}
    -Jg \sum_{\msf i} \mathbb b_{2,\msf i}^{\mc M}
\end{equation}
defined by the bonds
\begin{align}
    \mathbb b_{1,\msf i}^{\mc M} &=
    \includeTikz{0}{HamRepS31}{\HAM{}{}{}{1}{}{}{1}{}{}{}{}{\ub 2}{\ub 2}{\ub 2}{\ub 2}{\ub 0}}
    - \includeTikz{0}{HamRepS32}{\HAM{}{}{}{1}{}{}{1}{}{}{}{}{\ub 2}{\ub 2}{\ub 2}{\ub 2}{\ub 1}}
    , \\
    \mathbb b_{2,\msf i}^{\mc M} &=
    \includeTikz{0}{HamRepS33}{\HAM{}{}{}{1}{}{}{1}{}{}{}{}{\ub 2}{\ub 2}{\ub 2}{\ub 2}{\ub 0}}
    + \includeTikz{0}{HamRepS34}{\HAM{}{}{}{1}{}{}{1}{}{}{}{}{\ub 2}{\ub 2}{\ub 2}{\ub 2}{\ub 1}}
    - \includeTikz{0}{HamRepS35}{\HAM{}{}{}{1}{}{}{1}{}{}{}{}{\ub 2}{\ub 2}{\ub 2}{\ub 2}{\ub 2}} . \nonumber
\end{align}

\medskip \noindent
$\bul$ Let $\mc M = \msf{Rep}(\mathbb Z_1) \simeq \mathsf{Vec}$, where we denote the unique simple object of $\mathsf{Vec}$ as $\mathbb 1$. The non-trivial fusion rules are given by $\mathbb 1 \cat \ub 0 \simeq \mathbb 1 \cat \ub 1 \simeq \mathbb 1$ and $\mathbb 1 \cat \ub2 \simeq 2 \cdot \mathbb 1$. The $\F{\cat}$-symbols are given by the Clebsch-Gordan coefficients of $\mc S_3$, the relevant entries of which being
\begin{align*}
    \big(\F{\cat}^{\mathbb 1 \ub 2 \ub 2}_{\mathbb 1}\big)^{\ub 0,11}_{\mathbb 1,12} = 
    \big(\F{\cat}^{\mathbb 1 \ub 2 \ub 2}_{\mathbb 1}\big)^{\ub 0,11}_{\mathbb 1,21} 
    &= \frac{1}{\sqrt{2}}
    \\
    - \big(\F{\cat}^{\mathbb 1 \ub 2 \ub 2}_{\mathbb 1}\big)^{\ub 1,11}_{\mathbb 1,12} 
    =
    \big(\F{\cat}^{\mathbb 1 \ub 2 \ub 2}_{\mathbb 1}\big)^{\ub 1,11}_{\mathbb 1,21} = \frac{1}{\sqrt{2}}
    &= \frac{1}{\sqrt{2}}
    \\
    \big(\F{\cat}^{\mathbb 1 \ub 2 \ub 2}_{\mathbb 1}\big)^{\ub 2,12}_{\mathbb 1,11} = \big(\F{\cat}^{\mathbb 1 \ub 2 \ub 2}_{\mathbb 1}\big)^{\ub 2,11}_{\mathbb 1,22} 
    &= 1 \, .
\end{align*}
The total Hilbert space is given by $\mc H \simeq \bigotimes_{\msf i}\mathbb C^2$ with $\mathbb C^2 \simeq \Hom_{\mc M}(\mathbb 1 \cat \ub 2,\mathbb 1)$, and the Hamiltonian reads
\begin{align}
    \mathbb H &= J \sum_{\msf i} (X_{\msf i-\frac{1}{2}} X_{\msf i+\frac{1}{2}} + Y_{\msf i-\frac{1}{2}} Y_{\msf i+\frac{1}{2}} + g Z_{\msf i-\frac{1}{2}} Z_{\msf i+\frac{1}{2}}) \, ,
\end{align}
which we again recognize as the spin-1/2 Heisenberg XXZ model.

\medskip \noindent
$\bul$ Let $\mc M = \mathsf{Rep}(\mathbb Z_3)$, whose simple objects we denote by $\ub 0_{\mathbb Z_3}$, $\ub 1_{\mathbb Z_3}$ and $\ub 1_{\mathbb Z_3}^*$. The fusion rules are given by $A \cat \ub 0 \simeq A \cat \ub 1 \simeq A$ and $A \cat \ub2 \simeq \bigoplus_{B \not\simeq A} B$ where $A,B$ are simple objects in $\mathsf{Rep}(\mathbb Z_3)$. The relevant $\F{\cat}$-symbols are given by
\begin{align*}
    \nn
    \big(\F{\cat}^{A \ub 2 \ub 2}_{A}\big)^{\ub 0,11}_{B,11} 
    &= 
    \big(\F{\cat}^{\ub 0_{\mathbb Z_3} \ub 2 \ub 2}_{\ub 0_{\mathbb Z_3}}\big)^{\ub 1,11}_{\ub 1_{\mathbb Z_3},11} = 
    \big(\F{\cat}^{\ub 1_{\mathbb Z_3} \ub 2 \ub 2}_{\ub 1_{\mathbb Z_3}}\big)^{\ub 1,11}_{\ub 0_{\mathbb Z_3},11} = \frac{1}{\sqrt{2}}
    \\ \nn
    &= \big(\F{\cat}^{\ub 2_{\mathbb Z_3} \ub 2 \ub 2}_{\ub 2_{\mathbb Z_3}}\big)^{\ub 1,11}_{\ub 0_{\mathbb Z_3},11} = 
    - \big(\F{\cat}^{\ub 0_{\mathbb Z_3} \ub 2 \ub 2}_{\ub 0_{\mathbb Z_3}}\big)^{\ub 1,11}_{\ub 2_{\mathbb Z_3},11} 
    \\ \nn
    &= - \big(\F{\cat}^{\ub 1_{\mathbb Z_3} \ub 2 \ub 2}_{\ub 1_{\mathbb Z_3}}\big)^{\ub 1,11}_{\ub 2_{\mathbb Z_3},11} =
    - \big(\F{\cat}^{\ub 2_{\mathbb Z_3} \ub 2 \ub 2}_{\ub 2_{\mathbb Z_3}}\big)^{\ub 1,11}_{\ub 1_{\mathbb Z_3},11} \, , \\ 
    \big(\F{\cat}^{A \ub 2 \ub 2}_{B}\big)^{\ub 2,11}_{C,11} &= 1 \, ,
\end{align*}
where $A,B,C \in \mc I_{\msf{Rep}(\mathbb Z_2)}$ are all required to be distinct simple objects. We have
\begin{equation}
    \includeTikz{2.2}{objectMod}{}  
    = \!\! \bigoplus_{A \in \mc I_{\msf{Rep}(\mathbb Z_3)}} \!\!\! \includeTikz{2.2}{objectModA}{\object{mod}{A}}
\end{equation}
so that the local Hilbert space is isomorphic to $\mathbb C^3$. Importantly however, the global Hilbert space is subject to the constraint that no two neighbouring sites can be labeled by the same simple object. Putting everything together, the Hamiltonian is given by 
\begin{align}
    \label{eq:3coloring}
    \nn
    \mathbb H = 2J \sum_{\msf i} (V_{\msf i} + V_{\msf i}^\dag) 
    - \frac{Jg}{3} \sum_{\msf i}  ( &U_{\msf i-1}U_{\msf i+1}^\dag - U_{\msf i-1}U_{\msf i}U_{\msf i+1} 
    \\[-1em] &+ \text{h.c.})
\end{align}
with $\omega = e^{2\pi i/3}$, $\omega UV = VU$, and
\begin{equation*}
    U = \begin{pmatrix}
        1 & 0 & 0\\
        0 & \omega & 0\\
        0 & 0 & \omega^2
    \end{pmatrix}\, , \q
    V = \begin{pmatrix}
        0 & 1 & 0\\
        0 & 0 & 1\\
        1 & 0 & 0
    \end{pmatrix} \, .
\end{equation*}
While this model is a priori only defined on the constrained Hilbert space, the Hamiltonian in eq. \eqref{eq:3coloring} can be extended to act on the full space $\mc H_{\rm ext} = \bigotimes_{\msf i} \mathbb C^3$. The enlarged model is only dual to the spin-$1/2$ XXZ model on the subspace $\mc H \in \mc H_{\rm ext}$, meaning that for the enlarged model this duality is emergent. Due to its duality to the spin-$1/2$ XXZ model on $\mc H$, the enlarged model is integrable on a subspace, making it an example of a quasi-exactly solvable model \cite{Ushveridze1994,BatistaOrtiz2000,MoudgalyaMotrunich2022}. In this subspace, the model displays an emergent ${\rm U}(1)$ symmetry generated by $i(U_{\msf i}U_{\msf i+1}^{\dag}-U_{\msf i}^{\dag}U_{\msf i+1})/\sqrt{3}$, obtained as the dual of the generator $Z_{{\msf i}+\frac{1}{2}}$ of the ${\rm U}(1)$ symmetry in the XXZ model. It would be difficult to identify this symmetry without the duality transformation to the XXZ model, as this requires the identification of the correct subspace; this is manifest in our approach.

\medskip \noindent
$\bul$ Let $\mc M = \msf{Rep}(\mathbb Z_2)$, whose simple objects are denoted by $\ub 0_{\mathbb Z_2}$ and $\ub 1_{\mathbb Z_2}$. The non-trivial fusion rules are given by $\ub 0_{\mathbb Z_2} \cat \ub 1 \simeq \ub 1_{\mathbb Z_2}, \ub 1_{\mathbb Z_2} \cat \ub 1 \simeq \ub 0_{\mathbb Z_2}$ and $\ub 0_{\mathbb Z_2} \cat \ub 2 \simeq \ub 1_{\mathbb Z_2} \cat \ub 2 \simeq \ub 0_{\mathbb Z_2} \oplus \ub 1_{\mathbb Z_2}$. The relevant $\F{\cat}$-symbols are
\begin{align}
    \nn
    \big(\F{\cat}^{A \ub 2 \ub 2}_{A}\big)^{\ub 0,11}_{B,11} &= \big(\F{\cat}^{A \ub 2 \ub 2}_{(A \cat \ub 1)}\big)^{\ub 1,11}_{B,11} = \frac{1}{\sqrt{2}}\, ,
    \\
    \big(\F{\cat}^{A \ub 2 \ub 2}_{B}\big)^{\ub 2,11}_{C,11} &= 
    \begin{cases}
        \frac{1}{\sqrt{2}} \q {\rm if} \; A \otimes B \otimes C \simeq 0_{\mathbb Z_2} \\
        \frac{-1}{\sqrt{2}} \q {\rm if} \; A \otimes B \otimes C \simeq 1_{\mathbb Z_2}
    \end{cases} \!\!\! ,
\end{align}
where $A,B,C \in \mc I_{\msf{Rep}(\mathbb Z_2)}$ and $A \otimes B \otimes C$ refers to the monoidal structure of $\msf{Rep}(\mathbb Z_2)$ as a fusion category. We have 
\begin{equation}
    \includeTikz{2.2}{objectMod}{}  
    = \!\! \bigoplus_{A \in \mc I_{\msf{Rep}(\mathbb Z_2)}} \!\!\! \includeTikz{2.2}{objectModA}{\object{mod}{A}} \, ,
\end{equation}
which together with the fusion rules gives the Hilbert space $\mc H \simeq \bigotimes_{\msf i}(\mathbb C \oplus \mathbb C) \simeq \bigotimes_{\msf i}\mathbb C^2$. The Hamiltonian becomes
\begin{align}
    \label{eq:RepZ2}
    \mathbb H = J \sum_{\msf i} (Z_{\msf i-1} Z_{\msf i+1} + Z_{\msf i-1} X_{\msf i} Z_{\msf i+1}
    - g X_{\msf i}) \, .
\end{align}
The duality between this Hamiltonian and the spin-1/2 Heisenberg XXZ model was also obtained in \cite{bondAlgebra}, but using a bond algebra based on $\mc{D} = \msf{Rep}(\mathbb Z_2)$. The Hamiltonian \eqref{eq:RepZ2} can also be understood as the half-integer sector of the spin-1 $\rm{su}(2)_4$ chain considered in \cite{PhysRevB.87.235120}.

\medskip \noindent
$\bul$ Let $\mc M = \msf{Rep}(\mc S_3)$ be the regular module category. We have $\includeTikz{2.2}{objectMod}{} = \includeTikz{2.2}{objectModUb0}{\object{mod}{\ub 0}} \oplus \includeTikz{2.2}{objectModUb1}{\object{mod}{\ub 1}} \oplus \includeTikz{2.2}{objectModUb2}{\object{mod}{\ub 2}}$, and the Hamiltonian acts on a Hilbert space of the form
\begin{equation}
    \begin{split}
        \mc H
        \stackrel{\rm eff.}{=} \; & \bigoplus_{\{A\}} \mathbb C \Big[ \!\!\!
        \includeTikz{0}{chainRepS3}{\chain{1}{1}{1}{A_{\msf i-2}}{A_{\msf i-1}}{A_{\msf i}}{A_{\msf i+1}}{\ub 2}{\ub 2}{\ub 2}} \!\!\! \Big]
    \end{split}
\end{equation}
spanned by all admissible configurations of simple objects $\{A\}$. It is convenient to think of such allowed configurations as paths on the \emph{adjacency graph}
\begin{equation}
    \includeTikz{0}{adjacency4}{\adjacency{4}{\ub 0}{\ub 2}{\ub 1}{}{}{}}
\end{equation}
associated with the fusion rules $N_{A_{\msf i} \ub 2}^{A_{\msf i + 1}}$. The $\F{\cat}$-symbols are given by the $\F{}$ symbols of $\msf{Rep}(\mc S_3)$. The Hamiltonian does not admit a particularly nice form, but has been studied before as an anyonic spin chain based the two-dimensional representation of $\mc S_3$ \cite{PhysRevB.94.085138}, or equivalently as the integer sector of an $\rm{su}(2)_4$ spin-1 chain \cite{PhysRevB.87.235120}. Importantly, the dualities between this model and the previous ones are all examples of non-abelian dualities. The duality to the XXZ chain has been observed before in \cite{PhysRevB.94.085138}.


\subsection{$\mathsf{Rep}({\rm U}_q(\mathfrak{sl}_2))$: quantum IRF-Vertex models}
\label{Vertexmodel}

\noindent
In a seminal work, Pasquier \cite{1988CMaPh.118..355P} derived an intertwiner between the interaction-round-the-face (IRF) and the vertex models using different representations of the Temperley-Lieb algebra. Interestingly, his construction is effectively equivalent to the categorical approach advocated in this paper. This example is slightly beyond the framework employed so far as the input data does not quite meet all the requirements to be a fusion category. However, our construction still largely applies allowing us to define dual models with categorical symmetries.
Let $\mathcal{D}=\mathsf{Rep}({\rm U}_q(\mathfrak{sl}_2))$ be the representation category of the quantum group defined as the $q$-deformed universal envelopping algebra ${\rm U}_q$ of the Lie algebra $\mathfrak{sl}_2$. We will restrict to the case where $q$ is \emph{not} a root of unity. Isomorphism classes of simple objects in $\mc D$ are labeled by half-integer spins $j \in \frac{1}{2} \mathbb N$, with fusion rules given by
\begin{equation}
    j_1 \otimes j_2 \simeq \bigoplus_{j = |j_1 - j_2|}^{j_1 + j_2} j_3 \, .
\end{equation}
The $F$-symbols of this category are well known and those required for our derivations can be found in \cite{1988CMaPh.118..355P,kirillov1990representations}. In the limit $q \to 1$, these boil down to the so-called $6j$-symbols of ${\rm SU}(2)$. As an example, we consider the model
\begin{equation}
    \mathbb H = \sum_{\msf i} \mathbb b_{\msf i}^{\mc M} \, ,
\end{equation}
defined by the bond
\begin{equation}
    \mathbb b_{\msf i}^{\mc M} = \includeTikz{0}{HamIRF}{\HAM{}{}{}{1}{}{}{1}{}{}{}{}{\nicefrac{1}{2}}{\nicefrac{1}{2}}{\nicefrac{1}{2}}{\nicefrac{1}{2}}{0}} .
\end{equation}

\medskip \noindent
$\bul$ Let $\mc M = \mathsf{Rep}({\rm U}_q(\mathfrak{sl}_2))$ be the regular module category. We have $\includeTikz{2.2}{objectMod}{} = \bigoplus_{j \in \frac{1}{2}\mathbb N} \includeTikz{2.2}{objectModj}{\object{mod}{j}}$ and the Hilbert space is given by summing over over all possible spin configurations such that neighbouring sites are labelled by spins that differ by $\frac{1}{2}$:
\begin{equation}
    \begin{split}
        \label{eq:decompIRF}
        \mc H
        \stackrel{\rm eff.}{=} \; & \bigoplus_{\{j\}} \mathbb C \Big[ \!\!\!
        \includeTikz{0}{chainIRF}{\chain{1}{1}{1}{j_{\msf i-2}}{j_{\msf i-1}}{j_{\msf i}}{j_{\msf i+1}}{\nicefrac{1}{2}}{\nicefrac{1}{2}}{\nicefrac{1}{2}}} \!\!\! \Big].
    \end{split}
\end{equation}
The Hamiltonian is then defined by
\begin{equation}
    \la j_{\msf i - 1},j'_\msf i,j_{\msf i + 1}| \mathbb b_{\msf i}^{\mc M} |j_{\msf i - 1},j_\msf i,j_{\msf i + 1} \ra = \delta_{j_{\msf i-1},j_{\msf i+1}} \frac{\sqrt{\lfloor j_\msf i \rfloor_q \lfloor j'_\msf i \rfloor_q}}{\lfloor j_{\msf i-1} \rfloor_q}
\end{equation}
where $\lfloor j \rfloor_q$ is defined as $\lfloor j \rfloor_q = (q^j - q^{-j})/(q - q^{-1})$. This Hamiltonian is the quantum analogue of the IRF model.

\medskip \noindent
$\bul$ Let $\mc M = \mathsf{Vec}$, where we again denote the unique simple object of $\mathsf{Vec}$ as $\mathbb 1$. The fusion rules are given by $\mathbb 1 \cat j \simeq (2j+1) \cdot \mathbb 1$ and the $\F{\cat}$-symbols can be found in \cite{1988CMaPh.118..355P,kirillov1990representations}. In the limit $q \to 1$, these boil down to the Clebsch-Gordan coefficients of ${\rm SU}(2)$. The total Hilbert space is now given by $\mc H \simeq \bigotimes_{\msf i}\mathbb C^2$ with $\mathbb C^2 \simeq \Hom_{\mc M}(\mathbb 1 \cat \frac{1}{2},\mathbb 1)$ and the Hamiltonian becomes
\begin{align}
    \mathbb H = -\frac{1}{2} \sum_{\msf i} 
    &\big( X_{\msf i-\frac{1}{2}}X_{\msf i+\frac{1}{2}} + Y_{\msf i-\frac{1}{2}}Y_{\msf i+\frac{1}{2}} 
    \\
    \nn
    + \frac{q + q^{-1}}{2}&(Z_{\msf i-\frac{1}{2}}Z_{\msf i+\frac{1}{2}} - \mathbb 1) + \frac{q - q^{-1}}{2}(Z_{\msf i-\frac{1}{2}} - Z_{\msf i+\frac{1}{2}})\big) \, ,
\end{align}
which is the (1+1)d quantum analogue of the 2d classical \emph{6-vertex} model. Note that it coincides with the XXZ model up to an additive constant and a boundary term. The $\mc M = \mathsf{Rep}({\rm U}_q(\mathfrak{sl}_2))$ model is dual to this model on subspace $\mc H$ in eq.~\eqref{eq:decompIRF}, which makes it an emergent duality.


\subsection{$\mc H_3$: Haagerup anyonic spin chain}
\label{Haagerup}

\noindent
Most of the dualities considered thus far involve fairy simple symmetric operators and most of them have been extensively studied in the past using more conventional methods. We shall now consider a much less trivial class of models based on the \emph{Haagerup subfactor} $\mc H_3$ fusion category that cannot be derived from (quantum) group theory \cite{haagerup1994principal,asaeda1999exotic}, bringing our approach to full fruition. This means in particular that the bonds cannot easily be written in terms of more conventional operators such as Pauli matrices stemming from (quantum) group theory, and are best defined directly in terms of $F$-symbols. In addition, the Hilbert spaces of these models do not admit a tensor product structure, and therefore all the dualities presented in this section are emergent. The relevant categorical data is taken from \cite{grossman2012quantum,barter2021computing}. Let $\mc D = \mc H_3$ with simple objects $\{\mathbb 1, \alpha, \alpha^2, \rho, \alpha\rho, \alpha^2\rho\}$. The quantum dimensions are given by $d_{\mathbb 1} = d_{\alpha} = d_{\alpha^2} = 1$ and $d_{\rho} = d_{\alpha\rho} = d_{\alpha^2\rho} = (3+\sqrt{13})/2$. Defining $\chi := \rho \oplus \alpha\rho \oplus \alpha^2\rho$, the fusion rules read 
\begin{align*}
	\setlength{\tabcolsep}{0.5em}
	\begin{tabularx}{0.91\columnwidth}{c|ccccccc}
	    $\otimes$ & $\mathbb 1$ & $\alpha$ & $\alpha^2$ & $\rho$ & $\alpha\rho$ & $\alpha^2\rho$
	    \\
	    \midrule
	    $\mathbb 1$ & $\mathbb 1$ & $\alpha$ & $\alpha^2$ & $\rho$ & $\alpha\rho$ & $\alpha^2\rho$
	    \\
	    $\alpha$ & $\alpha$ & $\alpha^2$ & $\mathbb 1$ & $\alpha\rho$ & $\alpha^2\rho$ & $\rho$
	    \\
	    $\alpha^2$ & $\alpha^2$ & $\mathbb 1$ & $\alpha$ & $\alpha^2\rho$ & $\rho$ & $\alpha\rho$
	    \\
	    $\rho$ & $\rho$ & $\alpha^2\rho$ & $\alpha\rho$ & $\mathbb 1 \oplus \chi$ & $\alpha^2 \oplus \chi$ & $\alpha \oplus \chi$
	    \\
	    $\alpha\rho$ & $\alpha\rho$ & $\rho$ & $\alpha^2\rho$ & $\alpha \oplus \chi$ & $\mathbb 1 \oplus \chi$ & $\alpha^2 \oplus \chi$
	    \\
	    $\alpha^2\rho$ & $\alpha^2\rho$ & $\alpha\rho$ & $\rho$ & $\alpha^2 \oplus \chi$ & $\alpha \oplus \chi$ & $\mathbb 1 \oplus \chi$
	\end{tabularx} \, .
\end{align*}
Introducing the notation $\alpha^0 := \mathbb 1$, we remark that the simple objects $\{\alpha^i\}$ with $i \in \{0,1,2\}$ form a $\Vect_{\mathbb Z_3}$-subcategory. We have the following subset of non-vanishing $F$-symbols:
\begin{gather*}
    \big(\F{}^{\alpha^i\rho\rho}_{\alpha^i} \big)_{(\alpha^i\rho),11}^{\mathbb 1,11} = 1
    \, , \q 
    \big(\F{}^{(\alpha^i\rho)\rho\rho}_{(\alpha^i\rho)} \big)_{\alpha^i,11}^{\mathbb 1,11} = d_{\rho}-3 \, ,
    \\
    \big(\F{}^{(\alpha^i\rho)\rho\rho}_{(\alpha^i\rho)} \big)_{(\alpha^j\rho),11}^{\mathbb 1,11} = \sqrt{d_{\rho}-3} \, .
\end{gather*}
As an example, we consider the model
\begin{equation}
    \mathbb H = - \sum_{\msf i} \mathbb b_{\msf i}^{\mc M} \, ,
\end{equation}
defined by the bond
\begin{equation}
    \mathbb b_{\msf i}^{\mc M} = \includeTikz{0}{HamHaagerup}{\HAM{}{}{}{1}{}{}{1}{}{}{}{}{\rho}{\rho}{\rho}{\rho}{\mathbb 1}} \, .
\end{equation}

\medskip \noindent
$\bul$ Let $\mc M = \mc H_3$ be the regular module category. The Hamiltonian acts on an effective total Hilbert space of the form
\begin{equation}
    \label{eq:decompH3}
    \mc H
    \stackrel{\rm eff.}{=} \bigoplus_{\{A\}} \mathbb C \Big[ \!\!\!
    \includeTikz{0}{chainHaagerup}{\chain{1}{1}{1}{A_{\msf i-2}}{A_{\msf i-1}}{A_{\msf i}}{A_{\msf i+1}}{\rho}{\rho}{\rho}} \!\!\! \Big],
\end{equation}
that decomposes over all admissible configurations of simple objects $\{A\}$. 
It is convenient to think of such allowed configurations as paths on the adjacency graph
\begin{equation}
    \includeTikz{0}{adjacency1}{\adjacency{1}{\mathbb 1}{\alpha}{\rho}{\Gamma}{\Lambda}{G}}
\end{equation}
associated with the fusion rules $N_{A_{\msf i}\rho}^{A_{\msf i+1}}$ with $A_{\msf i} \in \mc H_3$. The Hamiltonian can be derived from the $F$-symbols provided above, but does not admit a nice form so we refrain from writing it down explicitly. This model---or its classical statistical mechanical counterpart---were studied numerically in \cite{Huang:2021nvb} and \cite{Vanhove:2021zop}, showing evidence that these are critical. These studies provide first concrete indications that a CFT based on the Haagerup subfactor exists, going against the current belief that all CFTs can be built from standard methods and supporting the claim that one can associate a CFT to any modular fusion category. There is a conceptual subtlety associated to these models; as a fusion category, $\mc H_3$ is not modular and as such should not describe a CFT. This can be remedied by instead considering the double of $\mc H_3$, denoted as $\mc Z(\mc H_3)$, as the input to a lattice model. The problem with models based on $\mc Z(\mc H_3)$ is that their local Hilbert space is very large, making them numerically intractable. Using our formalism however, it can be shown that any model built from $\mc H_3$ is dual to a $\mc Z(\mc H_3)$ model by considering $\mc H_3$ as a module category of $\mc Z(\mc H_3)$. In this way, we obtain models for which it can be shown, numerically, that they are critical, and due to their duality relation implies criticality for the associated $\mc Z(\mc H_3)$ model.

\medskip \noindent
$\bul$ Let $\mc M = \mc M_{3,1}$ be the $\mc H_3$-module category with simple objects $\{\Gamma,\Gamma\alpha,\Gamma\alpha^2,\Lambda\}$. The fusion rules are provided by $(\Gamma\alpha^i) \cat \alpha^j \simeq (\Gamma\alpha^{i+j})$ where $i,j \in \{0,1,2\}$ and $i+j$ is modulo 3, $\Lambda \cat \alpha^i \simeq \Lambda$ and
\begin{align*}
	\setlength{\tabcolsep}{0.5em}
	\begin{tabularx}{0.97\columnwidth}{c|ccc}
	    $\cat$ & $\rho$ & $\alpha\rho$ & $\alpha^2\rho$
	    \\
	    \midrule
        $\Gamma$ & $\Gamma\alpha \oplus \Gamma\alpha^2 \oplus \Lambda$ & $\Gamma \oplus \Gamma\alpha \oplus \Lambda$ & $\Gamma \oplus \Gamma\alpha^2 \oplus \Lambda$
        \\
        $\Gamma\alpha$ & $\Gamma \oplus \Gamma\alpha \oplus \Lambda$ & $\Gamma \oplus \Gamma\alpha^2 \oplus \Lambda$ & $\Gamma\alpha \oplus \Gamma\alpha^2 \oplus \Lambda$
        \\
        $\Gamma\alpha^2$ & $\Gamma \oplus \Gamma\alpha^2 \oplus \Lambda$ & $\Gamma\alpha \oplus \Gamma\alpha^2 \oplus \Lambda$ & $\Gamma \oplus \Gamma\alpha \oplus \Lambda$
        \\
        $\Lambda$ & $\Upsilon$ & $\Upsilon$ & $\Upsilon$
	\end{tabularx} \, ,
\end{align*}
where $\Upsilon := \Gamma \oplus \Gamma\alpha \oplus \Gamma\alpha^2 \oplus \Lambda$. A subset of $\F{\cat}$-symbols is given by
\begin{gather*}
    \big(\F{\cat}^{(\Gamma\alpha^i)\rho\rho}_{(\Gamma\alpha^i)} \big)_{\Lambda,11}^{\mathbb 1,11} = \sqrt{7-2d_{\rho}} \, ,
    \\ 
    \big(\F{\cat}^{\Lambda\rho\rho}_{\Lambda} \big)_{(\Gamma\alpha^i),11}^{\mathbb 1,11} = \sqrt{\frac{4-d_{\rho}}{3}}\, ,
    \\
    \big(\F{\cat}^{(\Gamma\alpha^i)\rho\rho}_{(\Gamma\alpha^i)} \big)_{(\Gamma\alpha^j),11}^{\mathbb 1,11} = \big(\F{\cat}^{\Lambda\rho\rho}_{\Lambda} \big)_{\Lambda,11}^{\mathbb 1,11} = \sqrt{d_{\rho}-3} \, ,
\end{gather*}
where $i,j \in \{0,1,2\}$ are such that the fusion spaces are all non-trivial. The effective Hilbert space has the same form as eq.~\eqref{eq:decompH3} but admissible
configurations are now identified with paths on the adjacency graph
\begin{equation}
    \includeTikz{0}{adjacency2}{\adjacency{2}{\mathbb 1}{\alpha}{\rho}{\Gamma}{\Lambda}{G}}
\end{equation}
associated with the fusion rules $N_{A_{\msf i}\rho}^{A_{\msf i+1}}$ with $A_{\msf i} \in \mc M_{3,1}$. The explicit Hamiltonian is then obtained from the $\F{\cat}$-symbols provided above, but does not admit a concise presentation.

\medskip \noindent
$\bul$ Let $\mc M = \mc M_{3,2}$ be the $\mc H_3$-module category with simple objects $\{\Omega,\Omega\rho\}$. The fusion rules are $\Omega \cat \alpha^i \simeq \Omega$, $\Omega \cat \alpha^i\rho \simeq \Omega\rho$, $\Omega\rho \cat \alpha^i \simeq \Omega\rho$ and $\Omega\rho \cat \alpha^i\rho \simeq \Omega \oplus (3 \cdot \Omega\rho)$. A subset of $\F{\cat}$-symbols is given by
\begin{align*}
    \big(\F{\cat}^{\Omega\rho\rho}_{\Omega} \big)_{(\Omega\rho),11}^{\mathbb 1,11} &= 1 \, , \q
    \big(\F{\cat}^{(\Omega\rho)\rho\rho}_{(\Omega\rho)} \big)_{\Omega,11}^{\mathbb 1,11} = d_{\rho}-3 \, ,
    \\
    \big(\F{\cat}^{(\Omega\rho)\rho\rho}_{(\Omega\rho)} \big)_{(\Omega\rho),11}^{\mathbb 1,11} &= 
    \big(\F{\cat}^{(\Omega\rho)\rho\rho}_{(\Omega\rho)} \big)_{(\Omega\rho),23}^{\mathbb 1,11}  \\
     &=     \big(\F{\cat}^{(\Omega\rho)\rho\rho}_{(\Omega\rho)} \big)_{(\Omega\rho),32}^{\mathbb 1,11}= \sqrt{d_{\rho}-3} \, ,
\end{align*}
where the multiplicites in the last three entries are the only ones for which this $\F{\cat}$-symbol is non-vanishing. The Hilbert space is given by
\begin{equation*}
    \mc H
    \stackrel{\rm eff.}{=} \! \bigoplus_{\{A\}}\bigoplus_{\{n\}} \mathbb C \Big[\hspace{-0.4em}
    \includeTikz{0}{chainIRFH3}{\chain{n_{\msf i - \frac{3}{2}}}{n_{\msf i - \frac{1}{2}}}{n_{\msf i + \frac{1}{2}}}{A_{\msf i-2}}{A_{\msf i-1}}{A_{\msf i}}{A_{\msf i+1}}{\rho}{\rho}{\rho}} \hspace{-0.4em}\Big] ,
\end{equation*}
where admissible configurations are identified with paths on the adjacency graph
\begin{equation}
    \includeTikz{0}{adjacency3}{\adjacency{3}{\mathbb 1}{\alpha}{\rho}{\Gamma}{\Lambda}{G}}
\end{equation}
associated with the fusion rules $N_{A_{\msf i}\rho}^{A_{\msf i+1}}$ with $A_{\msf i} \in \mc M_{3,2}$. As before, the explicit Hamiltonian is obtained from the $\F{\cat}$-symbols provided above.

%% file: _Discussion.tex
\section{Discussion and outlook}
\label{Section IV}

\noindent
\emph{We proposed a systematic framework to establish dual maps between {\rm (1+1)d} quantum models displaying categorical symmetries. We illustrated our construction with examples that are simple and yet non-trivial. Dualities as constructed above are all realized as non-local transformations on the Hilbert space, implemented by matrix product operator (MPO) intertwiners with non-trivial virtual bond dimension.}

\subsection{On the classification of duality maps}

\noindent
One of the key merits of our approach is that,  given a Hamiltonian in the framework of category theory, a classification of \emph{exact non-trivial} dualities emerges. These duality maps are required to (i) preserve the spectra of the dual Hamiltonians up to degeneracies, (ii) preserve the physical locality of the dual Hamiltonians, and (iii) act non-trivially on generic operators or physical degrees of freedom. The first two conditions imply that we want a map between two local descriptions of analogous underlying physics in two different languages. We require condition (iii) because we want to consider non-trivial maps, i.e., dualities that transform the thermodynamic phase of the model system on which they act (such as strong-coupling/weak-coupling relations).

We have considered the general case where the symmetries of the model system are described by a fusion category $\mc C$, in which case the order operators that characterize the phase of the model are described by the monoidal center $\mc Z(\mc C)$. Then, possible non-trivial actions of a duality on the order operators amount to braided equivalences of the form $\mc Z(\mc C) \rightarrow \mc Z(\mc C')$. It is understood that the existence of such an equivalence is the same as requiring $\mc C$ and $\mc C'$ to be Morita equivalent \cite{etingof2016tensor}. In other words, every representative in the Morita class of $\mc C$ yields a duality map. Statements regarding exhaustiveness of our approach follows from mathematical results pertaining to Morita equivalence: representatives of the Morita class of $\mc C$ are in one-to-one correspondence with the different choices of module categories over $\mc C$. Our approach assigns a Hamiltonian to each such choice of module category, and therefore exhausts all dualities of this kind. The classification of this type of duality is then equivalent to the classification of module categories over a fusion category. For several fusion categories of interest, this classification is known.

\subsection{Non-abelian dualities}

\noindent
Let us briefly comment on folklore results pertaining to so-called non-abelian dualities, which questions the possibility of having dualities in models with non-abelian symmetries \cite{RevModPhys.52.453}. Clearly, the notion of non-abelian duality needs to be refined, since the relevant symmetry for a duality is the one used to construct the bond algebra \cite{COBANERA2013574}. For instance, the Kennedy-Tasaki transformation of the spin-1 Heisenberg model is based on its $\mathbb Z_2 \times \mathbb Z_2$-symmetry and as such is an abelian duality, despite the presence of a larger (non-abelian) ${\rm SU}(2)$ symmetry in the model. A non-abelian duality thus requires a bond algebra based on a non-abelian symmetry, but additionally has to be such that the duality cannot be obtained from an abelian bond algebra. An example of the latter case is provided in sec.~\ref{sec:RepS3}, where the bond algebra is described by the fusion category $\mc{D} = \msf{Rep}(\mc S_3)$. The duality between $\mc{M}=\msf{Vec}$ and $\mc{M}=\msf{Rep}(\mathbb Z_2)$ can also be obtained from a bond algebra based on $\msf{Rep}(\mathbb Z_2)$ (see ref.~\cite{bondAlgebra}). In contrast, the duality between $\mc{M} = \msf{Vec}$ and $\mc{M} = \msf{Rep}(\mc S_3)$ cannot be obtained from an abelian bond algebra so that it is a true non-abelian duality. We thus define a non-abelian duality as a duality where the fusion category describing the bond-algebra is based on a non-abelian group and at least one of the module categories involved is also based on a non-abelian group.

Typically, dual symmetries associated to a non-abelian duality are non-invertible and most naturally described using the language of fusion categories. The categorical approach advocated in this paper takes these fusion categories as a starting point and is therefore particularly apt to deal with such non-abelian dualities. In fact, this demonstrates the suitability of the categorical approach to completely solve the non-abelian duality conundrum, at least for the one-dimensional case. Importantly, these non-abelian dualities are typically not self-dualities, as the corresponding dual symmetry realizations are in general distinct.

\subsection{Extensions}

\noindent
Within the same mathematical framework, our study can be extended in several directions: Firstly, we can account for choices of boundary conditions and establish how such choices interact with the duality relations. Secondly, we can systematically define the order and disorder operators of the (1+1)d models. As evoked in the main text, these are related to the anyons of the topological phase that shares the same input category theoretical data, and are described by the center $\mc Z(\mc C)$ of the fusion category $\mc C$ that describes the symmetries. Moreover, we can readily compute how these operators transform under the duality relations via the MPO intertwiners, which provides a lattice implementation of all possible braided equivalences $\mc Z(\mc C) \rightarrow \mc Z(\mc C')$. Thirdly, our construction naturally applies for Hamiltonians with longer range interactions as well. Indeed, relying on the recoupling theory of the underlying tensors, we can readily define bonds simultaneously acting on a larger number of sites. Fourthly, in our examples, we mainly focused on establishing exact duality transformations involving the complete Hilbert space of the theory. Exact dualities between theories may also \emph{emerge} in certain subspaces of the whole Hilbert space, such as the low-energy sector of a given model Hamiltonian \cite{cobanera2010unified,bondAlgebra}. Engineering \emph{emergent dualities} can be considered a theoretical tool to construct {\it quasi-exactly solvable models} \cite{Ushveridze1994,BatistaOrtiz2000,MoudgalyaMotrunich2022} starting from integrable ones, as illustrated in sec.~\ref{sec:RepS3}. Finally, duality relations of classical statistical mechanics models in terms of their transfer matrices can be investigated using the same tools. These studies will be reported in a follow-up manuscript.

The last example we considered in the main text, namely the IRF-Vertex correspondence, suggests possible extensions beyond the current framework. In particular, it seems possible to relax some of the defining properties of a spherical fusion category whilst still being able to consistently define dual models. For instance, it would be interesting to derive Hamiltonian models with symmetries associated with categories of representations over Lie groups (see for instance \cite{Etingof2003ModuleCO} for a discussion about module categories over $\msf{Rep}({\rm SL}(2))$). There are, in principle, subtleties associated to relaxing the finiteness constraint of fusion categories and working with \emph{tensor categories} instead. We expect that for models of physical interest, with compact Lie group symmetries, these subtleties become irrelevant and our framework still applies.

\subsection{Bulk-boundary correspondence}

\noindent
Dualities have been considered in the context of boundary theories of topological quantum field theories \cite{Thorngren:2019iar,Freed:2018cec,PhysRevResearch.2.043086}. One can think of a (1+1)d quantum field theory as living on the boundary of a (2+1)d topologically ordered system. For a given bulk topological order, one can typically define distinct boundary theories which through the bulk-boundary correspondence all encode the same physics. For a class of lattice models known as quantum doubles, this strategy was used to obtain certain (1+1)d quantum lattice models on their boundaries that are dual to one another \cite{PhysRevResearch.2.033417,Albert:2021vts}. 
It is important to note, however, that the formalism developed in our paper is significantly different since our dualities apply to generic Hamiltonians (gapped or gapless) which are not required to be the boundary theory of a higher-dimensional topologically ordered system. Most importantly, our construction is inherently algebraic and relies purely on the symmetry of the bond algebra to extract the relevant categorical structures. Nonetheless, tensor networks provide a very natural and precise language for constructing boundary theories \cite{PhysRevLett.112.036402}, and we expect one can extend bulk-boundary constructions to more general string-net models. Indeed, since it is known that gapped boundaries of string-net models are classified by module categories \cite{kongBdries,Fuchs:2012dt,KONG201762}, we foresee a relation to our classification of dualities. 

\subsection{Higher space dimensions}

\noindent
An even more tantalizing direction consists in generalizing our construction to (2+1)d quantum models. Formally, this amounts to considering a \emph{categorification} of the present construction, whereby the input category theoretical data are provided by a spherical fusion 2-category and a choice of a finite semi-simple module 2-category over it \cite{douglas2018fusion,gaiotto2019condensations}. The first step of such a generalization then amounts to defining the higher-dimensional analogues of the tensors given in eq.~\eqref{eq:FSymbols}. In the simple case where the input spherical fusion category is that associated with the (3+1)d toric code, we distinguish two tensor network representations that satisfy symmetry conditions w.r.t. string- and membrane-like operators, respectively \cite{delcamp2020tensor,williamson2020stability}. Furthermore, it was shown in \cite{delcamp2020tensor} that the corresponding intertwining tensor network realizes the duality between the (2+1)d transverse field Ising and Wegner's $\mathbb Z_2$-gauge theory \cite{PhysRev.60.252,RevModPhys.51.659, Fisher2004, RevModPhys.52.453,doi:10.1063/1.1665530}. Mimicking the example in sec.~\ref{sec:Ising}, we should be able to use these tensor network representations to explicitly reconstruct these Hamiltonians within our formalism.

A general framework for defining such higher-dimensional tensors and deriving the corresponding symmetry conditions is outlined in \cite{CDmod2} with an emphasis on generalizations of the (3+1)d toric code for arbitrary finite groups. Akin to the lower-dimensional analogues, we distinguish two canonical tensor network representations associated with module 2-categories labelling the so-called rough and smooth gapped boundaries of the topological models. Moreover, we also obtain duality maps as the intertwining operators between these tensor network representations, recovering results obtained in \cite{PhysRevX.5.011024}. Following the notations of eq.~\eqref{eq:gauging}, these duality maps are of the form
\begin{equation*}
    \; \includeTikz{0}{gaugingHigherD1}{\gaugingHigherD{1}}  \; = \;  \includeTikz{0}{gaugingHigherD2}{\gaugingHigherD{2}} \; .
\end{equation*}
Just as their one-dimensional analogues relate paramagnetic eigenstates to long-range ordered GHZ states (see end of sec.~\ref{sec:Ising}), these operators map symmetry protected topologically ordered states to states exhibiting intrinsic topological order \cite{PhysRevB.86.115109}. It will be very interesting to consider those questions from the point of view of fusion 2-categories, consider duality relations beyond that described above, and explicitly construct (2+1)d Hamiltonian models related by such duality relations.

%% file: main.bbl
%